\documentclass[%
 reprint,
 superscriptaddress,
 numerical,
 showpacs,
 amsmath,amssymb,
 aps,
 longbibliography,
 prd, 
 floatfix
]{revtex4-1}
\usepackage{graphicx}

\usepackage[%
  colorlinks=true,
  urlcolor=blue,
  linkcolor=blue,
  citecolor=blue
]{hyperref}

\usepackage[pdftex]{color}
\usepackage[usenames,dvipsnames,svgnames,table]{xcolor}

\usepackage{mathdots}

\usepackage{txfonts}

\usepackage{textcomp}

\newcommand{\Ch}{\mathrm{Ch}}

\DeclareMathOperator{\sgn}{sgn}
\DeclareMathOperator{\Pf}{Pf}
\DeclareMathOperator{\Tr}{Tr}

\begin{document}

\title{Second-Order Bulk-Boundary Correspondence in Rotationally Symmetric Topological Superconductors from Stacked Dirac Hamiltonians}

\author{Elis Roberts}
\affiliation{T.C.M. Group, Cavendish Laboratory, University of Cambridge, J.J. Thomson Avenue, Cambridge, CB3 0HE, United Kingdom}
\author{Jan Behrends}
\affiliation{T.C.M. Group, Cavendish Laboratory, University of Cambridge, J.J. Thomson Avenue, Cambridge, CB3 0HE, United Kingdom}
\author{Benjamin B\'{e}ri}
\affiliation{T.C.M. Group, Cavendish Laboratory, University of Cambridge, J.J. Thomson Avenue, Cambridge, CB3 0HE, United Kingdom}
\affiliation{DAMTP, University of Cambridge, Wilberforce Road, Cambridge, CB3 0WA, United Kingdom}

\begin{abstract}
Two-dimensional second-order topological superconductors host zero-dimensional Majorana bound states at their boundaries.
In this work, focusing on rotation-invariant crystalline topological superconductors, we establish a bulk-boundary correspondence linking the presence of such Majorana bound states to bulk topological invariants introduced by Benalcazar \textit{et al}.
We thus establish when a topological crystalline superconductor protected by rotational symmetry displays second-order topological superconductivity.
Our approach is based on stacked Dirac Hamiltonians, using which we relate transitions between topological phases to the transformation properties between adjacent gapped boundaries.
We find that in addition to the bulk rotational invariants, the presence of Majorana boundary bound states in a given geometry depends on the interplay between weak topological invariants and the location of the rotation center relative to the lattice.
We provide numerical examples for our predictions and discuss possible extensions of our approach.
\end{abstract}

\maketitle

\section{Introduction}

The topological classification of phases of matter is one of the cornerstones of modern condensed-matter physics~\cite{Schnyder:2008ez,Kitaev:2009bg,Ryu:2012en}.
Depending on their dimensionality and the presence of antiunitary symmetries, gapped noninteracting Hamiltonians may fall into topologically distinct sectors characterized by sets of topological invariants.
Crystalline symmetries enrich the classification of topological insulators and superconductors, giving rise to a wider class of materials, so-called crystalline topological insulators~\cite{Fu:2007ei,Fu:2011ia,Alexandradinata:2014jd,Slager:2013iv,Morimoto:2013cw,Chiu:2015ex,Kruthoff:2017jj,Bradlyn:2017fy,Po:2017ci}.
The interplay of crystalline and antiunitary symmetries makes the topological classification a challenging task, as there are for example 230~space groups in three dimensions, allowing for a plethora of symmetry-protected topological phases partially characterized by various symmetry indicators~\cite{Fu:2007ei,Teo:2013cp,Benalcazar:2014hb,Kruthoff:2017jj,Bradlyn:2017fy,Po:2017ci,Liu:2014kk,Fang:2012dn,Fang:2013jk,Fang2017}.

One main goal of the symmetry classification of topological insulators and superconductors is to establish a correspondence between the invariants defined in the bulk and in-gap states that arise at the surfaces~\cite{Hatsugai:1993fc,Hasan:2010ku,Qi:2011hb}.
In crystalline topological insulators, this bulk-boundary correspondence links the bulk invariants to gapless modes at surfaces that respect the underlying spatial symmetries~\cite{Fu:2007ei,Fu:2011ia,Alexandradinata:2014jd}.

Spatial symmetries may also give rise to so-called higher-order topological insulators and superconductors.
These phases have gapped boundaries, but host ``higher-order boundary modes'': gapless boundary-excitations of codimension greater than one, e.g., bound to their hinges or corners~\cite{Benalcazar:2017ks,Benalcazar:2017cn,Langbehn:2017jn,Schindler:2018hi}.
Higher-order topological phases have been studied in systems protected by order-two symmetries (e.g., reflection and inversion symmetry~\cite{Benalcazar:2017ks,Benalcazar:2017cn,Langbehn:2017jn,Kunst:2018fi,Geier:2018ev,Trifunovic:2019hi}), rotational invariance~\cite{Song:2017ev,Schindler:2018hi,Benalcazar:2019bs}, and combinations of the above~\cite{VanMiert:2018cb,Bultinck:2019cn,Schindler:2018hl}.
Gapless hinge and corner excitations may also appear in interacting models~\cite{Dwivedi:2018kp,You:2019kr}, Floquet phases~\cite{Rodriguez:2019dj,Chaudhary2019} and can coexist with gapless surface states~\cite{Ghorashi:2019cj}.
Higher-order topology does not necessarily rely on an underlying regular lattice~\cite{Agarwala:2019}, but can be also found in quasicrystals respecting certain spatial symmetries~\cite{Varjas:2019jd,Chen:2020jo}.
Corner modes have been found experimentally in various metamaterials, including phononic lattices~\cite{Serra:2018dj}, engineered electronic lattices~\cite{Kempkes:2019cl}, topolectrical~\cite{Imhof:2018gs} and microwave circuits~\cite{Peterson:2018kz}.
Strong experimental evidence further suggests the existence of hinge modes in bismuth~\cite{Schindler:2018hl}.

In higher-order topological phases, the presence and robustness of boundary modes depends on how the underlying spatial symmetries transform the degrees of freedom of neighboring surfaces into another.
This raises the question of how to relate this transformation of neighboring surfaces to topological invariants defined in the bulk.
Establishing this relationship amounts to deriving a bulk-boundary correspondence in a manner that keeps the role of the defining symmetries transparent.
This has been the guiding principle behind recent work relating symmetry indicators to higher-order boundary modes
in insulators~\cite{Khalaf:2018hq,Schindler:2019jk}, and it has also been a key element in the work of Trifunovic and Brouwer establishing the bulk-boundary correspondence for higher-order topological phases with order-two symmetries in the absence of weak (i.e., lower dimensional) invariants~\cite{Geier:2018ev,Trifunovic:2019hi}.
Here we describe how such a bulk-boundary correspondence program can be carried out beyond these cases, focusing on two-dimensional (2D) crystalline superconductors with $n$-fold rotational symmetry (i.e., $C_n$ symmetry), and allowing for nonvanishing weak invariants.
Establishing a link between edge transformation properties and bulk invariants provides an illuminating perspective complementary to counting arguments based on bulk defect classifications~\cite{Teo:2013cp,Benalcazar:2014hb}, and gives results consistent with examples based on very recent extensions of symmetry indicators to the superconducting classes~\cite{Shiozaki2019,Geier2019}.

Our approach is based on an effective description in terms of stacked Dirac models~\cite{Liu:2014kk,Khalaf:2018hq}.
Using this, we show that rotational invariance dictates a relationship between adjacent surfaces and that this may give rise to protected second-order boundary modes in the form of Majorana bound
states.
We express this bulk-boundary correspondence in terms of the bulk invariants for rotationally symmetric crystalline superconductors developed in Ref.~\onlinecite{Benalcazar:2014hb} and an additional contribution signifying the combined effects of weak topological invariants and the physical rotation center.
While our considerations are general, for the purposes of a detailed exposition we will be focusing on $C_4$-symmetric systems: of the $C_2$, $C_3$, $C_4$, and $C_6$ symmetries possible in 2D crystals, the $C_4$-symmetric case is the one displaying the richest combination of stacked Dirac and second-order topological superconducting features.
(We shall comment on applying our methods to the other cases in the Appendices.)
To demonstrate the validity of our stacked Dirac approach, we also illustrate our results on several concrete lattice models.

In what follows, for brevity we shall refer to the second order Majorana bound states we find as corner modes, even though rotational symmetry does not, strictly speaking, require them to be at the geometrical corners of the system: Their position can be moved in a rotation-symmetric manner e.g., by adding suitable Kitaev chains to the boundary~\cite{Schindler:2018hi,Teo:2013cp}.
However, such a deformation merely shifts the Majorana bound states around the boundary without altering their position relative to each other and, as such, it cannot gap out the Majoranas.
In what follows, the term corner mode should thus be understood up to such Kitaev chain deformations.

This paper is organized as follows:
After briefly summarizing the symmetry classification of rotationally invariant superconductors~\cite{Benalcazar:2014hb} in Sec.~\ref{sec:benalcazar_classification}, we introduce our stacked Dirac model based approach in Sec.~\ref{sec:stacked_dirac}.
We present an effective edge theory and consider the most general mass terms that gap out the edge modes.
To relate the bulk description to the boundaries, we relate the topologically distinct rotation properties of the boundary mass term to the bulk invariants in Sec.~\ref{sec:bulk_invariants}.
We show some explicit examples in Sec.~\ref{sec:examples} and conclude in Sec.~\ref{sec:conclusion}.
In the Appendices, we clarify the role of the unit cell and explicitly derive the edge Hamiltonian, as well as outline how this approach is applied to $ C_2 $ and $ C_6 $ symmetry.

\section{Bulk Topological Classification}
\label{sec:benalcazar_classification}

We work with 2D topological crystalline superconductors in class D.
These are particle-hole (PH) symmetric systems which admit an effectively single-particle Bogoliubov-de-Gennes (BdG) Hamiltonian.
Working in momentum space and denoting this BdG Hamiltonian by $ H(\mathbf{k}) $ at momentum $ \mathbf{k} $ in the Brillouin zone (BZ), PH symmetry is given by
\begin{equation}
\Xi H(\mathbf{k}) \Xi^{-1} = - H(-\mathbf{k}),
\label{eq:PH_symmetry}
\end{equation}
where $ \Xi $ is an antiunitary operator satisfying $ \Xi^2 = +1 $.

The presence of an additional $n$-fold rotational symmetry $ C_n $ allows for a richer topological classification~\cite{Benalcazar:2014hb,Fang2017,Song:2017ev} than if the only symmetry was PH symmetry.
We now review the classification scheme devised by Benalcazar \textit{et al.\@} for classifying crystalline superconductors with rotational symmetry~\cite{Benalcazar:2014hb}.
This scheme was an early example of a symmetry indicator approach to classifying crystalline topological phases, which is to use the symmetry representations of occupied bands at high symmetry points in the BZ~\cite{Fu:2007ei,Fang:2012dn,Fang:2013jk,Teo:2013cp}.

\begin{figure}
\centering
\includegraphics[width=\linewidth]{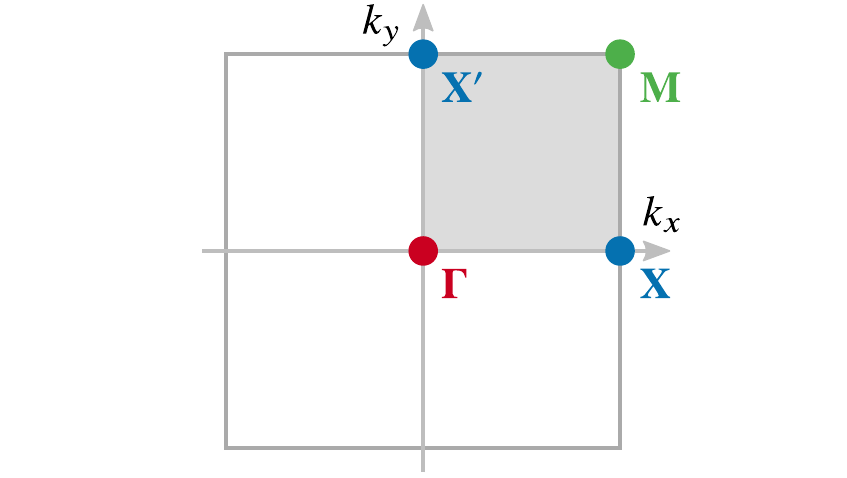}
\caption{The Brillouin zone for $ C_4 $-symmetric models. There are two fourfold fixed points labeled $ \boldsymbol{\Gamma} $ and $ \mathbf{M} $, and two twofold fixed points $ \mathbf{X} $ and $ \mathbf{X'} $ that transform into each other upon a fourfold rotation. The shaded region indicates the fundamental domain that generates the entire BZ.}
\label{fig:BZ}
\end{figure}

\begin{figure}
\centering
\includegraphics[width=\linewidth]{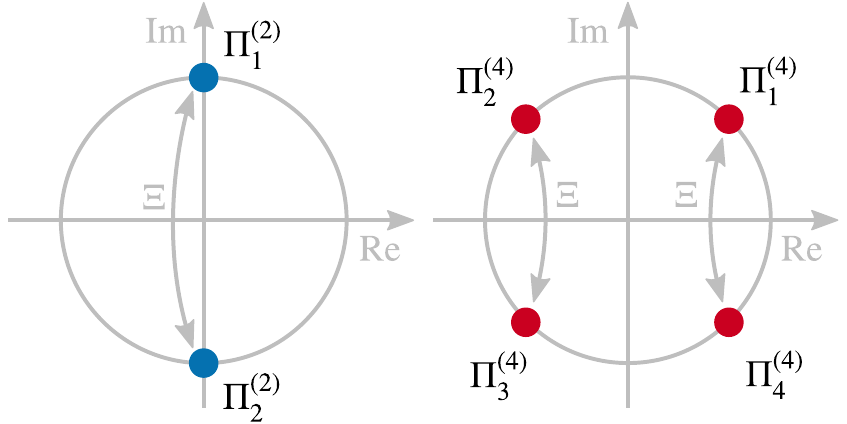}
\caption{The rotation eigenvalues for $ C_2 $ and $ C_4 $ symmetry, respectively. The PH operator relates complex conjugate pairs of rotation eigenvalues (while also switching between positive and negative energy bands).}
\label{fig:rotation_eigenvalues}
\end{figure}

The rotational symmetry of the model is expressed through the relation
\begin{equation}
r_n H(\mathbf{k}) r_n^{\dagger} = H(R_n \mathbf{k}),
\label{eq:bulk_symmetry}
\end{equation}
where $ r_n $ is an $n$-fold rotation operator obeying $ r_n^n=-1 $ and $ R_n $ is the $\text{SO}(2)$ matrix for $n$-fold rotations in the 2D plane.
Since $ r_n $ conserves charge, it commutes with the PH symmetry operator $ [\Xi, r_n] = 0$~\cite{Benalcazar:2014hb}.
(There are some subtleties to this statement if Cooper pairs have nonzero angular momentum, which we discuss in Appendix~\ref{subsec:pairing_symmetry}.)
The $ r_n^n=-1 $ requirement comes from the fact that $ r_n $ is a single particle operator acting on a particle with half-odd-integer spin (a fermion), for which a Berry phase of $ -1 $ is acquired under a full $ 2 \pi $ rotation.
Since these are \emph{crystalline} superconductors, the BZ contains certain high-symmetry points (HSPs) $ \boldsymbol{\Pi}^{(n)} $, which are invariant under rotation $ R_n \boldsymbol{\Pi}^{(n)} = \boldsymbol{\Pi}^{(n)} $ up to a reciprocal lattice vector.
At these points, the rotational symmetry is simply $ [r_n, H(\boldsymbol{\Pi}^{(n)})]=0$, and as such the momentum eigenstates can be chosen as eigenstates of the rotation operator.
This allows us to label each state at $ \boldsymbol{\Pi}^{(n)} $ with its rotation eigenvalue
\begin{align}
\Pi_p^{(n)} = e^{i\pi(2p-1)/n},& & \text{for } p=1,2,\ldots n.
\label{eq:label_eigenvalues}
\end{align}
For example, in a $ C_4 $-symmetric BZ (shown in Figure~\ref{fig:BZ}), there are both fourfold and twofold fixed points, whose rotation eigenvalues are shown in Figure~\ref{fig:rotation_eigenvalues}.

We now outline how these rotation eigenvalues are used to topologically classify gapped superconductors in two dimensions.
We start by defining a trivial superconductor as one that can be connected to a superconductor in the atomic limit without closing the gap and while respecting the same crystalline and PH symmetries throughout.
Here the `atomic limit' is understood as a symmetry-respecting array of zero-dimensional superconductors~\cite{Geier2019}.
For crystal structures compatible with a symmetric boundary, as we note is required for a generic bulk-boundary correspondence (see Appendix~\ref{sec:unit_cell_restriction}), a unit cell can be chosen such that an atomic limit superconductor's ground state wave function has no momentum-dependent features.
With this more restricted definition of topological equivalence (because it involves obeying an additional unitary symmetry), the boundary between two inequivalent phases does not necessarily possess edge states, as we explore in this paper.
For the ground state wave function to have no momentum-dependent features, the rotation eigenvalues of the negative energy states (which, within the BdG description, are all occupied in the ground state) must be the same at all HSPs in the BZ.
This motivates the definition of the topological invariants as 
\begin{equation}
 [\Pi_p^{(n)}] \equiv \#\Pi_p^{(n)} - \#\Gamma_p^{(n)},
\label{eq:invariant_definitions}
\end{equation}
where $ \#\Pi_p^{(n)} $ is the number of negative energy BdG bands with eigenvalue $ \Pi_p^{(n)} $.
Intuitively, these are chosen because occupancies of rotation eigenvalues will not change unless there is a gap closing, and taking the difference relative to a reference momentum [chosen as $\boldsymbol{\Gamma} $ in Eq.~\eqref{eq:invariant_definitions}] is required for the invariants to be stable under the addition of  trivial bands.
Under this definition, a $C_n$-symmetric superconductor is topological if $[\Pi_p^{(n)}]$ is nonzero for any $ p $.

A complete topological characterization requires establishing the set of independent $[\Pi_p^{(n)}]$.
They are not all independent because rotational symmetry constrains the rotation eigenvalues at $C_n$-related points in the BZ to be the same (e.g., the $C_2$ eigenvalues of the $C_4$-related $\mathbf{X}$ and $\mathbf{X'}$ in the fourfold case, shown in Fig.~\ref{fig:BZ}).
PH symmetry places further restrictions on these invariants, since if the rotation eigenvalue of a state is $ \Pi_p^{(n)} $, its PH-conjugate state has eigenvalue $ \Pi_p^{(n)*} = \Pi_{n-p+1}^{(n)} $.
That is to say, the number of occupied eigenvalues $ \Pi_p^{(n)} $ is equal to the number of \emph{unoccupied} eigenvalues $ \Pi_{n-p+1}^{(n)} $, which implies
\begin{equation}
 [\Pi_p^{(n)}] = - [\Pi_{n-p+1}^{(n)}].
 \label{eq:PH_invariant_relations}
\end{equation}

For $ C_4 $-symmetric systems there are three independent rotation invariants \cite{Benalcazar:2014hb},
\begin{subequations}
\begin{align}
 [X]   & \equiv \#X_1 - (\#\Gamma_1 + \#\Gamma_3),\\
 [M_1] & \equiv \#M_1 - \#\Gamma_1, \\
 [M_2] & \equiv \#M_2 - \#\Gamma_2,
\end{align}\label{eq:rotation_invariants}\end{subequations}
which, in conjunction with the Chern number $ \Ch $, fully classify the bulk topology in this symmetry class.

\subsection{Importance of Rotation Center}
\label{subsec:rotation_center}

\begin{figure}
\centering
\includegraphics[width=\linewidth]{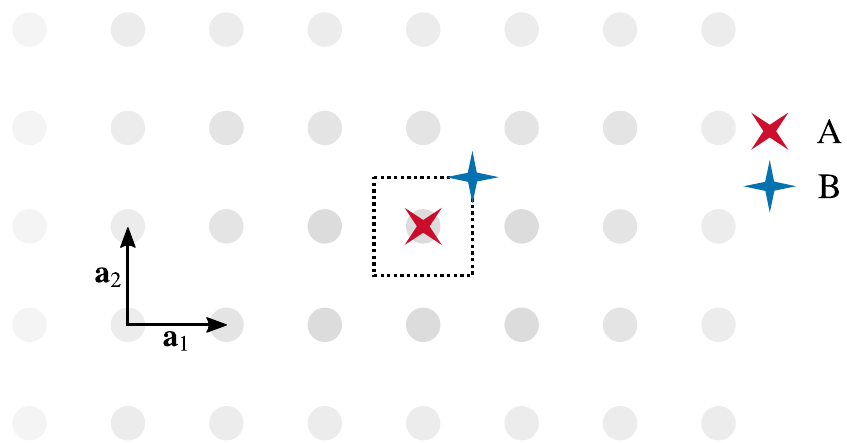}
\caption{Two options (A and B) for the rotation center in an infinite $ C_4 $-symmetric lattice. The dotted box shows a primitive unit cell with its associated lattice site in the middle. Case B has its rotation operator shifted by $ \mathbf{c} = \frac{1}{2} (\mathbf{a}_1 + \mathbf{a}_2) $}
\label{fig:rot_center_options}
\end{figure}

In the previous section, we started with the rotational symmetry relation Eq.~\eqref{eq:bulk_symmetry}, but a system with periodic boundary conditions can have many centers of rotation~\cite{Fang:2013jk,Mondragon-Shem2019}, as exemplified in Fig.~\ref{fig:rot_center_options}.
Although operators implementing rotation about different centers are easily related through composition with translation operators, the classification of periodic Hamiltonians summarized above relies on a momentum-independent rotation operator \cite{Benalcazar:2014hb}, which can only be true for one of the rotation centers.
Since a finite system with boundaries may only satisfy rotational symmetry about one of the rotation centers, the physical symmetry operator relating different edges of a finite system may be \emph{different} (but closely related) to the symmetry operator used to classify periodic Hamiltonians in Ref.~\onlinecite{Benalcazar:2014hb}.
In this section we explicitly relate these distinct rotation operators in the case of $ C_4 $ symmetry, which allows for two rotation centers that we dub A and B.

\subsubsection{Rotation Center A}

We now explicitly derive the rotation operator in momentum-space for case A depicted in Fig.~\ref{fig:rot_center_options}, in similar spirit to Ref.~\onlinecite{Fang:2013jk}.
Let lattice sites be situated at $\mathbf{R}=n_{1}\mathbf{a}_{1}+n_{2}\mathbf{a}_{2}$, where $n_{i}\in \mathbb{Z}$ are integer coefficients of primitive lattice vectors $ \mathbf{a}_i $.
Associated with each lattice site are orbitals $\alpha$ located at atomic positions $\mathbf{d}_{\alpha}$ within a unit cell, such that many orbitals may share the same atomic position.
Consider the position of a particular orbital, given by $\mathbf{r}\equiv\mathbf{R}+\mathbf{d}_{\alpha}$.
Let $ R_A $ be defined as a pure rotation $ R_n $ about the origin which coincides with a lattice site.
If this operation is to be a symmetry, then an atom located at $ \mathbf{r} $ must be mapped to another atomic site so that
\begin{equation}
R_A:~ \mathbf{r} \to R_n \mathbf{r} = R_n (\mathbf{R}+\mathbf{d}_{\alpha}) = \mathbf{R'}+\mathbf{d}_{\beta},
\end{equation}
for some other lattice point $\mathbf{R'}$ and atomic site $\mathbf{d}_{\beta}$.
A key point to note is that for certain lattices one \emph{cannot} choose a basis in the unit cell such that $\mathbf{R'}=R_n \mathbf{R} $ for all $ \mathbf{d}_{\alpha} $, as we soon explain in more detail.
In second-quantized notation the operator $\hat{R}_A$ changes the position of each atomic orbital as
\begin{equation}
\hat{R}_A^{}\hat{c}_{\alpha}^\dagger (\mathbf{R}+\mathbf{d}_\alpha) \hat{R}_A^{-1}
=\hat{c}_{\beta}^\dagger (\mathbf{R'}+\mathbf{d}_\beta) \mathcal{R}_{\beta\alpha},
\end{equation}
which includes a unitary matrix $\mathcal{R}_{\beta\alpha}$ (with implicit summation over orbitals $ \beta $) to account for rotation amongst atomic orbitals, whose elements $ \mathcal{R}_{\beta\alpha} $ are only nonzero when $ \mathbf{d}_{\alpha} \to \mathbf{d}_{\beta} $.
Our Fourier transform convention has the periodic phase taken with respect to lattice sites, such that momentum space operators are given by~\footnote{This Fourier transform convention implies a certain choice of basis functions~\cite{Alexandradinata:2014ju,Alexandradinata:2016kb}. For details on the basis choice cf.\ Appendix~\ref{sec:unit_cell_restriction}.}
\begin{equation}
\hat{c}_{\alpha}^\dagger (\mathbf{k})=\frac{1}{\sqrt{N}} \sum_{\mathbf{R}} \hat{c}_{\alpha}^\dagger (\mathbf{R}+\mathbf{d}_\alpha) \exp(i\mathbf{k}\cdot\mathbf{R}),
\end{equation}
which transform as~\cite{Fang:2013jk}
\begin{align}
\hat{R}_A^{} & \hat{c}_{\alpha}^\dagger (\mathbf{k}) \hat{R}_A^{-1}
   = \frac{1}{\sqrt{N}} \sum_{\mathbf{R}} \hat{R}_A \hat{c}_{\alpha}^\dagger (\mathbf{R}+\mathbf{d}_\alpha)\hat{R}_A^{-1}\exp(i\mathbf{k}\cdot\mathbf{R}) \nonumber \\
 & =\frac{1}{\sqrt{N}}\sum_{\mathbf{R}} \hat{c}_{\beta}^\dagger (\mathbf{R'}+\mathbf{d}_\beta) \mathcal{R}_{\beta\alpha} \exp(i\mathbf{k}\cdot\mathbf{R}) \nonumber \\
 & =\frac{1}{\sqrt{N}}\sum_{\mathbf{R'}} \hat{c}_{\beta}^\dagger (\mathbf{R'}+\mathbf{d}_\beta) \mathcal{R}_{\beta\alpha}e^{i\mathbf{k}\cdot(R_n^{T}\mathbf{R'}+R_n^{T}\mathbf{d}_{\beta}-\mathbf{d}_{\alpha})} \nonumber \\
 & = \hat{c}_{\beta}^\dagger (R_n \mathbf{k}) \mathcal{R}_{\beta\alpha} e^{i (R_n \mathbf{k})\cdot(\mathbf{d}_{\beta}-R_n \mathbf{d}_{\alpha})}. \label{eq:momentum_rotation}
\end{align}
This shows that the basis spinors $ \hat{\xi}_{\mathbf{k}}^\dagger = \bigoplus_{\alpha} \left(\hat{c}_{\alpha}^\dagger (\mathbf{k}),\hat{c}_{\alpha} (-\mathbf{k}) \right)$ of the second-quantized Hamiltonian
\begin{equation}
\hat{H} = \int \frac{d^2 \mathbf{k}}{(2\pi)^2}\,\hat{\xi}_{\mathbf{k}}^{\dagger} H(\mathbf{k}) \hat{\xi}_{\mathbf{k}}^{},
\end{equation}
transform as
\begin{align}
\hat{R}_A^{} \hat{\xi}_{\mathbf{k}}^{\dagger} \hat{R}_A^{-1}
&= \bigoplus_{\alpha} \left( \hat{R}_A^{} \hat{c}_{\alpha}^{\dagger}(\mathbf{k}) \hat{R}_A^{-1},\, \hat{R}_A^{} \hat{c}_{\alpha} (-\mathbf{k}) \hat{R}_A^{-1} \right) \nonumber \\
&= \bigoplus_{\alpha} \left( \hat{c}_{\beta}^{\dagger} (R_n \mathbf{k}) \mathcal{R}_{\beta\alpha}, \hat{c}_{\beta} (-R_n \mathbf{k}) \mathcal{R}_{\beta\alpha}^{*} \right) e^{i (R_n\mathbf{k}) \cdot(\mathbf{d}_{\beta}-R_n\mathbf{d}_{\alpha})} \nonumber \\
&\equiv \hat{\xi}_{R_n \mathbf{k}}^{\dagger} r_n (\mathbf{k}).
\end{align}
(Note that Refs.~\onlinecite{Ono2019,Geier2019} consider more general phase choices for the matrix $\mathcal{R}^*$ in the hole sector depending on the symmetry of the superconducting pairing term, which we discuss in Appendix~\ref{subsec:pairing_symmetry}.)
Rotational invariance of the second-quantized Hamiltonian $\hat{R}_A^{} \hat{H} \hat{R}_A^{-1} = \hat{H}$ implies that the Bloch Hamiltonian needs to satisfy
\begin{equation}
r_n^{}(\mathbf{k}) H(\mathbf{k}) r_n^{\dagger}(\mathbf{k}) = H (R_n \mathbf{k}).
\label{eq:bulk_symmetry_momentum}
\end{equation}

In general, the momentum-dependent part $ e^{i(R_n \mathbf{k})\cdot(\mathbf{d}_{\beta}-R_n^{T}\mathbf{d}_{\alpha})} $ of $ r_n^{}(\mathbf{k}) $ is not a complex phase universal to all orbitals because $ \mathbf{d}_{\beta} $ may lie in a different unit cell than $ R_n^{} \mathbf{d}_{\alpha} $ for certain $ \alpha $.
This occurs, for example, when atoms are situated at the edge of a unit cell.
If it is impossible to define a $ C_n $-symmetric unit cell without atoms on the edges of the cell, then the rotation operator is generally not momentum-independent and the classification scheme of Ref.~\onlinecite{Benalcazar:2014hb} (in its current form) does not hold; cf.\ Appendix~\ref{sec:unit_cell_restriction} for details.
Lattices of the same type also present an impediment for formulating a bulk-boundary correspondence:
When the atomic sites lie at the edge of a unit cell, it is impossible to tile a finite rotationally symmetric system without resorting to an extensive number of partial unit cells at the boundary.
Any classification scheme for such a system would be non-generic, as it needs to take into account the lattice termination.
When all atoms lie wholly within the unit cell, which respects rotational invariance individually, we can indeed have $ R_n \mathbf{d}_{\alpha} = \mathbf{d}_{\beta} $ for all orbitals $ \alpha $ such that $ e^{i (R_n \mathbf{k}) \cdot(\mathbf{d}_{\beta}-R_n \mathbf{d}_{\alpha})} = 1 $, recovering Eq.~\eqref{eq:bulk_symmetry} with $ r_n (\mathbf{k}) \to r_n $.
Henceforth, when simply referring to the unit cell, we shall be working with this restricted $C_n$-symmetric unit cell notion that allows for a well-defined bulk-boundary correspondence.

\subsubsection{Rotation Center B}

Now consider a different operation $ R_B $ which consists of a pure rotation $ R_n $ about a different center which is shifted by a vector $ \mathbf{c} $.
Different lattices have different options for $ \mathbf{c} $ as long as rotation about $ \mathbf{c} $ maps lattice sites to other lattice sites.
In the case of $C_4$ symmetry there is only the option of $ \mathbf{c} = \frac{1}{2}(\mathbf{a}_1 + \mathbf{a}_2) $, shown in Fig.~\ref{fig:rot_center_options}.
Lattice sites are still situated at $ \mathbf{R}=n_{1}\mathbf{a}_{1}+n_{2}\mathbf{a}_{2} $, and we again consider a particular orbital located at position $\mathbf{r}=\mathbf{R}+\mathbf{d}_{\alpha}$.
The rotation $R_B$ then changes each position
\begin{align}
 R_B: \mathbf{r} \to R_n (\mathbf{r}-\mathbf{c}) + \mathbf{c}
 &= R_n (\mathbf{R} + \mathbf{d}_\alpha) + (1-R_n) \mathbf{c} \\
 &= \mathbf{R'} + \mathbf{d}_\beta + (1-R_n) \mathbf{c},
\end{align}
i.e., it can be considered a combination of the rotation $R_A$ about the origin and an additional translation by $(1-R_n) \mathbf{c}$.
In second-quantized notation, the rotation changes the creation operators
\begin{equation}
\hat{R}_B\hat{c}_{\alpha}^\dagger (\mathbf{R} + \mathbf{d}_\alpha) \hat{R}_B^{-1}
= \hat{c}_{\beta}^\dagger (\mathbf{R'}+\mathbf{d}_\beta +(1 - R_n) \mathbf{c}) \mathcal{R}_{\beta\alpha},
\end{equation}
which includes the additional translation by a lattice vector $ (1 - R_n) \mathbf{c} $.
It is important to note that even with this shift, the transformation of orbitals into each other is the same as before, i.e., $ \mathcal{R}_{\alpha \beta} $ is the same as it was for $ R_{A} $.
With the same Fourier transform convention, we see that the momentum space operators now transform as
\begin{align}
 & \hat{R}_B^{} \hat{c}_{\alpha}^\dagger (\mathbf{k})\hat{R}_B^{-1}
 =\frac{1}{\sqrt{N}}\sum_{\mathbf{R}}\hat{R}_B\hat{c}_{\alpha}^\dagger (\mathbf{R}+\mathbf{d}_\alpha)\hat{R}_B^{-1}\exp(i\mathbf{k}\cdot\mathbf{R}) \nonumber \\
 & =\frac{1}{\sqrt{N}}\sum_{\mathbf{R}} \hat{c}_{\beta}^\dagger (\mathbf{R'}+\mathbf{d}_\beta +(1 - R_n) \mathbf{c}) \mathcal{R}_{\beta\alpha} e^{i\mathbf{k}\cdot\mathbf{R}} \nonumber \\
 & =\frac{1}{\sqrt{N}}\sum_{\mathbf{R'}} \hat{c}_{\beta}^\dagger (\mathbf{R'}+\mathbf{d}_\beta +(1 - R_n) \mathbf{c}) \mathcal{R}_{\beta\alpha} e^{i\mathbf{k}\cdot(R_n^T (\mathbf{R'}+ \mathbf{d}_{\beta})-\mathbf{d}_{\alpha})} \nonumber \\
 & =\frac{1}{\sqrt{N}}\sum_{\mathbf{R'}} \hat{c}_{\beta}^\dagger (\mathbf{R'} +\mathbf{d}_\beta) \mathcal{R}_{\beta\alpha} e^{i(R_n\mathbf{k}) \cdot \mathbf{R'}} e^{i \mathbf{k} \cdot ( R_n^T \mathbf{d}_{\beta}-\mathbf{d}_{\alpha}- (R_n^T - 1) \mathbf{c})} \nonumber \\
 & = \hat{c}_{\beta}^\dagger (R_n \mathbf{k}) \mathcal{R}_{\beta\alpha} e^{i (R_n\mathbf{k})\cdot(\mathbf{d}_{\beta}-R_n \mathbf{d}_{\alpha})}  e^{-i \mathbf{k} \cdot (R_n^T - 1)\mathbf{c}}.
 \label{eq:shifted_momentum_rotation}
\end{align}
Comparing Eq.~\eqref{eq:shifted_momentum_rotation} to Eq.~\eqref{eq:momentum_rotation}, we see that the rotation operators are related by a momentum-dependent phase
\begin{equation}
\hat{R}_B^{}\hat{c}_{\alpha}^\dagger (\mathbf{k}) \hat{R}_B^{-1} = \hat{R}_A \hat{c}_{\alpha}^\dagger (\mathbf{k}) \hat{R}_A^{-1}\, e^{-i \mathbf{k} \cdot (R_n^T - 1)\mathbf{c}}.
\end{equation}
This extra phase \emph{is} $\alpha$-independent, so that the basis spinors transform as
\begin{align}
\hat{R}_B^{} \hat{\xi}_{\mathbf{k}}^{\dagger} \hat{R}_B^{-1} &=  \hat{\xi}_{R_n \mathbf{k}}^{\dagger} r_n e^{- i \mathbf{k} \cdot (R_n^T - 1)\mathbf{c}} \\
&\equiv \hat{\xi}_{R_n \mathbf{k}}^{\dagger} r_{n,\mathbf{c}} (\mathbf{k}),
\label{eq:shifted_rotation}
\end{align}
where we introduce a new notation for the rotation operator such that $ r_n \equiv r_{n,\mathbf{c}=\mathbf{0}} $.
Of note is that the momentum-independence of $ r_n $ necessarily implies a momentum-dependence for $ r_{n,\mathbf{c}\neq\mathbf{0}}(\mathbf{k}) $.
The symmetry relation of the Bloch Hamiltonian is indifferent to this complex phase and is still given by Eq.~\eqref{eq:bulk_symmetry_momentum}.

\subsubsection{Physical Rotation Operator}

When a superconducting Hamiltonian is terminated, only one of $ \hat{R}_A $ or $ \hat{R}_B $ can be a symmetry of the whole system $ \hat{H} $ since both rotation centers are mutually incompatible.
Thus, eigenstates of $ \hat{H} $ are simultaneously eigenstates of either $ \hat{R}_A $ or $ \hat{R}_B $.
For the bulk (not terminated) system, at HSPs $ \boldsymbol{\Pi}^{(n)} $ in momentum space, one has $ r_{n,\mathbf{c}}(\boldsymbol{\Pi}^{(n)}) = \pm \, r_n $ because $ e^{i \boldsymbol{\Pi}^{(n)} \cdot (R_n^{T} - 1)\mathbf{c}} = \pm 1 $ for any valid rotation center $ \mathbf{c} $, though $ r_{n,\mathbf{c}}(\boldsymbol{\Gamma}) = r_n $ always.
When we construct an effective bulk theory in the next section, references to the rotation operator are always to the operator $ r_n $ used to classify bulk Hamiltonians, but when we proceed to deriving the rotational symmetry of the edge theory we need to consider the physical rotation operator $ r_{n,\mathbf{c}}(\mathbf{k}) $.

\section{Stacked Dirac Models and Boundary Theory}
\label{sec:stacked_dirac}

We seek a mapping from the full classification of Ref.~\onlinecite{Benalcazar:2014hb} summarized in Sec.~\ref{sec:benalcazar_classification} to the second-order boundary signature.
We consider superconductors without conventional gapless edge states, therefore we focus on the $ \Ch = 0 $ case of vanishing Chern
number.
As stated in the Introduction, our exposition is focused on $C_4$ symmetry; the modifications required to treat $C_2$ and $C_6$ cases are discussed in Appendix~\ref{sec:other_symmetries}.
Since Majorana modes must always come in pairs, a $ C_3 $-symmetric system is not able to sustain unpaired Majoranas on its three corners, so we ignore this case entirely.

Our approach is the construction of a continuum model which allows us to describe interfaces between systems with different topological invariants, reminiscent of a Jackiw-Rebbi approach~\cite{Jackiw:1976ky,Su:1979cb}.

\subsection{Stacked Dirac Models}
\label{subsec:stacked_dirac}

We determine the boundary signature for each topological phase based on a description near the gap closing transitions that change the topology.
The previously defined invariants $[X]$, $[M_1]$ and $[M_2]$ only change for gap closings at HSPs $ \boldsymbol{\Pi}^{(n)} $, though $ \Ch $ also changes for gap closings at any generic momenta.
Due to $C_4$ symmetry, gap closings at generic momenta $\mathbf{k}_0 $ (not HSPs) must come in multiplets of four (at $R_4^j \mathbf{k}_0$ with $j=0,1,2,3$), which changes the Chern number by $\pm 4$.
As these gap closings at generic momenta can be smoothly shifted to a high-symmetry point, henceforth we consider that all gap closings occur at the HSPs $\boldsymbol\Pi^{(n)}$.

Near a transition at a HSP $\boldsymbol{\Pi}_\alpha$, a natural description is provided by a massive 2D Dirac Hamiltonians $\mathcal{H}_{\boldsymbol{\Pi}_\alpha}^{\alpha}(\mathbf{k})$, with a sign change of the mass across the interface modeling a boundary between regions with different values of their bulk topological invariants.
(The momentum $\mathbf{k}$ here is understood relative to $\boldsymbol{\Pi}_\alpha $.)
We will then link the rotation properties encoded in the rotation invariants of Eq.~\eqref{eq:rotation_invariants} to properties of these Dirac fermions.
Working with a Dirac model means that our anticipated bulk-boundary correspondence will be in terms of the \emph{difference} between topological phases, which indeed is the most general scenario to which a bulk- boundary correspondence can apply \cite{Hasan:2010ku}.
For any change in topological phase there are multiple possible stacked Dirac realizations, but we will show that the boundary signature follows from a feature common to all of these realizations.

The effective model $ \mathcal{H}(\mathbf{k}) $ is the direct addition of all these Dirac Hamiltonians, which we refer to as a ``stack'' of Dirac models $ \mathcal{H}^{\alpha}_{\boldsymbol{\Pi}_{\alpha}}(\mathbf{k}) $, written as
\begin{equation}
H(\mathbf{k}) \to \mathcal{H}(\mathbf{k}) = \bigoplus_{\alpha} \mathcal{H}^{\alpha}_{\boldsymbol{\Pi}_{\alpha}}(\mathbf{k}).
\label{eq:Dirac_stack}
\end{equation}
Physically, this corresponds to stacking many systems together and leaving them decoupled, but with the overall system remaining 2D.
We have introduced a (redundant) label $ \boldsymbol{\Pi}_{\alpha} $ to emphasize the origin of each Dirac Hamiltonian for clarity.
Each Dirac model has the same chirality \footnote{The $ \mathbf{k} \cdot \boldsymbol{\sigma}$ term may always be brought to this form because the relative sign of $ k_1 $ and $ k_2 $ is altered by a basis rotation $ \mathcal{H}^{\alpha}(\mathbf{k}) \to \sigma_1 \mathcal{H}^{\alpha}(\mathbf{k}) \sigma_1 $ for which we would change the sign assigned to $ m_{\alpha} $.} and is described by a Hamiltonian of the form
\begin{equation}
\mathcal{H}^{\alpha}_{\boldsymbol{\Pi}_{\alpha}}(\mathbf{k}) = v_{\alpha} \mathbf{k} \cdot\boldsymbol{\sigma}+ m_{\alpha}\sigma_{3}
\label{eq:Dirac_single}
\end{equation}
respecting PH symmetry [Eq.~\eqref{eq:PH_symmetry}] with $\Xi = \sigma_1 \mathcal{K}$, where $ \boldsymbol{\sigma} = (\sigma_1, \sigma_2) $ is a 2D vector of Pauli matrices.
Each Hamiltonian has its own (possibly distinct) positive velocity $ v_{\alpha} $ (chosen to be isotropic for simplicity), and the parameters $ m_{\alpha} $ control the band separations of each Dirac model.
Other off-diagonal mass terms are in principle allowed by symmetry, but to streamline our discussions we choose to include these later among the allowed terms for the edge theory~\footnote{Allowing symmetric terms of the form $ \sigma_3 \otimes M $, for example, where $ M = M^T $ and $ OMO^T = \bigoplus_{\alpha} m_{\alpha} $ does not change the resulting edge theory but its derivation (Appendix~\ref{sec:derivation_edge_theory}) requires a different ansatz.}.

In working with this continuum picture, we can always envision having folded the HSPs back to $ \boldsymbol{\Gamma} $: This is always possible through an infinitesimal perturbation that reduces translational symmetry to a symmetry under translations of two lattice vectors~\cite{Fang:2019kk}.
It may happen that such reduction of translation symmetry only occurs near the edge, but to treat the bulk and the boundary on
the same footing we consider the 2D effective model as if its translation invariance had been reduced throughout.
Nevertheless, the Dirac Hamiltonians inherit their properties from the conventional rotation invariants which \emph{do} distinguish between different HSPs, relying on the underlying crystalline symmetry.
(For example, for a pair of Dirac Hamiltonians describing gap closings at $\mathbf{X}$ and $\mathbf{X'} $, we could allow for deformations of $v_x$ velocities relative to $v_y$ such that $\mathcal{H}^{\alpha}_{\mathbf{X}}$ and $\mathcal{H}^{\alpha+1}_{\mathbf{X'}}$ are each only twofold symmetric, but are related to \emph{each other} via a fourfold rotation.)
For this reason, one may prefer to think of $\mathbf{k}$ as the (small) momentum relative to the respective HSP, even if $\mathbf{k}$ becomes the (small) absolute momentum about $\boldsymbol{\Gamma}$ in the folded picture.

\subsection{Rotation Eigenvalues and Signed Representations}

In this subsection we describe how the rotation eigenvalues of bulk bands pick out irreducible representations of the rotation operator $ r_4^{\alpha} $ for each Dirac model in the stack.
In a $ C_n $-symmetric BZ, the HSPs may be categorized as either being mapped onto themselves (i.e., fixed) under $n$-fold rotation (e.g., $ \boldsymbol{\Gamma}$ and $ \mathbf{M} $ for $ C_4 $) or as being mapped to other HSPs (forming an orbit) under $n$-fold rotation (e.g., the twofold fixed points $ \mathbf{X} $ and $ \mathbf{X'} $ that map into each other under fourfold rotation).
We treat these two cases slightly differently.
As before, we exemplify our approach on $C_4$-symmetric systems.
We start with the 4-fold fixed points.

\subsubsection{Rotation Invariant Momenta}

Fourfold rotational symmetry of a Dirac Hamiltonian in the stack means it must satisfy
\begin{equation}
r_4^{\alpha}\, \mathcal{H}^{\alpha}_{\boldsymbol{\Pi}_{\alpha}}(\mathbf{k})\, r_4^{\alpha\dagger} = \mathcal{H}^{\alpha}_{\boldsymbol{\Pi}_{\alpha}}(R_4 \mathbf{k}),
\label{eq:bulk_dirac_symmetry}
\end{equation}
where $ \boldsymbol{\Pi}_{\alpha} \in \{ \boldsymbol{\Gamma}, \mathbf{M} \} $.
Recalling that our effective Hamiltonian is written in terms of Pauli matrices, it satisfies
\begin{equation}
e^{-i\sigma_{3}\pi/4} \mathcal{H}^{\alpha}_{\boldsymbol{\Pi}_{\alpha}} (\mathbf{k}) e^{i\sigma_{3}\pi/4}
= \mathcal{H}^{\alpha}_{\boldsymbol{\Pi}_{\alpha}}(R_4 \mathbf{k}) .
\end{equation}
This lets us identify $ r_{4}^{\alpha} \propto e^{-i\sigma_{3}\pi/4} $ up to a complex phase.
Insisting that the rotation operator commutes with the PH operator $ \Xi = \sigma_1 \mathcal{K} $ leaves only a freedom in the \emph{sign} of the representation, however, so that $ r_{4}^{\alpha} = \eta_{\alpha} e^{-i\sigma_{3}\pi/4} $, where $ \eta_{\alpha} = \pm 1 $.
This sign, in particular sign differences between representations for different $\mathcal{H}^\alpha$, has physical consequences on the edge of the model, which we show below.
(A similar approach has been used by Khalaf \textit{et al.\@}~\cite{Khalaf:2018hq}.)
These representations are referred to as ``signed representations'' when their sign is important~\cite{Khalaf:2018hq}.
Interestingly, for $C_4$-symmetric points, each rotation eigenvalue directly corresponds to a representation sign \emph{and} a sign for the bulk mass term.
Crucially these two parameters are not independent: As seen in Table~\ref{tab:rotation_eigenvalue_correspondence}, listing the four possibilities shown in Fig.~\ref{fig:rotation_eigenvalues} for the occupied rotation eigenvalues at a $C_4$-symmetric point gives the signed representation $ r_4^{\alpha} $ \emph{and} the sign of the mass $ m_{\alpha} $ for each Dirac Hamiltonian in the stack.

\begin{table}
{\setlength{\tabcolsep}{1em}
  \begin{tabular}{cc|ccc}
    \toprule
    \multicolumn{2}{c|}{Occupied $\Pi_p^{(4)}$} & $ r_4^{\alpha} $ & $ m_{\alpha} $ & $ \eta_{\alpha} $\\
    \colrule
    $ e^{i\pi/4}$ & $\Gamma_1,\, M_1 $ & $+\,e^{-i\sigma_{3}\pi/4}$ & $>0$ & $+1$ \\
    $ e^{-i\pi/4}$ & $\Gamma_4,\, M_4 $ & $+\,e^{-i\sigma_{3}\pi/4}$ & $<0$ & $+1$ \\
    $ e^{i3\pi/4}$ & $\Gamma_2,\, M_2 $ & $-\,e^{-i\sigma_{3}\pi/4}$ & $<0$ & $-1$ \\
    $ e^{-i3\pi/4}$ & $\Gamma_3,\, M_3 $ & $-\,e^{-i\sigma_{3}\pi/4}$ & $>0$ & $-1$ \\
    \botrule
  \end{tabular}}
  \caption{The correspondence of the rotation eigenvalue of the negative energy band to the bulk mass and rotation representation at the fourfold symmetric points $ \boldsymbol{\Gamma} $ and $ \mathbf{M} $.}
  \label{tab:rotation_eigenvalue_correspondence}
\end{table}

\subsubsection{Momenta Transforming into Each Other}
\label{sec:momenta_transforming_into_each_other}

For Dirac Hamiltonians originating from twofold fixed points such as $ \boldsymbol{\Pi}_{\alpha} \in \{ \mathbf{X},\, \mathbf{X'} \}$ in a $ C_4 $-symmetric BZ, we instead have twofold rotational symmetry
\begin{equation}
r_2^{\alpha}\, \mathcal{H}^{\alpha}_{\boldsymbol{\Pi}_{\alpha}}(\mathbf{k})\, r_2^{\alpha\dagger} = \mathcal{H}^{\alpha}_{\boldsymbol{\Pi}_{\alpha}}(R_2 \mathbf{k}).
\end{equation}
In a similar fashion to how we deduced $ r_4^{\alpha} $, we could deduce that $ r_2^{\alpha} \propto e^{-i\sigma_{3}\pi/2} = -i \sigma_3 $, of which two choices $ r_2^{\alpha} = \eta_{\alpha} e^{-i\sigma_{3}\pi/2} $ with $ \eta_{\alpha} = \pm 1 $ commute with PH symmetry.
(We refer to the $ \eta_{\alpha} = - 1 $ case as the negatively signed representation.)
One notices here, however, that specifying the occupied rotation eigenvalue does not uniquely pick out a representation sign \emph{and} a mass sign as it did for the fourfold fixed points.
This is because $ \mp i \sigma_3 \to \pm i \sigma_3 $ exchanges its two diagonal elements, while changing the sign of the bulk mass would change which band has negative energy; changing both at once thus leaves the occupied rotation eigenvalue unchanged.
The sign of the bulk mass has implications for the edge states that appear on the boundary (specifically their direction of propagation), which will be taken into account when ensuring that we construct Dirac models describing differences between topological phases with the same Chern number~\footnote{That the sign of the mass at $ \mathbf{X}^{(\mathbf{\prime})}$ is not set by the occupied rotation eigenvalues is also one source of the ``surface-state ambiguity'' tabulated in Ref.~\onlinecite{Khalaf:2018hq}, which is where the surface signature of a nontrivial bulk is not uniquely determined from the symmetry indicators. Once we specify that $ \Delta \Ch = 0 $ (using information beyond symmetry indicators alone), there will be no ambiguity in the surface-state of this system.}.

As we want to know how the system (particularly the boundary) behaves under a $ \pi / 2 $-rotation, we need to use the underlying fourfold symmetry of the system.
In the original lattice model, momentum states at $ \mathbf{X} +\mathbf{k}$ are mapped to $ \mathbf{X^\prime} +R_4\mathbf{k}$ under a fourfold rotation $R_4$ and vice versa.
In terms of our stacked Dirac picture, such $C_4$ symmetry dictates that the Dirac Hamiltonians originating from these points be related by unitary transformations $ U_\mathbf{X^\prime}^{}$, that is, $\mathcal{H}^{\alpha}_{\mathbf{X}}(R_{4}\mathbf{k})=U_{\mathbf{X'}}^{}\mathcal{H}^{\alpha+1}_{\mathbf{X'}}(\mathbf{k})U_{\mathbf{X'}}^{\dagger}$ and $\mathcal{H}^{\alpha+1}_{\mathbf{X'}}(R_{4}\mathbf{k})=U_{\mathbf{X}}^{}\mathcal{H}^{\alpha}_{\mathbf{X}}(\mathbf{k})U_{\mathbf{X}}^{\dagger}$.
Here, we chose to place the Dirac Hamiltonians for $\mathbf{X}$ and $\mathbf{X'}$ in
neighboring sub-blocks $\alpha$ and $\alpha+1$, respectively.
In terms of the resulting $ 4 \times 4 $ Hamiltonian,
\begin{equation}
\mathcal{H}^{\oplus}(\mathbf{k}) \equiv \mathcal{H}^{\alpha}_{\mathbf{X}}(\mathbf{k}) \oplus \mathcal{H}^{\alpha+1}_{\mathbf{X'}}(\mathbf{k}),
\end{equation}
the only form of this fourfold symmetry compatible with our convention [Eq.~\eqref{eq:Dirac_single}] of momenta and Pauli matrices appearing in $k_j\sigma_j$ combinations and having positive velocities is
\begin{equation}
\mathcal{H}^{\oplus}(R_4 \mathbf{k}) = \left[e^{-i\sigma_{3}\pi/4}\mathcal{H}^{\alpha+1}_{\mathbf{X'}}( \mathbf{k}) e^{i\sigma_{3}\pi/4}\right]
\oplus\left[e^{-i\sigma_{3}\pi/4}\mathcal{H}^{\alpha}_{\mathbf{X}}( \mathbf{k})e^{i\sigma_{3}\pi/4}\right],
\label{eq:C4symm_XX'}
\end{equation}
which also holds if velocities are anisotropic at $\mathbf{X}$ and $\mathbf{X}^\prime$ in a $C_4$ related manner.
The unitary relation between $\mathcal{H}^{\alpha}_{\mathbf{X}}(R_{4}\mathbf{k})$ and $\mathcal{H}^{\alpha+1}_{\mathbf{X}^\prime}(\mathbf{k})$ together with our convention of identical Dirac Hamiltonian chiralities also implies $ m_\alpha=m_{\alpha+1} $.
The symmetry relation \eqref{eq:C4symm_XX'} can be compactly expressed as
\begin{equation}
\mathcal{H}^{\oplus}(R_4 \mathbf{k}) = r_4^{\oplus}\,\mathcal{H}^{\oplus}(\mathbf{k})\, r_4^{\oplus \dagger},
\end{equation}
where the requirement of PH symmetry $ [r_4^{\oplus}, \Xi\oplus\Xi] =0 $ leaves two choices
\begin{equation}
r_4^{\oplus} = e^{-i\sigma_{3}\pi/4} \otimes \tau_1 \quad \text{or} \quad r_4^{\oplus} = e^{-i\sigma_{3}\pi/4} \otimes i\tau_2,
\end{equation}
apart from an overall sign that will later be seen to be inconsequential.
Here, we denote the space of stacked Dirac Hamiltonians $\alpha,\alpha+1$ by $\tau_\mu$.

Squaring these two representations gives $ (r_4^{\oplus})^2 = + e^{-i\sigma_{3}\pi/2} \otimes \mathbb{I}_2 $ or $ (r_4^{\oplus})^2 = - e^{-i\sigma_{3}\pi/2} \otimes \mathbb{I}_2 $, respectively, which is consistent with the two signed options for $ r_2^{} $ above and implies $ \eta_{\alpha} = \eta_{\alpha+1} $.
Unlike for the truly fourfold fixed points, the representation and sign of $ m_{\alpha} = m_{\alpha + 1} $ is not uniquely determined from occupied rotation eigenvalue---instead the correspondence is between the combination of occupied rotation eigenvalue and mass to rotation representation, shown in Table~\ref{tab:X-X_rotation_eigenvalue_correspondence}.

\begin{table}
{\setlength{\tabcolsep}{1em}
  \begin{tabular}{ccc|cc}
    \toprule
    \multicolumn{2}{c}{Occupied $ \Pi^{(2)}_{p} $} & $m_\alpha$ & $ r_4^{\oplus} $ & $\eta_\alpha$ \\
    \colrule
    $e^{i\pi/2}$ & $X_1$ & $>0$ & $e^{-i\sigma_{3}\pi/4}\otimes\tau_1$ & $+1$ \\
    $e^{-i\pi/2}$ & $X_2$ & $<0$ & $e^{-i\sigma_{3}\pi/4}\otimes\tau_1$ & $+1$ \\
    $e^{i\pi/2}$ & $X_1$ & $<0$ & $e^{-i\sigma_{3}\pi/4}\otimes i\tau_2$ & $-1$ \\
    $e^{-i\pi/2}$ & $X_2$ & $>0$ & $e^{-i\sigma_{3}\pi/4}\otimes i\tau_2$ & $-1$ \\
    \botrule
  \end{tabular}}
  \caption{The correspondence of the rotation eigenvalue of the negative energy band and bulk mass to the rotation representation for the $ \mathbf{X} / \mathbf{X'} $ points. Also shown is the sign of the twofold rotation representation defined as $r_2^\alpha = \eta_\alpha e^{-i\sigma_{3}\pi/2} $. }
  \label{tab:X-X_rotation_eigenvalue_correspondence}
\end{table}

\subsection{Dirac Stacks for Topological Interfaces}
\label{subsec:explicit_Dirac}

We now outline how a stacked Dirac model can be constructed to describe a transition of between $ C_n $-symmetric superconductors with different topological invariants.
As stated above, the stacked Dirac models capture \emph{differences} between topological phases.
Consider two regions $ \mathcal{I} $ and $ \mathcal{O} $ with BdG Hamiltonians $ H_{\mathcal{I}}(\mathbf{k}) $ and $ H_{\mathcal{O}}(\mathbf{k}) $, respectively, understood as being the regions inside ($\mathcal{I}$) and outside ($\mathcal{O}$) our system of interest.
Each system has independent occupancies $ \# \Pi_{p}^{(n)} $, meaning that we can define differences in occupancies:
\begin{equation}
\Delta \# \Pi_{p}^{(n)} \equiv \# \Pi_{p}^{(n)}|_{\mathcal{I}}^{} - \# \Pi_{p}^{(n)}|_{\mathcal{O}}^{}.
\end{equation}
For a meaningful description in terms of stacked Dirac models, we require the rotation operators in both regions $\mathcal{O}$ and $\mathcal{I}$ to be the same.
This is always possible through the addition of trivial bands to either region, which can safely be added since they do not change the topological invariants.
Once the rotation operator is the same in both systems, each must have the same total number of each rotation eigenvalue, so
$\# \Pi_{p}^{(n)}|_{\mathcal{I}}^{} + \bar{\#} \Pi_{p}^{(n)}|_{\mathcal{I}}^{} = \# \Pi_{p}^{(n)}|_{\mathcal{O}}^{} + \bar{\#} \Pi_{p}^{(n)}|_{\mathcal{O}}^{} $,
where $ \bar{\#} $ counts unoccupied states.
Using PH symmetry which relates occupied and unoccupied states, we see that these differences are not all independent [cf.\ Eq.~\eqref{eq:PH_invariant_relations}]:
\begin{equation}
\Delta \# \Pi_{p}^{(n)} = - \Delta \# \Pi_{n - p + 1}^{(n)}.
\label{eq:delta_relations}
\end{equation}
For example, in a $ C_4 $-symmetric system one has ten occupancies ($\Gamma_{1,2,3,4}$, $M_{1,2,3,4} $ and $ X_{1,2}=X'_{1,2} $) to consider, which are in this way reduced to five independent differences, chosen as $ \Delta \# X_1 $, $ \Delta \# \Gamma_1 $, $ \Delta \# \Gamma_2 $, $ \Delta \# M_1 $ and $ \Delta \# M_2 $.

For each independent difference $ \Delta \# \Pi_{p}^{(n)} $, one adds $ |\Delta \# \Pi_{p}^{(n)}| $ Dirac Hamiltonians to the stack with the appropriate rotation representations and masses.
Closing and reopening every gap by taking $ m_{\alpha} \to - m_{\alpha} $ for all $ \alpha $ then reproduces the transition $ H_{\mathcal{I}}(\mathbf{k}) \to H_{\mathcal{O}}(\mathbf{k}) $.

We now address the feature of Dirac Hamiltonians deriving from $ \mathbf{X} / \mathbf{X'} $, demonstrated in Table~\ref{tab:X-X_rotation_eigenvalue_correspondence}, which is that the occupation of either eigenvalues $ X_1 = i $ or $ X_2 = -i $ does not uniquely determine the sign of the bulk mass nor the rotation representation.
Thus, a given change $ \Delta \# X_1 $ may be realized through stacks of two different types of Dirac Hamiltonians.
Looking at Table~\ref{tab:X-X_rotation_eigenvalue_correspondence}, these two types can be distinguished by the sign $ \eta_{\alpha} $ of the twofold rotation operator $ r_2^{\alpha} = \eta_{\alpha} e^{-i\sigma_{3}\pi/2} $.
Let $ \Delta \# X_1^{\pm} $ denote the contributions to $ \Delta \# X_1 $ from Hamiltonians with $ \eta_{\alpha} = \pm 1 $, such that the overall change in occupation is $ \Delta \# X_1 = \Delta \# X_1^{+} + \Delta \# X_1^{-} $.
Note that this decomposition is specific to the construction of a Dirac model, rather than a direct property of the original Bloch Hamiltonians $ H_{\mathcal{I}}(\mathbf{k}) $ and $ H_{\mathcal{O}}(\mathbf{k}) $.
Distinguishing between $\Delta \#X_1^\pm $, as we now explain, allows us to construct a Dirac stack that does not change the Chern number when $ m_{\alpha} \to -m_{\alpha} $, as we require for the anomalous boundary states we wish to investigate.

Zero change in the Chern number implies that there should be an equal number of left- and right-moving modes at the $ \mathcal{I}-\mathcal{O} $ interface.
This is equivalent to the statement that there should be an equal number of Dirac Hamiltonians in the stack with positive and negative bulk masses (since we consider the scenario where all $m_\alpha $ change sign across the $ \mathcal{I}-\mathcal{O} $ boundary).
From Table~\ref{tab:rotation_eigenvalue_correspondence}, it is evident that these bulk masses are uniquely determined from changes in occupied rotation eigenvalues, which is not the case for Dirac Hamiltonians derived from gap closings at $ \mathbf{X}/\mathbf{X'} $.
Looking again at Table~\ref{tab:X-X_rotation_eigenvalue_correspondence}, and recalling that all the Dirac Hamiltonians have the same chirality, we see that for the same $ \Delta \# X_1^+ $ and $ \Delta \# X_1^- $ the contribution to $ \Delta \Ch$ is opposite because, for a given $\Pi_p^{(2)}$ being occupied, opposite signs of $\eta_\alpha$ imply opposite signs for $m_\alpha$.
Combining all these observations, we may rewrite the $ \Delta \Ch = 0 $ condition as
\begin{align}
0 &= \Delta \Ch \\
&= 2\Delta\#X_{1}^{-}-2\Delta\#X_{1}^{+}-\Delta\#\Gamma_{1}-\Delta\#M_{1}+\Delta\#\Gamma_{2}+\Delta\#M_{2} \\
&= -4\Delta\#X_1^{+}+2\Delta[X]-\Delta[M_{1}]+\Delta[M_{2}],
\label{eq:chern_condition}
\end{align}
Taken on their own, changes in rotation invariants $ [M_1] $, $ [M_2] $ and  $ [X] $ of Eq.~\eqref{eq:rotation_invariants} determine $ \Delta \Ch \mod 4 $ \cite{Fang:2012dn,Benalcazar:2014hb}, but with a specific stacked Dirac model realization we could equate $ \Delta \Ch = 0 $ exactly.
(The modulo $4$ ambiguity reappears if one does not have access to the $ \Delta \# X_1^{\pm} $ extra information because changing $ \Delta\# X_1^+ \to \Delta\# X_1^+ + 1 $ and $ \Delta\# X_1^- \to \Delta\# X_1^- - 1 $ would not affect $ \Delta[X] $ but would change $ \Delta \Ch \to \Delta \Ch -\,4 $.)

\subsection{Effective Edge Theory}

We are interested in what happens at the boundary between systems in different topological classes, which in our continuum model above occurs when the masses $ \{m_{\alpha}\} $ change sign.
Each bulk gap closing has an associated chiral edge mode localized at the boundary~\cite{Jackiw:1976ky,Hasan:2010ku}.
Its effective edge theory, allowing also for smooth (on the scale of the lattice spacing) variations in the local boundary direction, can be derived as described in Appendix~\ref{sec:effective_boundary_hamiltonian}.
The resulting stack of decoupled left- and right-moving chiral edge modes is described by the edge Hamiltonian 
\begin{equation}
h_{\mathbf{r},\mathbf{k}_\parallel} = h_{\mathbf{r},\mathbf{k}_\parallel}^{\rightarrow} \oplus h_{\mathbf{r},\mathbf{k}_\parallel}^{\leftarrow},
\quad \text{with} \quad
h_{\mathbf{r},\mathbf{k}_\parallel}^{s} = \bigoplus_{\alpha} h_{\mathbf{r},\mathbf{k}_\parallel}^{\alpha s},
\end{equation}
where $ s \in \{ \rightarrow,\,\leftarrow \}$ such that right- and left-moving modes
\begin{equation}
h_{\mathbf{r},\mathbf{k}_{\parallel}}^{\alpha\rightarrow} = + v_{\alpha} k_{\parallel}
\quad \text{and} \quad
h_{\mathbf{r},\mathbf{k}_{\parallel}}^{\alpha\leftarrow} = - v_{\alpha} k_{\parallel}
\end{equation}
have been placed in different sub-blocks \footnote{This $ 2 \times 2 $ block-diagonal structure is possible by choosing the original stacking order in Eq.~\eqref{eq:Dirac_stack} to be such that Dirac Hamiltonians with $ m_{\alpha} < 0 $ in the bulk appear first.}.
The subscript $ \parallel $ denotes a projection onto the direction $ \hat{\mathbf{n}}_{\parallel} $ along the edge (i.e., $ k_{\parallel} = \mathbf{k} \cdot \hat{\mathbf{n}}_{\parallel} $) and the subscript $ \mathbf{r} $, indicating the position along the boundary, is present to allow for the aforementioned smooth boundary variations.

\begin{figure}
\centering
\includegraphics[width=\linewidth]{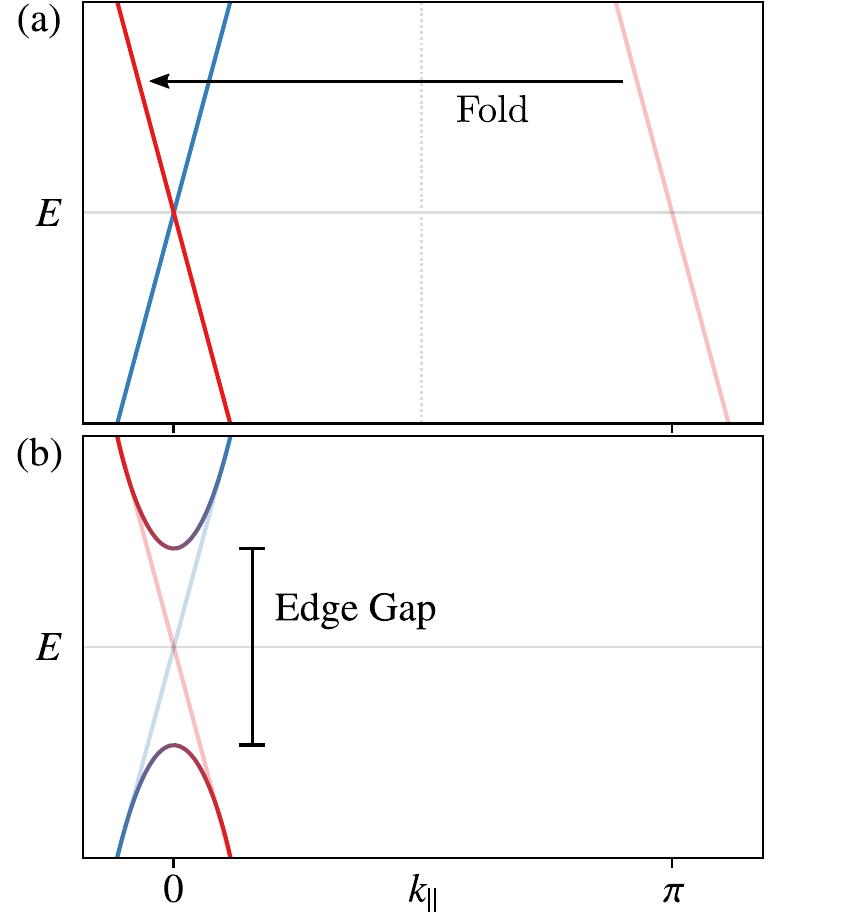}
\caption{Example spectrum along the edge of a 2D superconductor.
Each sign change of a bulk mass $ m_{\alpha} $ manifests itself as a chiral mode on the edge.
(a) The folding process which takes all HSPs to $ \boldsymbol{\Gamma} $, meaning that all edge modes are centered on $ k_{\parallel} = 0 $ in the continuum description.
(b) An edge mass term that couples a left-mover to a right-mover, opening up a gap on the edge.}
\label{fig:folding-edge-modes}
\end{figure}

Having limited ourselves to $ \Delta \Ch = 0 $ transitions, there are as many left-movers as right-movers in the stack.
Since gap closings happening at $ \mathbf{X} $ must also happen at $ \mathbf{X'} $ by rotational symmetry, the corresponding edge modes appear in pairs on the boundary with the same propagation direction (because their bulk masses and chiralities are the same).
We show an example spectrum for the edge Hamiltonian in Fig.~\ref{fig:folding-edge-modes}(a).

The edge Hamiltonian also possesses a PH symmetry that follows from the original PH symmetry of the bulk.
With a particular basis choice for the edge Hamiltonian (detailed in Appendix~\ref{sec:edge_PH-symmetry}), the PH operator is simply complex conjugation $ \mathcal{K} $ and the symmetry is given by
\begin{equation}
h_{\mathbf{r},\mathbf{k}_{\parallel}} = - \mathcal{K}\, h_{\mathbf{r},-\mathbf{k}_{\parallel}}\, \mathcal{K}.
\label{eq:edge_PH_symmetry}
\end{equation}

\subsubsection{Edge Projections of Rotation Representations}
\label{subsubsec:rotation_rep_edge}

We now describe how the sign of the bulk rotation representation is transmitted to the representations on the edge, while also recalling the fact that the rotation operator used to classify periodic Hamiltonians may be different to the rotation operator compatible with the boundary.
As shown in Sec.~\ref{sec:benalcazar_classification}, these operators are related in terms of the location $ \mathbf{c} $ of the rotation center within a unit cell as
\begin{equation}
r_{n,\mathbf{c}}^{\alpha} = \exp{[i \boldsymbol{\Pi}_{\alpha} \cdot (1 - R_n^{-1})\mathbf{c}]} \, r_{n}^{\alpha}.
\end{equation}
For $C_4$ symmetry, this is shown explicitly in Tables~\ref{tab:physical_representations} and \ref{tab:physical_X-X_representations}.
\begin{table}
{\setlength{\tabcolsep}{1em}
  \begin{tabular}{c|c|c}
    \toprule
    $ r_{4,\mathbf{0}}^{\alpha} $ & \multicolumn{2}{c}{$r_{4,(\mathbf{a}_1 + \mathbf{a}_2)/2}^{\alpha}$} \\
    {} & $ \boldsymbol{\Pi}_{\alpha} = \boldsymbol{\Gamma} $ & $\boldsymbol{\Pi}_{\alpha} = \mathbf{M} $ \\ \colrule
    $ +\,e^{-i\sigma_{3}\pi/4} $ & $ +\,e^{-i\sigma_{3}\pi/4} $ & $ -\,e^{-i\sigma_{3}\pi/4} $ \\
    $ -\,e^{-i\sigma_{3}\pi/4} $ & $ -\,e^{-i\sigma_{3}\pi/4} $ & $ +\,e^{-i\sigma_{3}\pi/4} $ \\
    \botrule
  \end{tabular}}
  \caption{The correspondence from $ r_4^{\alpha} \equiv r_{4,\mathbf{0}}^{\alpha}$ to the shifted rotation operator at the fourfold fixed points. The representation changes sign at $ \mathbf{M} $ but is unchanged at $ \boldsymbol{\Gamma} $.}
  \label{tab:physical_representations}
\end{table}

\begin{table}
{\setlength{\tabcolsep}{1em}
  \begin{tabular}{c|c}
    \toprule
    $ r_{4,\mathbf{0}}^{\oplus} $ & $ r_{4,(\mathbf{a}_1 + \mathbf{a}_2)/2}^{\oplus} $ \\\colrule
    $ e^{-i\sigma_{3}\pi/4} \otimes i\tau_2 $ & $ e^{-i\sigma_{3}\pi/4} \otimes \tau_1 $ \\
    $ e^{-i\sigma_{3}\pi/4} \otimes \tau_1 $ & $ e^{-i\sigma_{3}\pi/4} \otimes i\tau_2 $ \\
    \botrule
  \end{tabular}}
  \caption{The correspondence from $ r_4^{\oplus} \equiv r_{4,\mathbf{0}}^{\oplus} $ to the shifted rotation operator. These rotation representations only occur at $ \mathbf{X}/\mathbf{X'}$ points.}
  \label{tab:physical_X-X_representations}
\end{table}

It is from this operator that the rotation operator of the edge theory must be derived $ r_{n,\mathbf{c}}^{\alpha} \rightarrow u_{n,\mathbf{c}}^{\alpha} $.
The rotational symmetry for the edge Hamiltonian is then expressed through
\begin{equation}
h_{R_n\mathbf{r},R_n\mathbf{k}_{\parallel}} = u_{n,\mathbf{c}}^{}\, h_{\mathbf{r},\mathbf{k}_{\parallel}}\, u_{n,\mathbf{c}}^{T},
\label{eq:edge_rotation_symmetry}
\end{equation}
where $ u_{n,\mathbf{c}}^{} =\bigoplus_{\alpha} u_{n,\mathbf{c}}^{\alpha} $ is the direct sum of all the edge-projected rotation representations, and $u_{n,\mathbf{c}} $ consists only of real elements due to PH symmetry being simply complex conjugation.

For $ C_4 $-symmetric systems, $ u_{4,\mathbf{c}}^{} $ is block-diagonal with $ 1 \times 1 $ blocks for edge modes deriving from gap closings at $ \boldsymbol{\Gamma} $ or $ \mathbf{M} $, and $ 2 \times 2 $ blocks to transform between $ \mathbf{X} $- and $ \mathbf{X'}$-deriving edge modes.
These are derived explicitly for $ C_4 $ in Appendix~\ref{subsec:surface_rotation} and summarized in Table~\ref{tab:surface_representations}.

\begin{table}
  \begin{tabular}{cc|c}
    \toprule
    $ r_{4,\mathbf{c}}^{\alpha} $ & $\eta_{\alpha,\mathbf{c}}$ & $ u_{4,\mathbf{c}}^{\alpha} $ \\
    \colrule
    $ +\,e^{-i\sigma_{3}\pi/4} $ & $+1$ & $ +\,1 $ \\
    $ -\,e^{-i\sigma_{3}\pi/4} $ & $-1$ & $ -\,1 $ \\
    \botrule
  \end{tabular}
  \begin{tabular}{cc|cc}
    \toprule
    $ r_{4,\mathbf{c}}^{\oplus} $ & $\eta_{\alpha,\mathbf{c}}$ & $ u_{4,\mathbf{c}}^{\alpha} $ & $\det u_{4,\mathbf{c}}^{\alpha}$ \\
    \colrule
    $ e^{-i\sigma_{3}\pi/4} \otimes i\tau_2 $ & $-1 $ & $ i \tau_2 $ & $ +1 $\\
    $ e^{-i\sigma_{3}\pi/4} \otimes \tau_1 $ & $+1$ & $ \tau_1 $ & $-1$\\
    \botrule
  \end{tabular}
  \caption{The edge rotation representations $ u_{4,\mathbf{c}}^{\alpha} $ resulting from the bulk representation $r_{4,\mathbf{c}}^\alpha$ at $ \boldsymbol{\Gamma}$, $\mathbf{M}$, and $ r^{\oplus}_{4,\mathbf{c}} $ at $\mathbf{X}$, $\mathbf{X'} $. The signs of these representations are denoted by $\eta_{\alpha,\mathbf{c}} $, where $ \eta_{\alpha,\mathbf{0}} \equiv \eta_{\alpha} $.}
  \label{tab:surface_representations}
\end{table}

\subsection{Boundary Mass Terms}

In general, counterpropagating modes on the edges become gapped due to symmetry-allowed terms that couple these modes.
Such gapping terms (or \emph{mass} terms) $ \mu_{\mathbf{r}} $ couple left-moving to right-moving modes, appearing as off-diagonal terms in the (previously gapless) edge Hamiltonian
\begin{equation}
h_{\mathbf{r},\mathbf{k}_{\parallel}} = 
\begin{pmatrix}
h_{\mathbf{r},\mathbf{k}_{\parallel}}^{\rightarrow} + i \lambda_{\mathbf{r}}^{\rightarrow} & i\mu_{\mathbf{r}}^{} \\
-i\mu_{\mathbf{r}}^{T} & h_{\mathbf{r},\mathbf{k}_{\parallel}}^{\leftarrow} + i \lambda_{\mathbf{r}}^{\leftarrow} \end{pmatrix},
\label{eq:gapped_edge_hamiltonian}
\end{equation}
where we also included the (skew-symmetric) forward scattering matrices $\lambda_{\mathbf{r}}^{s}$ with $ s \in \{ \rightarrow, \leftarrow \}$.
(These terms are `forward scattering' in the sense that they scatter between modes moving in the same direction.)
Because of PH symmetry, $\mu_\mathbf{r}$ and $ \lambda_{\mathbf{r}}^{s}$ must be real.
To see why $\mu_{\mathbf{r}}$ can be interpreted as mass terms, it is illuminating to consider a simplified case where all $ 2p $ edge modes have the same velocity $ v_{\alpha} = 1 $ and to ignore all forward-scattering terms, giving the edge Hamiltonian
\begin{equation}
h_{\mathbf{r},\mathbf{k}_{\parallel}}^0 =
\begin{pmatrix}
k_{\parallel} \mathbb{I}_p & i \mu_{\mathbf{r}}\\
-i \mu_{\mathbf{r}}^{T} & -k_{\parallel} \mathbb{I}_p
\end{pmatrix}.
\end{equation}
Because $\mu_{\mathbf{r}}$ is a real matrix, it may be decomposed via a singular value decomposition (SVD) into $ \mu_{\mathbf{r}} = Y D W^T $ where $ Y $ and $ W $ are orthogonal matrices and $ D = \operatorname{diag}(\Delta_1,\Delta_2,\ldots,\Delta_p) $ is a diagonal matrix.
Using the SVD, the edge Hamiltonian can be factorized as
\begin{equation}
h_{\mathbf{r},\mathbf{k}_{\parallel}}^0 =
\begin{pmatrix}
Y & 0\\
0 & W
\end{pmatrix}
\begin{pmatrix}
k_{\parallel} \mathbb{I}_p & iD\\
-iD & -k_{\parallel} \mathbb{I}_p
\end{pmatrix}
\begin{pmatrix}
Y^T & 0\\
0 & W^T
\end{pmatrix} ,
\label{eq:tranformed_simplified_surface}
\end{equation}
i.e., it is unitarily equivalent to $p$ stacked one-dimensional massive Dirac Hamiltonians.
Each band separation is set by $\Delta_\alpha$ and the energy eigenvalues are $E_\alpha^2 = k_{\parallel}^2 + \Delta_\alpha^2$.
In an SVD, the matrices $Y$ and $W$ are typically chosen such that all $\Delta_\alpha \geq 0$.
Here, we fix $\det Y = \det W = 1$ by multiplying an appropriate number of rows of $Y$ (and $W$) by minus one, that is, an odd number of rows when initially $\det Y =-1$ ($\det W=-1$), and an even number of rows when initially $\det Y = +1$ ($\det W =+1$).
Keeping $h_{\mathbf{r},\mathbf{k}_{\parallel}}^0$ the same then requires changing the signs of the corresponding $\Delta_\alpha$ accordingly, such that $\sgn \det D = \prod_{\alpha} \sgn \Delta_{\alpha} = \pm 1$, where the minus sign arises when the number of sign changes in $Y$ and $W$ add up to an odd number.

When rotating from one edge to the neighboring one using Eq.~\eqref{eq:edge_rotation_symmetry}, the matrix $\mu_\mathbf{r}$ transforms as $\mu_\mathbf{r} \to \mu_{R\mathbf{r}}$.
A $\Delta_\alpha$ changing sign under this transformation expresses that there is a $\Delta_\alpha$ mass kink in the edge Dirac theory as we turn from one edge to the neighboring one. 
Such a mass kink binds a Majorana state~\cite{Fu:2008gu}; it gives rise to a corner Majorana mode in the system.
Considering all $\Delta_\alpha$, therefore, a sign change of $\det D$ along a corner results in an odd number of Majorana bound states, i.e., a single Majorana after the hybridization of pairs.
Since $\det Y = \det W =1$, the determinant of $ D $ changing sign is captured by a relative sign between $\sgn \det \mu_{\mathbf{r}}$ and $\sgn \det \mu_{R\mathbf{r}}$.

While this observation is illuminating, it relies on all edge modes dispersing in the same way and the absence of forward-scattering terms.
The next subsection will use a more robust characterization in terms of Pfaffians that works even in this more general scenario.

\subsubsection{Topologically Distinct Boundary Phases}
\label{subsubsec:topology_edge}

Having seen how back-scattering terms on the edge can be interpreted as mass terms in a one-dimensional theory, we now reframe this in terms of a familiar topological invariant for class D systems in one dimension---the Pfaffian invariant.
More precisely, this invariant is the product of Pfaffians at the HSPs in the BZ~\cite{Kitaev:2001gb,Budich:2013it}, where the Hamiltonian is skew-symmetric.
The continuum theory we use only captures changes of the topological invariant along a corner, but not any invariant itself.
Such a change in the topological invariant manifests in a sign change of the Pfaffian at $k_\parallel =0$, a point where the Hamiltonian is skew-symmetric (guaranteed by PH symmetry, $h_{\mathbf{r},\mathbf{k}_{\parallel} = \mathbf{0}} =-h^*_{\mathbf{r},\mathbf{k}_{\parallel} = \mathbf{0}}$).
Considering the Pfaffian at $ k_{\parallel}=0$ (but not at $k_{\parallel}=\pi $) is sufficient because having folded the HSPs (see Fig.~\ref{fig:folding-edge-modes}) all edge mass kinks involve edge gap closings in the proximity of $k_{\parallel}=0$.
Of these only the gap closings \emph{at} $k_{\parallel}=0$ are of importance: While forward scattering may cause some of the gap closings to split away symmetrically from $k_{\parallel}=0$, we need not account for these because they only result in \emph{pairs} of bound states which can hybridize and gap out.
More explicitly, we define $A_\mathbf{r} \equiv -i h_{\mathbf{r},\mathbf{k}_\parallel = \mathbf{0}}$, which is a real and skew-symmetric matrix,
and the indicator $\delta_\mathbf{r}$ using which we shall track changes in the edge invariant~\cite{Budich:2013it}
\begin{equation}
\delta_\mathbf{r} = \sgn \Pf A_\mathbf{r} = \sgn \left[ (-i)^p \Pf h_{\mathbf{r},\mathbf{k}_\parallel = \mathbf{0}} \right] .
\label{eq:invariant_corner}
\end{equation}
We can also verify that this gives the same result as our  simplified example $h^0_{\mathbf{r},\mathbf{k}_\parallel = \mathbf{0}}$ introduced above.
The Pfaffian of $A^0_{\mathbf{r}} = -i h^0_{\mathbf{r},\mathbf{k}_\parallel=\mathbf{0}}$ equals
\begin{align}
\Pf \left( A^0_{\mathbf{r}} \right)
&= 
\det \begin{pmatrix} Y & 0 \\ 0 & W \end{pmatrix} \Pf \begin{pmatrix} 0 & D \\ - D & 0 \end{pmatrix}
= \Pf \begin{pmatrix} 0 & D \\ -D & 0 \end{pmatrix} \\
&= (-1)^{p(p-1)/2} \det D,
\end{align}
where we used that $\det Y = \det W = 1$.
As the matrix dimension $p$ does not change around a corner, sign changes in $\Pf \left( A^0_{\mathbf{r}} \right)$ thus capture sign changes in $\det D $.
We emphasize, however, that Eq.~\eqref{eq:invariant_corner} goes beyond the counting argument for $h^0_{\mathbf{r},\mathbf{k}_\parallel=\mathbf{0}}$, as it also takes into account forward-scattering terms and allows different velocities $v_\alpha$.

\section{Bulk Rotation Invariants and Corner Majoranas}
\label{sec:bulk_invariants}

The bulk rotational symmetry has direct implications for the Pfaffian invariant $\delta_{\mathbf{r}}$ that distinguishes topological phases along the edge.
Using the rotational symmetry relation of Eq.~\eqref{eq:edge_rotation_symmetry}, which also holds for $A_\mathbf{r}$, we use a Pfaffian identity to assess the difference in topology for neighboring edges as follows:
\begin{align}
 \delta_{R_n \mathbf{r}}
 &= \sgn \Pf( A_{R_n \mathbf{r}} )
 = \sgn \Pf( u_{n,\mathbf{c}}^{} A_{\mathbf{r}} u_{n,\mathbf{c}}^T)
 = \sgn \Pf( A_{\mathbf{r}} ) \det u_{n,\mathbf{c}}^{} \nonumber \\
 &= \delta_{\mathbf{r}} \det u_{n,\mathbf{c}}^{},
\end{align}
where we used that $\det u_{n,\mathbf{c}}^{} = \pm 1$ since $ u_{n,\mathbf{c}}^{} $ is orthogonal.
Thus, the Pfaffian invariants for edges related by $ R_n $ are the same only if $ \det u_{n,\mathbf{c}}^{} = +1 $.

For $ \det u_{n,\mathbf{c}}^{} = -1 $, neighboring edges are topologically distinct and consequently must harbor an odd number of Majorana states bound between them.
The topological index $ \Upsilon_{\mathbf{c}}^{(n)} = 0,1$ equal to the number (modulo 2) of Majorana zero modes localized between neighboring edges is therefore~\footnote{$ \det u_{n,\mathbf{c}}^{} =+ 1 $ corresponding to the case without boundary Majorana bound states is also consistent with the fact that if $ \det u_{n,\mathbf{c}}^{} =+ 1 $, then for a circular boundary the transformation $ h_{\mathbf{r},\mathbf{k}} \to h_{R_n \mathbf{r},R_n \mathbf{k}} $ could be achieved via many infinitesimal orthogonal transformations (connected to the identity).
There would therefore exist a continuous deformation between the two Hamiltonians that does not close any gaps and respects all the symmetries, thereby making them topologically equivalent.}
\begin{equation}
(-1)^{\Upsilon_{\mathbf{c}}^{(n)}}=\det u_{n,\mathbf{c}}.
\label{eq:Upsilon_index_relation}
\end{equation}
Note that the sign of $ \det u_{n,\mathbf{c}}^{} = \prod_{\alpha} \det u_{n,\mathbf{c}}^{\alpha} $ flips with each addition onto the stack of a representation with $ \det u_{n,\mathbf{c}}^{\alpha} = -1 $.
It is also reassuring that had we (arbitrarily) chosen the basis states of the edge Hamiltonian to transform trivially under the \emph{negative} representation instead (amounting to redefining $ u_{n,\mathbf{c}}^{\alpha} \to - u_{n,\mathbf{c}}^{\alpha} $), then this index would be unchanged since $ \det u_{n,\mathbf{c}} \to (-1)^{2p} \det u_{n,\mathbf{c}}^{} = \det u_{n,\mathbf{c}}^{} $.

Another way to derive higher-order surface signatures for topological crystalline phases \cite{Khalaf:2018hq,Geier:2018ev,Trifunovic:2019hi,Juricic:2019} is the construction of a minimal set of mutually anticommuting mass matrices $ \Omega \equiv \{ \Omega_i \} $ (with $ \{ \Omega_i, \Omega_j \} = 2 \delta_{ij} $) that also anticommute with a (simplified) model surface Hamiltonian.
These mass terms are added to the surface in a symmetry-respecting manner, where the transformation properties of $ \Omega_i $ under crystalline symmetry operations dictate the possible existence of gapless regions on the boundary.
The presence of anomalous surface states is predicted from the cardinality of $ \Omega $.
However, such an approach is not immediately applicable to our edge Hamiltonian Eq.~\eqref{eq:gapped_edge_hamiltonian} because $ \mu_{\mathbf{r}} $ is a generic mass term that may not be constructed from $ \Omega $ as $ \mu_{\mathbf{r}} = \sum_{i} \epsilon_{\mathbf{r}}^{i} \Omega_i $ everywhere along the boundary
\footnote{Furthermore, a set $ \Omega $ will not always be closed under rotation $ \Omega_i \to u_{n,\mathbf{c}}^{} \Omega_i u_{n,\mathbf{c}}^T \notin \Omega $ when all the Dirac Hamiltonians in the stack are allowed to be different.}.
Our edge Hamiltonian may also contain forward scattering terms $ \lambda_{\mathbf{r}}^s $ that do not follow this prescription.
An advantage of our Pfaffian invariant approach is therefore that it swiftly demonstrates the relation between boundary Majoranas and rotation representations, without any special algebraic structure in the boundary Hamiltonian beyond that required by PH symmetry.

\subsection{Path Independence}
\label{subsec:trivial_bands}

It should be pointed out that because the original classification \cite{Benalcazar:2014hb} is in terms of a stable (i.e., robust under the addition of trivial bands) equivalence, many different stacked Dirac models can realize the same change in topological phase.
Conversely, it means that every phase change can be realized through many `paths' of gap closings in a phase diagram; for example a change in phase $ \Delta (\Ch, [X], [M_1], [M_2]) = (0, 0, 1, 1) $ could be realized through $ \Delta (\# \Gamma_1, \# \Gamma_2) = (-1,-1) $ or through $ \Delta (\# M_1, \# M_2) = (1,1) $.
For the $ \det u_{n,\mathbf{c}} $ index to be truly topological, it must be independent of the path of gap closings chosen to go from one phase to another.
We demonstrate that this is so focusing on the $C_4$-symmetric case below. 

In our $ C_4 $-symmetric Dirac construction, multiple possible paths arise because there are six independent parameters for gap closings [namely $\Delta(\#X_1^-, \#X_1^+, \#\Gamma_1, \# \Gamma_2, \#M_1, \#M_2)$] but only four constraints in the form of the four topological invariants $\Delta(\Ch, [X], [M_1],[M_2])$, leaving two degrees of freedom.
Because of the additive structure of the invariants, these two degrees of freedom define a plane in the space of $ \Delta( \# X_1^{-}, \# X_1^{+}, \# \Gamma_1, \# \Gamma_2, \# M_1, \# M_2) $.
This plane can be spanned by two gap closing patterns which do not change the phase, namely $ \Delta( \# X_1^{-}, \# \Gamma_1, \# M_1) = (1, 1, 1) $ and $ \Delta( \# X_1^{-}, \# \Gamma_2, \# M_2) = (-1, 1, 1) $.
The first set of of occupation number changes is consistent with adding a trivial Dirac (sub)stack with $\Gamma^{(4)}_p=e^{i\pi/4}$, and the second with adding a trivial Dirac stack with $\Gamma^{(4)}_p=e^{3i\pi/4}$, where both additions are to the inside ($\mathcal{I}$) region, with their PH conjugates appearing outside ($\mathcal{O}$).
One may verify that the Dirac (sub)stacks implementing these changes in occupation number have $ \det u_{4,\mathbf{c}}^{} = +1 $, both for $ \mathbf{c} = \mathbf{0} $ and $ \mathbf{c} = (\mathbf{a}_1 + \mathbf{a}_2) / 2 $.
This means that each plane in the parameter space has a definite value of $ \det u_{4,\mathbf{c}}^{} $.

We now explain why trivial bands cannot change $ \det u_{4,\mathbf{c}}^{} $.
In our Dirac model, adding PH conjugate pairs of trivial bands to $ \mathcal{I} $ and $ \mathcal{O} $ correspond to Dirac (sub)stacks that upon $m_\alpha \to -m_\alpha$ leave topological invariants unchanged.
Consider the minimal stack involving Dirac Hamiltonians $\mathcal{H}^1_{\boldsymbol{\Gamma}} $, $\mathcal{H}^2_{\mathbf{M}}$, and $ \mathcal{H}^{\oplus}=\mathcal{H}^3_{\mathbf{X}} \oplus \mathcal{H}^4_{\mathbf{X'}} $, at $ \boldsymbol{\Gamma}$ and $\mathbf{M} $ and $ \mathbf{X}/\mathbf{X'}$ respectively.
Requiring $ \Delta[M_i]=\Delta[X]=0$ sets $ M_p=\Gamma_p$ and $X_{p}=(\Gamma_p)^2$.
(In this paragraph we understand $ \Pi_p^{(4)}$ to mean the \emph{occupied} eigenvalue of this minimal stack.)
The value of $\Gamma_p$ sets the sign of $ m_1 $ and $ \eta_1 $, and consequently $ \sgn{m_{2}} = \sgn{m_{1}} $ and $ \eta_2 = \eta_1 $ (see Tables \ref{tab:rotation_eigenvalue_correspondence} and~\ref{tab:physical_representations}).
Ensuring $\Delta \Ch=0$ then requires us to choose a mass at $\mathbf{X}/\mathbf{X'}$ ($ m_3 = m_4 $) with opposite sign to $m_{1,2}$.
Together with $X_{p}=(\Gamma_p)^2$ this singles out a specific $\eta_3$ (Tables \ref{tab:X-X_rotation_eigenvalue_correspondence} and~\ref{tab:physical_X-X_representations}).
(Recall from Table~\ref{tab:surface_representations} that the signs $ \eta_{\alpha} $ are associated with a specific $ \det u_{4,\mathbf{0}}^{\alpha}$.)
Crucially, no matter what the value of $\Gamma_p$ is we always find $ \det u_{4,\mathbf{c}}^{1} \det u_{4,\mathbf{c}}^{2} \det u_{4,\mathbf{c}}^{3} = 1$, both for $\mathbf{c}=\mathbf{0} $ and $ \mathbf{c}=(\mathbf{a}_1+\mathbf{a}_2)/2 $.
Specifically: for $\mathbf{c}=\mathbf{0} $, $ \det u_{4,\mathbf{0}}^{1} \det u_{4,\mathbf{0}}^{2} = 1 $ and $\det u_{4,\mathbf{0}}^{3} = 1$ (Tables \ref{tab:rotation_eigenvalue_correspondence} and~\ref{tab:physical_representations}); and for $\mathbf{c}=(\mathbf{a}_1+\mathbf{a}_2)/2$, both $\det u_{4,(\mathbf{a}_1+\mathbf{a}_2)/2}^{2}$ and $\det u_{4,(\mathbf{a}_1+\mathbf{a}_2)/2}^{3}$ change sign (Tables \ref{tab:X-X_rotation_eigenvalue_correspondence} and~\ref{tab:physical_X-X_representations}) while $\det u_{4,(\mathbf{a}_1+\mathbf{a}_2)/2}^{1}$ stays unchanged.

The corner mode index is therefore independent of the exact sequence of gap closings leading to a particular topological phase.

\subsection{Constructing Topological Index for Corner States}

Having thus established the path independence, and thus the topological nature of our $\det u_{n,\mathbf{c}}$ index, we must be able to express it in terms of the bulk topological invariants. 
As seen from Table~\ref{tab:surface_representations}, only certain bulk rotation representations $ \{ r_{4,\mathbf{c}}^{\alpha}, r_{4,\mathbf{c}}^{\oplus} \} $ lead to edge rotation representations with $ \det u_{4,\mathbf{c}}^{\alpha} = -1 $.
In turn, these $ \{ r_{4,\mathbf{c}}^{\alpha}, r_{4,\mathbf{c}}^{\oplus} \} $ are characteristic of changes in the occupation number of certain rotation eigenvalues.
Thus, by tracking changes in occupation of a subset of rotation eigenvalues, one may deduce the number of edge modes with $ \det u_{n,\mathbf{c}}^{\alpha} = -1 $, giving us $ \det u_{n,\mathbf{c}}^{} $.
We expect a $ \mathbb{Z}_2 $-valued index $ \Upsilon_{\mathbf{c}}^{(n)} $ defined, as in Eq.~\eqref{eq:Upsilon_index_relation}, by $ (-1)^{\Upsilon_{\mathbf{c}}^{(n)}} \equiv \det u_{n,\mathbf{c}}^{} $, where $ \Upsilon_{\mathbf{c}}^{(n)} $ counts the number of Majorana modes between neighboring edges.

We now describe how such a relation is obtained in $C_4$-symmetric systems.
The central idea is to track how the changes $\Delta(\#X_1^-, \#X_1^+, \#\Gamma_1, \# \Gamma_2, \#M_1, \#M_2)$ influence $\det u_{n,\mathbf{c}}^{}$.
We start with the case of the rotation center being at $\mathbf{c}=\mathbf{0}$.
Consider the fourfold fixed point $ \boldsymbol{\Gamma} $: We see from Tables \ref{tab:rotation_eigenvalue_correspondence} and~\ref{tab:surface_representations} that for a change in occupation $ \Delta \# \Gamma_2 $, there will be $ | \Delta \# \Gamma_2 | $ Dirac Hamiltonians added to the stack that have $ \det u_{4,\mathbf{0}}^{\alpha} = -1 $.
On the other hand, the $ | \Delta \# \Gamma_1 | $ other Dirac Hamiltonians at $\boldsymbol{\Gamma} $ have $ \det u_{4,\mathbf{0}}^{\alpha} = 1 $ so need not be counted.
Similarly, we should also count $ | \Delta \# M_2 | $ but not $ | \Delta \# M_1 | $.
As for the Dirac Hamiltonian pair at the $ \mathbf{X}/\mathbf{X'} $ points, Tables \ref{tab:X-X_rotation_eigenvalue_correspondence} and~\ref{tab:surface_representations} show that we should count $ |\Delta \# X_1^{+}| $ because $ \det u_{4,\mathbf{0}}^{\alpha} = -\eta_{\alpha} $ for these Dirac Hamiltonians.

Thus, recalling that $\det u_{4,\mathbf{c}}^{} = \prod_\alpha \det u^{\alpha}_{4,\mathbf{c}} $, the index for $\mathbf{c}=\mathbf{0} $ is
\begin{equation}
\Upsilon_{\mathbf{0}}^{(4)} = \Delta \#\Gamma_{2} + \Delta \#M_{2} + \Delta \#X_{1}^{+}\mod2,
\end{equation}
where due to the modulo $2$ we could drop the absolute value symbols. 
When $ \mathbf{c} = \frac{1}{2} (\mathbf{a}_1 + \mathbf{a}_2) $, however, one can see from Tables \ref{tab:physical_representations} and~\ref{tab:physical_X-X_representations} that representations at $\mathbf{M}$ and $\mathbf{X},\mathbf{X'}$ acquire a minus sign, so that by analogous arguments
\begin{equation}
\Upsilon_{(\mathbf{a}_1 + \mathbf{a}_2) / 2}^{(4)} = \Delta \#\Gamma_{2} + \Delta \#M_{1} + \Delta \#X_{1}^{-}\mod2.
\end{equation}
This shifted index and the original are related through
\begin{align}
\Upsilon_{(\mathbf{a}_1 + \mathbf{a}_2) / 2}^{(4)} &= \Upsilon_{\mathbf{0}}^{(4)} + \Delta \#M_{1} + \Delta \#M_{2} + \Delta \# X_1 \mod2 \\
&= \Upsilon_{\mathbf{0}}^{(4)} + \Delta [M_1] + \Delta [M_2] + \Delta [X] \mod 2.
\label{eq:two_indices_relation}
\end{align}

The rotation center thus only influences the existence of Majoranas on the edge if $ \Delta \nu = \Delta [M_1]+ \Delta [M_2]+ \Delta [X] \mod 2 \neq 0$.
Recognizing that $\nu$ is precisely the weak topological invariant in $C_4$-symmetric systems \cite{Benalcazar:2014hb}, Eq.~\eqref{eq:two_indices_relation} can be seen to express the combined effect of the weak invariant and the rotation center announced in the Introduction. It remains to rewrite $ \Upsilon_{\mathbf{0}}^{(4)} $ purely in terms of rotation invariants, which is possible using $ \Delta \Ch = 0 $ derived earlier.
Substituting $ \Delta \# X_1^{+} $ from Eq.~\eqref{eq:chern_condition}, and using that $\Upsilon_{\mathbf{0}}^{(4)}=\Delta[M_{2}]+\Delta\#X_{1}^{+}\mod2$ we get
\begin{equation}
\Upsilon_{\mathbf{0}}^{(4)}=\frac{1}{4}\left(\Delta[M_{1}]+3\Delta[M_{2}]-2\Delta[X]\right) \mod 2.
\label{eq:index_center}
\end{equation}
Summarizing the bulk-boundary correspondence in one equation, we have
\begin{equation}
\Upsilon_{\mathbf{c}}^{(4)}=\frac{1}{4}\left(\Delta[M_{1}]+3\Delta[M_{2}]-2\Delta[X]\right) + \frac{1}{2\pi} \Delta \mathbf{G}_{\nu} \cdot \mathbf{c} \mod 2,
\label{eq:index_corner}
\end{equation}
where $ \Delta \mathbf{G}_{\nu} = \Delta \nu (\mathbf{b}_1+\mathbf{b}_2) $ is the weak index vector (in terms of reciprocal lattice vectors $ \mathbf{b}_{i} $ satisfying $ \mathbf{a}_{i} \cdot \mathbf{b}_{j} = 2\pi \delta_{i j} $).
This index between rotation invariants and boundary Majorana bound states is one of the central predictions of our stacked Dirac approach. 

Although the intermediate steps made use of `extra' information $ \Delta \# X_1^+ $ specific to the stacked Dirac model construction, the physical conclusion depends only on the topological invariants.
The way that similar indices have been derived before is to find example systems with corner modes (corner charge) and appeal to the linearity of indices in terms of the invariants \cite{Teo:2013cp,Benalcazar:2014hb,Benalcazar:2019bs} to reconstruct their form.
In contrast, here we have shown how any continuum description consistent with a given change of bulk topological invariants of rotationally symmetric topological superconductors encodes transformation properties of adjacent edge Hamiltonians and thus the topological index for corner Majorana modes.

\section{Examples}
\label{sec:examples}

\begin{figure}
\includegraphics[scale=1]{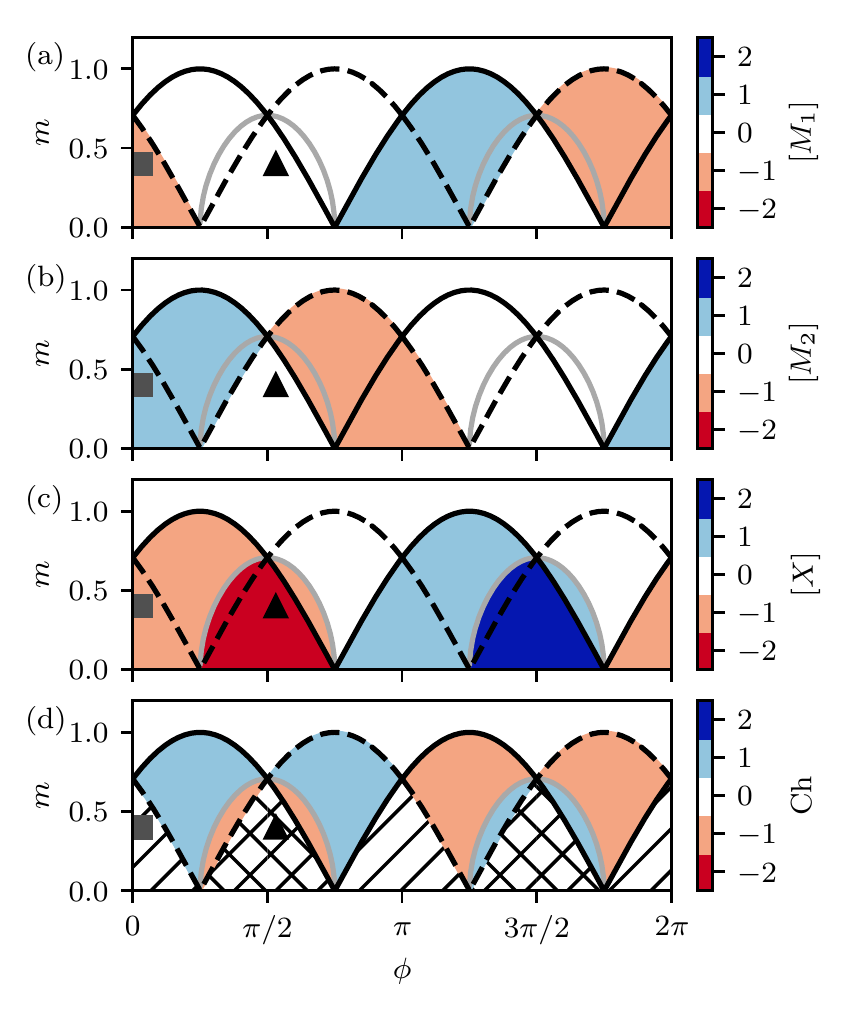}
\caption{Phase diagram for the lattice Hamiltonian~Eq.~\eqref{eq:general_benalcazar}. The black lines denote gap closings at $\boldsymbol\Gamma$, the black dashed lines at $\mathbf{M}$ and the gray lines at $\mathbf{X},\mathbf{X}'$.
In panel~(d), the black stripes in phases with $\mathrm{Ch}=0$ denote values of the topological index predicting corner modes.
Diagonal stripes $\Upsilon^{(4)}_\mathbf{0}=1$ and $\Upsilon^{(4)}_{(\mathbf{a}_1+\mathbf{a}_2)/2}=0$, and a crossed pattern $\Upsilon^{(4)}_\mathbf{0}=\Upsilon^{(4)}_{(\mathbf{a}_1+\mathbf{a}_2)/2} = 1$.
The gray square and black triangles mark the parameters used in Fig.~\ref{fig:eigenvalues}(a) and~(b), respectively.
}
\label{fig:phase_diagram}
\end{figure}

We illustrate our approach using a lattice model.
Consider a generalization of two models introduced by Benalcazar \textit{et al.\@} that are realized on a square lattice with primitive lattice vectors $\mathbf{a}_1 = a \hat{\mathbf{x}}$, $\mathbf{a}_2 = a \hat{\mathbf{y}}$~\cite{Benalcazar:2014hb}.
The combinations $\mathbf{a}_1'= \mathbf{a}_1 + \mathbf{a}_2$ and $\mathbf{a}_2' = -\mathbf{a}_1 +\mathbf{a}_2$ connect next-nearest-neighbor sites.
The BdG Hamiltonian
\begin{align}
 H(\mathbf{k})
 &=\begin{pmatrix}
   f_1 ( \mathbf{k}) \sigma_3 + g_1 (\mathbf{k}) \sigma_2 &  m ( \sigma_3 - i \sigma_0)  \\
 m (\sigma_3 + i \sigma_0 )    & f_2 ( \mathbf{k}) \sigma_3 + g_2 (\mathbf{k}) \sigma_2  \\ 
 \end{pmatrix}
 \label{eq:general_benalcazar}
\end{align}
with the onsite coupling $m$ and the two functions
\begin{align}
 f_i (\mathbf{k}) &= \cos \phi \cos (\mathbf{k}\cdot \mathbf{a}_i) + \sin \phi \cos (\mathbf{k}\cdot \mathbf{a}_i')  \\
 g_i (\mathbf{k}) &= \cos \phi \sin (\mathbf{k}\cdot \mathbf{a}_i) + \sin \phi \sin (\mathbf{k}\cdot \mathbf{a}_i')
\end{align}
describes a $C_4$-symmetric superconductor with PH symmetry $\Xi = \sigma_1 \mathcal{K} $ and fourfold rotation  
\begin{equation}
 r_4 =
 \begin{pmatrix}
   & - i \sigma_3 \\
   \sigma_0 & 
 \end{pmatrix}
\end{equation}
where $r_4^4 = -1$.
(All units of energy are absorbed into the Hamiltonian.)
As discussed in Sec.~\ref{sec:benalcazar_classification}, each gapped phase is characterized by a set four invariants, which we show in the phase diagram in Fig.~\ref{fig:phase_diagram}.
Changing the parameters $\phi \to \phi + \pi$ and $m \to -m$ results in $H \to -H$ hence in Fig.~\ref{fig:phase_diagram} we consider only positive $m$ values.
For $|m| >1$, the onsite coupling dominates and the Hamiltonian is trivial, independently of the parameter $\phi$.

\begin{figure}
\includegraphics[scale=1]{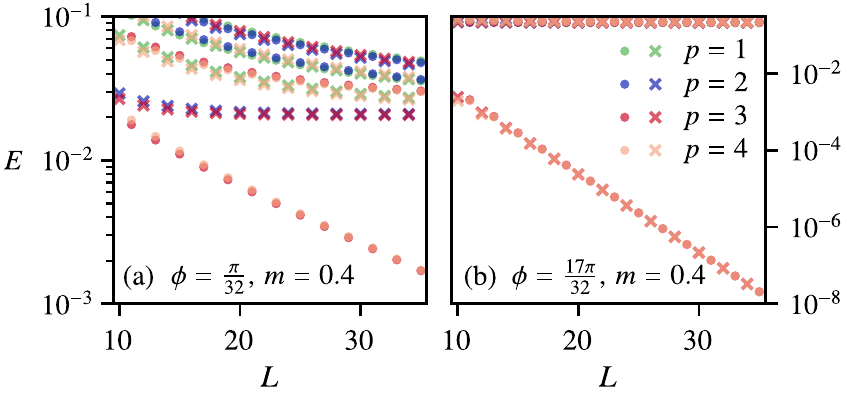}
\caption{Energy eigenvalues of the Hamiltonian defined in Eq.~\eqref{eq:general_benalcazar} with a boundary perturbation (with $t=0.025$) for a finite square lattice with $L \times L$ sites. The different colors denote the rotation eigenvalue $e^{i\pi/4 (2p-1)}$ and the different symbols distinguish different rotation centers (crosses and circles corresponding to even and odd $L$, respectively).
In panel (a), we show an example of phase I with $\phi = \pi/32$ and $m = 0.4$.
The system only supports gapless corner modes for odd $L$, which corresponds to a physical rotation center in the center of a unit cell.
In panel (b), we show an example of phase II with $\phi = 17\pi/32$ and $m = 0.4$. The corner modes remain for both rotation centers, i.e., both even and odd $L$.
We choose the logarithmic scale of the $y$ axis to visualize the exponential decrease in energy.}
\label{fig:eigenvalues}
\end{figure}

\begin{figure}
\centering
\includegraphics{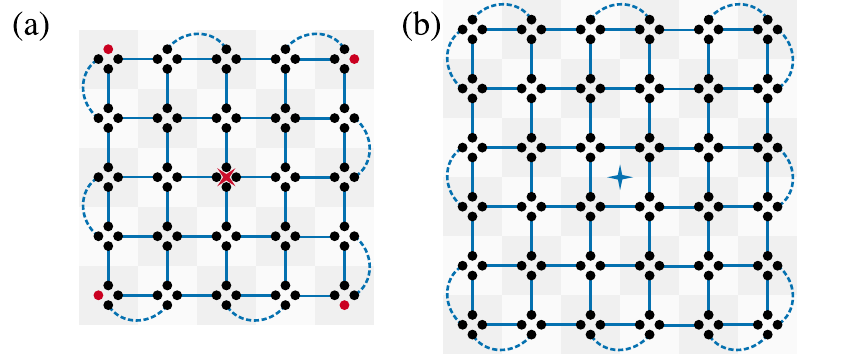}
\caption{Lattice model that demonstrates the importance of the rotation center. The Hamiltonian~\eqref{eq:general_benalcazar} at $\phi=m=0$ only contains terms, illustrated by the solid blue lines, that couple neighboring Majorana modes, illustrated by the black circles; cf.\ Ref.~\onlinecite{Benalcazar:2014hb}.
In this case, Majorana modes at the edge are completely decoupled from the bulk and do not contribute to the Hamiltonian, i.e., all edge modes have zero energy.
This degeneracy can be lifted by a density wave~\cite{Benalcazar:2014hb} modeled by coupling every second nearest-neighbor site on the edge (dashed blue lines).
(a) When the rotation center is in the center of a unit cell, any coupling that respects rotational invariance is bound to leave Majoranas at the corners uncoupled (red circles). This completely decoupled case is topologically equivalent to any case with a finite localization length of the corner modes, as for example considered in Fig.~\ref{fig:eigenvalues}.
(b) However, when the rotation center is at the corner of a unit cell, it is possible to open a surface gap without Majorana bound states.
}
\label{fig:even-odd}
\end{figure}

Two regimes are relevant for our classification: phase I around $(\phi,m)= (0,0)$ and phase II around $(\phi,m) = (\pi/2,0)$.
Phase I is characterized by $\Upsilon^{(4)}_\mathbf{0}=1$ and $\Upsilon^{(4)}_{(\mathbf{a}_1+\mathbf{a}_2)/2}=0$, thus, it only supports corner modes when the physical rotation center is in the center of a unit cell  (cf. Sec.~\ref{subsec:rotation_center} and Appendix~\ref{sec:unit_cell_restriction} for our notion of the unit cell).
To couple counterpropagating chiral edge modes, we add a density-wave-type boundary perturbation that respects rotation invariance but has periodicity of two unit cells.
Specifically, on each edge we couple every second pair of lattice sites via a nearest-neighbor hopping term $ i t c_j^\dagger \tau_z c_{j+1}^{} $ (where $\tau_{\mu}$ acts on the outer degree of freedom) to break translation invariance and open a gap. 
(In Fig.~\ref{fig:even-odd} we show the boundary perturbation together with the $m=\phi=0$ limit of the bulk system.)
We show the energy eigenvalues for square lattices with $L\times L$ sites in Fig.~\ref{fig:eigenvalues}(a).
When $L$ is odd, the rotation center is in the center of a unit cell, when $L$ is even, it is at its corner.
Corner modes therefore only arise when $L$ is odd.
Phase II, however, is characterized by $\Upsilon^{(4)}_\mathbf{0}=\Upsilon^{(4)}_{(\mathbf{a}_1+\mathbf{a}_2)/2}=1$, meaning that the presence of corner modes does not depend on the position of the rotation center, as we show in Fig.~\ref{fig:eigenvalues}(b).

In phase II, the surface gap closes when $m=0$.
Then, the corner modes delocalize along the edge and their energy in any finite system increases accordingly.
When tuning the parameters of the Hamiltonian to cross $m=0$, the localization length of the corner modes first increases when approaching $m=0$ and then decreases again with increasing surface gap size.
Thus, the presence of corner modes solely depends on bulk properties.
Corner modes may at most delocalize for fine-tuned points in parameter space, but they cannot be removed by attempting a surface-only topological phase transition via a surface gap closing. 

Using the above phases, more phases can be constructed by stacking different copies of this model.
For example, stacking phases I and II results in hybridization of the corner modes, such that $\Upsilon^{(4)}_\mathbf{0}=0$ and $\Upsilon^{(4)}_{(\mathbf{a}_1+\mathbf{a}_2)/2}=1$, i.e., only systems with the physical rotation center at the corners of the unit cell support corner modes.
Stacking the four primitive models introduced in Ref.~\onlinecite{Benalcazar:2014hb} enables us to construct models that realize all possible combinations of the bulk invariants.

\section{Conclusion}
\label{sec:conclusion}

In this work, we formulated a second-order bulk-boundary correspondence for $C_n$-symmetric 2D crystalline superconductors: We related the bulk topological invariants of Ref.~\onlinecite{Benalcazar:2014hb} to a topological index $\Upsilon^{(n)}_{\mathbf{c}}$ accounting for the presence of Majorana corner states in systems with $C_n$-symmetry-respecting boundaries.
The exact form of the corner index depends on the interplay of the weak invariants and the location of the physical rotation center with respect to the unit cell.
In particular, certain systems only support corner modes when the rotation center is in the center of a unit cell, while other systems require it to be at the unit cell boundary.
(These statements have no unit-cell-choice ambiguity: For a well-defined bulk-boundary correspondence we must use unit-cell choices described in Sec.~\ref{sec:benalcazar_classification} and Appendix~\ref{sec:unit_cell_restriction}.)
Our approach to identify the corner index is based on stacked Dirac Hamiltonians.
It is thus extendable to other crystal symmetries, as long as they can be represented for a minimal model of stacked Dirac Hamiltonians.

The index we find is consistent with previous classification schemes in rotationally symmetric superconductors.
For example, Teo and Hughes found an invariant for Majorana modes trapped at lattice defects that strongly resembles the indices given in Eqs.~\eqref{eq:index_center} and~\eqref{eq:index_corner}~\cite{Teo:2013cp,Benalcazar:2014hb}.
When predicting trapped Majorana modes, the Burgers vector of a lattice defect only matters when the weak invariant is nonzero, similar to the fact that $ \Upsilon_{\mathbf{c}=\mathbf{0}}^{(n)} $ and $ \Upsilon_{\mathbf{c}\neq\mathbf{0}}^{(n)} $ may only be different if the weak invariant is nonzero.

Invoking a counting argument, Ref.~\onlinecite{Teo:2013cp} noted that the invariants constructed for lattice defects can also be used to predict corner modes in finite systems.
Our work elucidates why this is so from an entirely different viewpoint:
We established how bulk invariants relate to the transformation properties of adjacent edge Hamiltonians, the latter having become the unifying perspective for constructing various examples of higher-order topological phases~\cite{Geier:2018ev,Khalaf:2018hq,Trifunovic:2019hi}.

We illustrated our results using lattice models.
In particular, we showed that the physical rotation center in finite systems may indeed give rise to different corner mode configurations.
Furthermore, we explicitly demonstrated that the bulk-boundary correspondence is robust against gap closings at the boundary, i.e., that the presence of corner modes is purely determined by bulk quantities that relate different edges to another.
In all lattice model examples, we identified corner modes using the scaling of the energies: For finite 2D square samples of size $L\times L$, the energy of the second-order bound states decays exponentially with $L$.

The latter scaling observation may be particularly helpful for future studies considering hybrid higher-order topology~\cite{Bultinck:2019cn}, expected to arise in our systems when we allow for nonzero Chern number.
In such cases, the quantized energy levels of the delocalized chiral edge modes are expected to show a $1/L$ decay with increasing $L$, in sharp contrast with the exponential decay of the second-order bound state energies. 

\begin{acknowledgments}
This work was supported by an EPSRC Studentship and the ERC Starting Grant No. 678795 TopInSy.
\end{acknowledgments}

\appendix

\section{Restrictions on Choice of Unit Cell}
\label{sec:unit_cell_restriction}

The bulk classification of Ref.~\onlinecite{Benalcazar:2014hb} of rotationally symmetric superconductors employed in the main text relies on having momentum-independent matrices $r_n$ that rotate the momentum-space tight-binding Hamiltonian, cf.\ Eq.~\eqref{eq:bulk_symmetry}.
Generally, the form of the tight-binding Hamiltonian depends on the choice of the basis functions.
In particular, using orbitals $\varphi_{\mathbf{R},\alpha} (\mathbf{r}-\mathbf{R}-\mathbf{d}_\alpha)$ for each orbital $\alpha$ at the position $\mathbf{R} + \mathbf{d}_\alpha$ with the Bravais lattice vector $\mathbf{R}$ and atomic position $\mathbf{d}_\alpha$ enables us to construct basis functions~\cite{Goringe:1997bx,Alexandradinata:2016kb}
\begin{equation}
 \bar{\phi}_{\mathbf{k},\alpha}^{}(\mathbf{r}) = \frac{1}{\sqrt{N}} \sum_\mathbf{R} e^{-i \mathbf{k}\cdot (\mathbf{R}+\mathbf{d}_\alpha)} \varphi_{\mathbf{R},\alpha} (\mathbf{r}-\mathbf{R}-\mathbf{d}_\alpha)
\end{equation}
where the sum goes over all $N$ unit cells at positions $\mathbf{R}$.
The resulting tight-binding Hamiltonian
\begin{equation}
 \bar{H}_{\alpha\beta} (\mathbf{k}) = \int d^2 \mathbf{r}\, \bar{\phi}_{\mathbf{k},\alpha}^* (\mathbf{r}) \hat{H} \bar{\phi}_{\mathbf{k},\beta}^{} (\mathbf{r})
\end{equation}
with the operator $\hat{H}$ acting on the basis functions is not periodic under a shift of a reciprocal lattice vector $\mathbf{G}$, but rather transforms~\cite{Alexandradinata:2014ju,Alexandradinata:2016kb}
\begin{equation}
 \bar{H} (\mathbf{k}+\mathbf{G}) = \mathcal{V}^\dagger (\mathbf{G}) \bar{H} (\mathbf{k}) \mathcal{V} (\mathbf{G}) .
\end{equation}
The unitary matrix $\mathcal{V} (\mathbf{G})$ takes into account the momentum-dependence of the different atomic sites at $\mathbf{d}_\alpha$ within each unit cell.
The matrix is diagonal with elements $ \mathcal{V}_{\alpha \beta}(\mathbf{G}) = e^{-i \mathbf{d}_\alpha \cdot \mathbf{G}} \delta_{\alpha\beta}$.

The benefit of this basis choice is that matrix representations of symmorphic symmetries, such as rotation, are always momentum-independent. For example, a rotation $\hat{R}$ that rotates to $R_n (\mathbf{R}+\mathbf{d}_\alpha) = \mathbf{R}' + \mathbf{d}_\beta$ changes the creation operators of the orbital at $\mathbf{R}+\mathbf{d}_\alpha$~\cite{Fang:2013jk}
\begin{equation}
 \hat{R} \bar{c}^\dagger_{\alpha} (\mathbf{R}+\mathbf{d}_\alpha) \hat{R}^{-1} = \bar{c}_\beta^\dagger (\mathbf{R}'+\mathbf{d}_\beta) \mathcal{R}_{\beta\alpha}
\end{equation}
where the unitary matrix  $\mathcal{R}_{\alpha\beta}$ accounts for rotation of atomic orbitals, as already used in the main text.
This implies for the momentum-space representation of the annihilation operator~\cite{Alexandradinata:2016kb}
\begin{align}
   \hat{R} \bar{c}^\dagger_{\alpha} (\mathbf{k}) \hat{R}^{-1}
=& \hat{R} \sum_\mathbf{R} \bar{c}^\dagger_\alpha (\mathbf{R}+\mathbf{d}_\alpha) e^{i\mathbf{k}\cdot (\mathbf{R}+\mathbf{d}_\alpha)} \hat{R}^{-1} \\
=& \sum_{\mathbf{R}} \bar{c}_\beta^\dagger (\mathbf{R}'+\mathbf{d}_\beta) \mathcal{R}_{\beta\alpha} e^{i\mathbf{k}\cdot (\mathbf{R}+\mathbf{d}_\alpha)}.
\end{align}
Using $\mathbf{R} + \mathbf{d}_\alpha = R_n^T (\mathbf{R}' + \mathbf{d}_\beta)$ and changing the summation from $\mathbf{R}\to \mathbf{R}'$ gives
\begin{align}
   \hat{R} \bar{c}^\dagger_{\alpha} (\mathbf{k}) \hat{R}^{-1}
=& \sum_{\mathbf{R}'} \bar{c}_\beta^\dagger (\mathbf{R}'+\mathbf{d}_\beta) e^{i(R_n \mathbf{k})\cdot (\mathbf{R}'+\mathbf{d}_\beta)} \mathcal{R}_{\beta\alpha} \\
=& \bar{c}_\beta^\dagger (R_n\mathbf{k}) \mathcal{R}_{\beta\alpha} .
\end{align}
The Hamiltonian $\hat{H} = \sum_{\mathbf{k}} \bar{c}_\alpha^\dagger (\mathbf{k}) \bar{H}_{\alpha\beta} (\mathbf{k}) \bar{c}_\beta (\mathbf{k})$ is invariant under the rotation $\hat{R}$, giving~\cite{Fang:2013jk}
\begin{align}
 \hat{R} \hat{H} \hat{R}^{-1}
 &= \sum_{\mathbf{k}} \bar{c}_\alpha^\dagger (R_n\mathbf{k}) \mathcal{R}_{\alpha\alpha'} \bar{H}_{\alpha'\beta'}  (\mathbf{k}) \mathcal{R}_{\beta\beta'} \bar{c}_\beta (R_n \mathbf{k}) \nonumber \\
 &= \sum_{\mathbf{k}} \bar{c}_\alpha^\dagger (R_n \mathbf{k}) \bar{H}_{\alpha\beta} (R_n \mathbf{k}) \bar{c}_\beta (R_n\mathbf{k}) = \hat{H}
\end{align}
which implies $\mathcal{R} \bar{H} (\mathbf{k}) \mathcal{R}^\dagger = \bar{H} (R_n \mathbf{k})$ for the tight-binding Hamiltonian. For superconducting BdG Hamiltonians, the structure of the Nambu spinors needs to be taken into account, which promotes the operator $\mathcal{R}$ to
\begin{equation}
 r_n = \begin{pmatrix} \mathcal{R} & \\ & \mathcal{R}^* \end{pmatrix},
\end{equation}
cf.\ Eq.~\eqref{eq:bulk_symmetry}.
These operators are always independent of momentum; cf.\ Ref.~\onlinecite{Alexandradinata:2016kb} for a more general discussion that includes both symmorphic and nonsymmorphic symmetries.

In the main text, we implicitly use a different set of basis functions that gives the tight-binding Hamiltonian
\begin{equation}
 H (\mathbf{k})
 = \mathcal{V} (\mathbf{k}) \bar{H} (\mathbf{k}) \mathcal{V}^\dagger (\mathbf{k}) .
\end{equation}
This different basis choice has the advantage that the tight-binding Hamiltonian is invariant upon a shift by a reciprocal lattice vector, especially that $\mathcal{H}$ is identical at certain HSPs $\boldsymbol{\Pi}^{(n)}$ and their rotated counterparts $R_n \boldsymbol{\Pi}^{(n)}$, e.g., at $\mathbf{M}$ and $R_4 \mathbf{M}$ in $C_4$-symmetric lattices.
However, as pointed out in the main text, the operator $r_n$ is generally momentum-dependent, in particular,
\begin{align}
H (R\mathbf{k})
 &= \mathcal{V} (R_n \mathbf{k}) r_n \mathcal{V}^\dagger (\mathbf{k}) H (\mathbf{k}) \mathcal{V} (\mathbf{k}) r_n^\dagger \mathcal{V}^\dagger (R_n \mathbf{k}) \\
 &= r_n' (\mathbf{k}) H (\mathbf{k}) {r_n'}^\dagger (\mathbf{k})
\end{align}
with $r_n' (\mathbf{k}) = \mathcal{V} (R_n \mathbf{k}) r_n \mathcal{V}^\dagger (\mathbf{k})$.
We restore Eq.~\eqref{eq:bulk_symmetry} when $r_n \mathcal{V}^\dagger (\mathbf{k}) = \mathcal{V}^\dagger (R_n \mathbf{k}) r_n$.
This relation is only true when each atomic position $\mathbf{d}_\alpha$ is rotated to a (not necessarily different) position $\mathbf{d}_\beta$ \emph{within the same unit cell}.
We realize this by computing the action of $r_n$ on $ \mathcal{V}^\dagger (\mathbf{k})$ explicitly.
The matrix $r_n$ shifts all atomic sites $\mathbf{d}_\alpha \to R_n \mathbf{d}_\alpha$ and transforms the internal degrees of freedom on each $\mathbf{d}_\alpha$ (for example, $p_x$ orbitals are transformed into $p_y$ orbitals under a $C_4$ rotation).
The matrix elements of $r_n$ can thus be written
\begin{equation}
 r_n^{\alpha\beta} = \tilde{r}_{\alpha\beta} \delta_{\mathbf{d}_\alpha, R_n \mathbf{d}_\beta},
\end{equation}
where $\tilde{r}$ transforms the internal degrees of freedom and the Kronecker delta ensures that all sites $\mathbf{d}_\beta$ are changed to $R_n \mathbf{d}_\beta$.
Then,
\begin{align}
 \left. r_n \mathcal{V}^\dagger (\mathbf{k}) \right|_{\alpha\beta}
 &= \tilde{r}_{\alpha\alpha'} \delta_{\mathbf{d}_\alpha, R_n \mathbf{d}_{\alpha'}} e^{i \mathbf{k} \cdot \mathbf{d}_{\alpha'}} \delta_{\alpha'\beta}
  = e^{i \mathbf{k} \cdot R_n^T \mathbf{d}_{\alpha}} \tilde{r}_{\alpha\beta} \delta_{\mathbf{d}_\alpha, R_n\mathbf{d}_\beta} \nonumber \\
 &= e^{i (R_n \mathbf{k}) \cdot \mathbf{d}_{\alpha}} \delta_{\alpha\alpha'} \tilde{r}_{\alpha'\beta} \delta_{\mathbf{d}_{\alpha'}, R_n \mathbf{d}_\beta}
  = \left. \mathcal{V}^\dagger (R_n \mathbf{k}) r_n \right|_{\alpha\beta} ,
\end{align}
i.e., the operator $r_n$ acting on the tight-binding Hamiltonian $H(\mathbf{k})$ is momentum-independent. This derivation relies on a rotationally invariant unit cell, since each unit cell must contain both atomic positions $\mathbf{d}_\alpha$ and $R\mathbf{d}_\alpha$.

\begin{figure}
\includegraphics{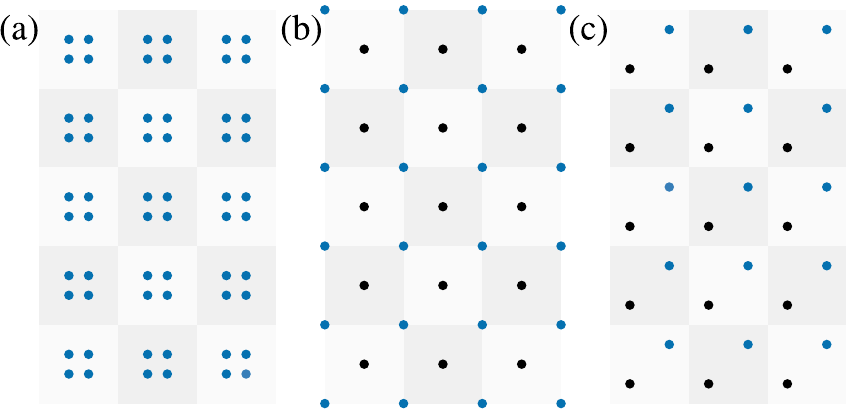}
\caption{Different $C_4$-symmetric lattices. (a) Each unit cell respects $C_4$ symmetry individually and is compatible with a finite lattice. (b) Each unit cell respects $C_4$ symmetry, but contains fractional atomic sites and is therefore incompatible with a finite lattice. (c) For the same lattice as in (b), we can define a different unit cell that is compatible with a finite system, but does not respect $C_4$ symmetry.}
\label{fig:appendix_unit_cell}
\end{figure}

Not every rotationally invariant lattice allows us to define unit cells that respect rotational invariance individually, as we demonstrate using Fig.~\ref{fig:appendix_unit_cell}.
The lattice shown in Fig.~\ref{fig:appendix_unit_cell}(a) is $C_4$-symmetric lattice with four atomic sites in each unit cell.
Two different choices of unit cells respect rotational invariance individually and are compatible with a finite system.
Similarly, the lattice shown in Fig.~\ref{fig:appendix_unit_cell}(b) and~(c) is $C_4$-symmetric, however, any finite system that respect $C_4$ symmetry is incompatible with a $C_4$-symmetric unit cell.
While the unit cell in Fig.~\ref{fig:appendix_unit_cell}(b) is rotationally invariant, it contains fractional atomic sites.
Any lattice boundary must therefore contain additional partial unit cells.
The unit cell choice in Fig.~\ref{fig:appendix_unit_cell}(c) is compatible with a finite system, but the unit cell itself is not $C_4$-symmetric, such that $\mathbf{d}_\alpha$ and $R \mathbf{d}_\alpha$ are not contained in each unit cell.
This gives some additional momentum-dependent contribution to $r_n' (\mathbf{k}) = \mathcal{V} (R \mathbf{k}) r_n \mathcal{V}^\dagger (\mathbf{k})$, which in turn spoils the bulk classification used in the main text that relies on $[r_n,H(\boldsymbol\Pi^{(n)})] = 0$~\cite{Benalcazar:2014hb}.
The momentum-dependent contribution is generally model-dependent, such that a classification is beyond the scope of this work.

\section{Derivation of Edge Theory}
\label{sec:derivation_edge_theory}

In this Appendix, we show explicitly how each pair of bands in the stack of Dirac Hamiltonians give rise to chiral edge modes, in a description that allows for smooth (on the scale of the lattice spacing) variations of the boundary.
We follow a similar prescription to Refs.~\onlinecite{Geier:2018ev,Khalaf:2018hq}, in which we project onto the low-energy subspace of states localized to the edge of the material.
This gives rise to explicit forms of the rotation operator and PH operator on the edge---although overall signs are generally basis-dependent, certain signatures that determine the presence of corner modes are independent of the choice of basis; cf.\ Appendix~\ref{sec:surface_projections}.

\subsection{Effective Boundary Hamiltonian}
\label{sec:effective_boundary_hamiltonian}

For each Dirac model in the stack, allow the mass term to vary spatially $ m_{\alpha} \to m_{\alpha}(\mathbf{r})$ and decompose momentum into components parallel and perpendicular to the boundary.
For the unit vector $ \hat{\mathbf{n}}_{\mathbf{r}}=(\cos \varphi,\sin \varphi) $ normal to the edge (which varies as a function of position $ \mathbf{r} $ along the boundary), decompose $ \mathbf{k}= k_{\parallel}\hat{\mathbf{n}}_{\parallel} + k_{\perp}\hat{\mathbf{n}}_{\mathbf{r}} $ and take $ \hat{\mathbf{n}}_{\parallel} = (-\sin \varphi,\cos \varphi) $ to follow the edge in a counterclockwise direction such that $ k_{\parallel} $ may be positive or negative.
(We also define $ \mathbf{k}_{\parallel} = k_{\parallel} \hat{\mathbf{n}}_{\parallel} $ for convenience.)
Let $ \lambda $ be a coordinate along the edge normal, where $\lambda=0$ denotes the position of the boundary where $ m_{\alpha}(\lambda=0) = 0 $ changes sign.
When a transition is realized through many simultaneous gap closings, all gap closings $ m_{\alpha}(\mathbf{r}) = 0 $ happen at the same boundary.
In this notation, each Hamiltonian in the stack reads
\begin{equation}
 \mathcal{H}^{\alpha}(\mathbf{k})=
 m_{\alpha}(\lambda)\sigma_{3}+
 v_{\alpha} \mathbf{k}_{\parallel}\cdot\boldsymbol{\sigma}
 -i v_{\alpha}\hat{\mathbf{n}}_{\mathbf{r}}\cdot\boldsymbol{\sigma}\partial_{\lambda}
\end{equation}
with $v_\alpha >0$ as also used in the main text.
In principle, since the normal vectors $\hat{\mathbf{n}}_\mathbf{r}$ and $\hat{\mathbf{n}}_\parallel$ depend on the position $\mathbf{r}$ along the boundary, the momentum operator $\mathbf{k}_\parallel$ does not commute with them.
As we only want to consider slowly varying normal vectors, we can neglect this nonzero commutator.
Similarly, this approach does not cover sharp changes of the normal vectors~\cite{Khalaf2019}, which appear directly at the corners of a sample.
This does not limit our analysis, as the description away from these sharp changes remains valid and allows to observe differences between smooth edges.

When a mass term changes sign as $\sgn (m_\alpha (\lambda)) =\sgn (\lambda) $ along the transition, chiral modes localized to the edge may be found using the ansatz
\begin{equation}
\Psi_\alpha (\mathbf{k}_{\parallel},\lambda)=e^{-\int_{0}^{\lambda}d\lambda'm_{\alpha}(\lambda') / v_{\alpha}} \psi_\alpha (\mathbf{k}_{\parallel}).
\label{eq:ansatz}
\end{equation}
Substituting this ansatz into the Hamiltonian, we obtain
\begin{equation}
 \left( 2 m_\alpha (\lambda) \sigma_3 P_+ + v_\alpha \mathbf{k}_\parallel \cdot \boldsymbol\sigma \right) \psi_\alpha  (\mathbf{k}_{\parallel})
 =  E_{\mathbf{k}_\parallel}  \psi_\alpha  (\mathbf{k}_{\parallel}),
\end{equation}
with the projector $P_\pm = \frac{1}{2} ( 1 \pm i \sigma_3 \hat{\mathbf{n}}_\mathbf{r} \cdot \boldsymbol\sigma ) = \frac{1}{2} ( 1 \mp \hat{\mathbf{n}}_\parallel \cdot \boldsymbol\sigma )$.
The wave function $ \psi_\alpha (\mathbf{k}_{\parallel})$ is only a $\lambda$-independent solution when $P_+ \psi_\alpha  (\mathbf{k}_{\parallel}) = 0$.
As $P_+ \psi_\alpha  (\mathbf{k}_{\parallel}) = 0$ implies $P_-  \psi_\alpha  (\mathbf{k}_{\parallel}) = \psi_\alpha (\mathbf{k}_{\parallel})$, the solution satisfies
\begin{equation}
 v_\alpha \mathbf{k}_\parallel \cdot \boldsymbol\sigma  \psi_\alpha (\mathbf{k}_{\parallel}) = v_\alpha k_\parallel \psi_\alpha (\mathbf{k}_{\parallel}) .
\end{equation}
Had the bulk mass changed in the opposite way as $ \sgn (m_{\alpha}(\lambda)) = - \sgn(\lambda) $, the ansatz would have a different sign in the exponent, and the solutions $P_-\psi_\alpha (\mathbf{k}_{\parallel})=0$ would propagate in the opposite direction.

Here we switch to a more convenient basis, generated by $V_{\mathbf{r}\rightarrow}$ such that
\begin{subequations}\begin{align}
V_{\mathbf{r}\rightarrow}^\dagger P_{+} V_{\mathbf{r}\rightarrow}^{}
&=\frac{1}{2}(1-\sigma_{3}),\\
V_{\mathbf{r}\rightarrow}^\dagger (v_{\alpha} \mathbf{k}_{\parallel}\cdot\boldsymbol{\sigma}) V_{\mathbf{r}\rightarrow}^{}
&=v_{\alpha} k_{\parallel}\sigma_{3}.
\label{eq:basis_rotation_symbolic}
\end{align}\end{subequations}
This can be achieved by choosing $ V_{\mathbf{r}\rightarrow}^{} \propto \exp(i\frac{\pi}{4}\hat{\mathbf{n}}_{\mathbf{r}}\cdot\boldsymbol{\sigma}) $, where we shall fix this constant of proportionality below using PH symmetry.
This allows the $ 1 \times 1 $ edge Hamiltonian to be easily procured by applying a projector $ p_+ = (1, 0)^T $ to pick out the correct subspace.
Explicitly performing these steps, we end up with a low-energy edge Hamiltonian for right-movers
\begin{subequations}\begin{align}
 h_{\mathbf{r},\mathbf{k}}^{\alpha\rightarrow}
 &\equiv p_{+}^{T} V_{\mathbf{r}\rightarrow}^{\dagger} \mathcal{H}^{\alpha}(\mathbf{k})V_{\mathbf{r}\rightarrow}^{} p^{}_{+}\\
 &= v_{\alpha} k_{\parallel}.
\end{align}\label{eq:ham_right_moving}\end{subequations}
Performing similar steps for left-moving ans\"{a}tze, differing in the choice of basis $ V_{\mathbf{r}\leftarrow}^{} $ and projected component $ p_{-} = (0, 1)^T $---though one still has $ V_{\mathbf{r}\leftarrow}^{} \propto \exp(i\frac{\pi}{4}\hat{\mathbf{n}}_{\mathbf{r}}\cdot\boldsymbol{\sigma}) $---we obtain a similar edge Hamiltonian dispersing with opposite velocity
\begin{subequations}\begin{align}
h_{\mathbf{r},\mathbf{k}}^{\beta\leftarrow}
&\equiv p_{-}^{T} V_{\mathbf{r}\leftarrow}^{\dagger} \mathcal{H}^{\beta}(\mathbf{k}) V_{\mathbf{r}\leftarrow}^{} p_{-} \\
&= - v_{\beta} k_{\parallel}.
\end{align}\label{eq:ham_left_moving}\end{subequations}

\subsection{Surface Projections of Rotation Representations}
\label{sec:surface_projections}

The advantage of the transformation that projects on boundary modes [Eqs.~\eqref{eq:ham_right_moving} and ~\eqref{eq:ham_left_moving}] is that it allows to track the transformation of edge modes, as we show in this section.
We first discuss how to fix a basis requiring PH symmetry before computing the edge projections of rotation and PH symmetry.

\subsubsection{Choice of Basis}
\label{subsec:basis_choice}

The transformation $V_{\mathbf{r}\rightarrow}$ that rotates the projector $P_+$ into $ V_{\mathbf{r}\rightarrow}^\dagger P_{+}V_{\mathbf{r}\rightarrow} = \frac{1}{2}(1-\sigma_{3}) $ is only defined up to a phase, $ V_{\mathbf{r}\rightarrow} \propto \exp(i\frac{\pi}{4}\hat{\mathbf{n}}_{\mathbf{r}}\cdot\boldsymbol{\sigma}) $.
Here, we fix this phase by requiring that the eigenstates $\psi_{\alpha} (\mathbf{k}) = V_{\mathbf{r}\rightarrow} p_+$ respect PH symmetry, i.e., $ \Xi \psi_{\alpha}(-\mathbf{k}_{\parallel}) = \psi_{\alpha}(\mathbf{k}_{\parallel}) $ with $\Xi = \sigma_1 \mathcal{K}$.
Further using $p_+ = (1,0)^T$ gives
\begin{equation}
V_{\mathbf{r}\rightarrow}= e^{-i (\pi/4 + \varphi/2)} \exp \left(i\frac{\pi}{4}\hat{\mathbf{n}}_{\mathbf{r}}\cdot\boldsymbol{\sigma} \right) .
\end{equation}
Note that this is the same as requiring
\begin{equation}
 r_n^{\star} V_{\mathbf{r}\rightarrow} p_+ = V_{R_n\mathbf{r}\rightarrow} p_+,
 \label{eq:basis_rotation_transformation_new}
\end{equation}
where $ r_n^{\star} = e^{-i\sigma_3\pi/n} $ is the positively signed rotation representation.
For left-moving modes, the basis rotation operator is chosen as
\begin{equation}
V_{\mathbf{r}\leftarrow}= e^{-i (\pi/4 - \varphi/2)} \exp \left( i\frac{\pi}{4}\hat{\mathbf{n}}_{\mathbf{r}}\cdot\boldsymbol{\sigma} \right)
\end{equation}
for the same reasons.

\subsubsection{Surface Rotation Representations from Bulk}
\label{subsec:surface_rotation}

Using the notation introduced above, we show how to derive the edge rotation representation from the bulk representation along the lines of Ref.~\onlinecite{Khalaf:2018hq}.
For some gap closings, e.g., at $\boldsymbol\Gamma$ and $\mathbf{M}$ in $C_4$-invariant systems, each Dirac Hamiltonian from the stack respects rotational invariance via $r_n^\alpha \mathcal{H}^\alpha (\mathbf{k}) r_n^{\alpha\dagger} = \mathcal{H}^\alpha (R_n \mathbf{k})$; cf.\ Eq.~\eqref{eq:bulk_dirac_symmetry}.
The edge Hamiltonian for right-moving states, Eq.~\eqref{eq:ham_right_moving}, thus transforms as
\begin{align}
h_{\mathbf{r},\mathbf{k}}^{\alpha\rightarrow}
 =& p_{+}^{T} V_{\mathbf{r}\rightarrow}^\dagger \mathcal{H}^{\alpha} (\mathbf{k}) V_{\mathbf{r} \rightarrow}^{} p_{+}^{}\\
 =& p_{+}^{T} V_{\mathbf{r}\rightarrow}^\dagger r_{n,\mathbf{c}}^{\alpha\dagger}  V_{R_n \mathbf{r}\rightarrow}^{} V_{R_n\mathbf{r}\rightarrow}^{\dagger} \mathcal{H}^{\alpha} (R_n \mathbf{k}) V_{R_n \mathbf{r}\rightarrow}^{} V_{R_n\mathbf{r}\rightarrow}^{\dagger} r_{n,\mathbf{c}}^\alpha V_{\mathbf{r} \rightarrow}^{} p_{+}^{}, \nonumber
\end{align}
where we inserted $1=V_{R_n \mathbf{r}\rightarrow}^{} V_{R_n\mathbf{r}\rightarrow}^{\dagger}$.
Using that $p_+^T p_+^{} = 1$ and $ [p_+^{} p_+^T , V_{\mathbf{r}\rightarrow}^\dagger r_{n,\mathbf{c}}^{\alpha\dagger} V_{R_n\mathbf{r}\rightarrow}] = 0 $, we obtain
\begin{align}
h_{\mathbf{r},\mathbf{k}}^{\alpha\rightarrow}
 =& p_{+}^{T} p_{+}^{} p_{+}^{T} V_{\mathbf{r}\rightarrow}^\dagger r_{n,\mathbf{c}}^{\alpha\dagger} V_{R_n\mathbf{r}\rightarrow}^{} V_{R_n \mathbf{r} \rightarrow}^{\dagger} \mathcal{H}^{\alpha} (R_{n}\mathbf{k}) \nonumber \\
& \times V_{R_n \mathbf{r}\rightarrow} V_{R_n\mathbf{r}\rightarrow}^{\dagger} r_{n,\mathbf{c}}^\alpha V_{\mathbf{r}\rightarrow}^{} p_{+}^{} p_{+}^{T} p_{+}^{} \nonumber \\
=& (p_{+}^{T} V_{\mathbf{r}\rightarrow}^{\dagger} r_{n,\mathbf{c}}^{\alpha\dagger} V_{R_n \mathbf{r}\rightarrow} p_{+}) \,h_{R_n \mathbf{r},R_n\mathbf{k}}^{\alpha\rightarrow}\,(p_{+}^{T} V_{R_n\mathbf{r}\rightarrow}^{\dagger} r_{n,\mathbf{c}}^\alpha V_{\mathbf{r}\rightarrow} p_{+}) \nonumber \\
\equiv & u_{n,\mathbf{c}}^{\alpha\dagger}\, h_{R_n \mathbf{r},R_n\mathbf{k}}^{\alpha\rightarrow}\,u_{n,\mathbf{c}}^\alpha.
\end{align}
The equivalent result for left-movers is the same with $(+,\rightarrow)$ replaced by $(-,\leftarrow)$.
Using Eq.~\eqref{eq:basis_rotation_transformation_new}, we see that $ r_{n,\mathbf{c}}^\alpha = \eta_{\alpha} e^{-i\sigma_{3}\pi/n} $ implies
\begin{equation}
 u_{n,\mathbf{c}}^\alpha = p_{+}^{T} V_{R_n \mathbf{r}\rightarrow}^{\dagger} r_{n,\mathbf{c}}^\alpha V_{\mathbf{r}\rightarrow}^{} p_{+}^{} = \eta_{\alpha}.
\end{equation}
Thus each $u_{n,\mathbf{c}}^\alpha$ is simply a sign. Since the sign itself is basis-dependent, only differences in sign can be of physical importance, as we discussed in the main text.

\subsubsection{Other High Symmetry Points}

When a Dirac Hamiltonian at a HSP does not transform into itself, but to another HSP under rotation, the rotation representation must account for this.
For example, in a $C_4$-symmetric system, the Dirac Hamiltonian at $\mathbf{X}$ transforms to $\mathbf{X'}$ and vice versa.
As discussed in the main text, these Dirac Hamiltonians must be combined into a $ 4 \times 4 $ Hamiltonian
\begin{equation}
\mathcal{H}^{\oplus}(\mathbf{k}) \equiv \mathcal{H}^{\alpha}_{\mathbf{X}}(\mathbf{k}) \oplus \mathcal{H}^{\alpha+1}_{\mathbf{X'}}(\mathbf{k}) .
\end{equation}
Denoting the space of the two stacked Dirac Hamiltonians by $\tau_\mu$, two choices of rotation representation are consistent with PH symmetry,
\begin{equation}
 r_{4,\mathbf{c}}^{\oplus} = \begin{cases} e^{-i \sigma_3 \pi/4} \otimes i \tau_2 \\ e^{-i \sigma_3 \pi/4} \otimes \tau_1 , \end{cases}
\end{equation}
where we neglect an inconsequential possibility for an overall sign.
The projection onto an edge is a straightforward generalization of the approach we discussed above.
Both projector $p_\pm$ and basis rotation $V_{\mathbf{r} s_\alpha}$ (with $s_\alpha \in \{ \leftarrow,\rightarrow \}$) must be stacked.
While the projector is stacked via
\begin{align}
 p_+^T = \begin{pmatrix} 1 & 0 & 0 & 0 \\ 0 & 0 & 1 & 0 \end{pmatrix}, & &
 p_-^T = \begin{pmatrix} 0 & 1 & 0 & 0 \\ 0 & 0 & 0 & 1 \end{pmatrix},
\end{align}
where the inner degree of freedom corresponds to the $\sigma_\mu$ space and the outer degree of freedom to the $\tau_\nu$ space, two choices to stack $V_{\mathbf{r} s_\alpha}$ are consistent with PH symmetry, $V_{\mathbf{r} s_\alpha} \oplus (s_v V_{\mathbf{r} s_\alpha})$ with the sign $s_v = \pm1$.
The projection $h^{\oplus s_\alpha}_{\mathbf{r},\mathbf{k}}$ of the Hamiltonian onto the edge at $\mathbf{r}$ is thus a $2 \times 2$ matrix.
As the mass terms at $\mathbf{X}$ and $\mathbf{X'}$ must have the same sign, the two modes of $h^{\oplus s_\alpha}_{\mathbf{r},\mathbf{k}}$ are always copropagating, $h^{\oplus\rightarrow}_{\mathbf{r},\mathbf{k}} = v_\alpha k_\parallel \tau_0$ and $h^{\oplus\leftarrow}_{\mathbf{r},\mathbf{k}} = -v_\alpha k_\parallel \tau_0$.

Following the derivation in Sec.~\ref{subsec:surface_rotation}, the resulting representation of the rotation on the edge is
\begin{equation}
 u_{4,\mathbf{c}}^{\alpha} = \begin{cases}
 s_v i \tau_2 & \text{ for } r_{4,\mathbf{c}}^{\oplus} = e^{-i \sigma_3 \pi/4} \otimes i \tau_2 \\
 s_v \tau_1 & \text{ for } r_{4,\mathbf{c}}^{\oplus} = e^{-i \sigma_3 \pi/4} \otimes \tau_1 .
 \end{cases}
\end{equation}
Since the overall sign $s_v$ does not change $\det u_{4,\mathbf{c}}^{\alpha} = \pm 1$, it can be safely neglected as we do in the main text.

\subsection{Edge Projection of Particle-Hole Operator}
\label{sec:edge_PH-symmetry}

Since the phase of the chiral edge modes $\psi_{\alpha} (\mathbf{k}_\parallel)$ is chosen such these states respect bulk PH symmetry, the edge projection of PH symmetry simply becomes complex conjugation.
We realize this by writing the edge Hamiltonian in second-quantized notation for a position $ \mathbf{r} $ on the boundary
\begin{equation}
 \hat{h}_{\mathbf{r}} = \sum_{\mathbf{k}_\parallel,\alpha\beta} \hat{\gamma}_\alpha^\dagger (\mathbf{k}_\parallel)\, h_{\mathbf{r},\mathbf{k}_\parallel}^{\alpha\beta}\, \hat{\gamma}_\beta^{} (\mathbf{k}_\parallel)
\end{equation}
with operators $\hat{\gamma}_\alpha^{} (\mathbf{k}_\parallel) = \psi_\alpha^* (\mathbf{k}_\parallel) \cdot \hat{\xi}_\alpha (\mathbf{k}_{\parallel})$ and the Nambu spinor $\hat{\xi}_\alpha (\mathbf{k}) = (\hat{c}_\alpha^{} (\mathbf{k}),\hat{c}_\alpha^\dagger (-\mathbf{k}))$. 
As the wave function respects PH symmetry, $\psi_\alpha^* (\mathbf{k}_\parallel) = \sigma_x \psi_\alpha (-\mathbf{k}_\parallel)$, we realize that the annihilation operator $\hat{\gamma}_\alpha (\mathbf{k}_\parallel)$ equals the corresponding creation operator at its negative momentum,
\begin{align}
 \hat{\gamma}_\alpha (\mathbf{k}_\parallel)
 &= \psi_\alpha (-\mathbf{k}_\parallel) \cdot (\sigma_x \hat{\xi}_\alpha (\mathbf{k}_\parallel))=  \psi_\alpha (-\mathbf{k}_\parallel) \cdot \hat{\xi}_\alpha^\dagger (-\mathbf{k}_\parallel) \\
 &= \hat{\gamma}_\alpha^\dagger (-\mathbf{k}_\parallel),
\end{align}
where we used that $\sigma_x \hat{\xi}_\alpha (\mathbf{k}) = \hat{\xi}_\alpha^\dagger (-\mathbf{k})$.
This implies that the $\hat{\gamma}_\alpha (\mathbf{k}_\parallel)$ are in fact (Fourier-transformed) Majorana fields.
Further using $\Tr[h_{\mathbf{r},\mathbf{k}_\parallel=\mathbf{0}}]=0$, the second-quantized Hamiltonian reads
\begin{equation}
 \hat{h}_{\mathbf{r}} = -\sum_{\mathbf{k}_\parallel,\alpha\beta} \hat{\gamma}_\alpha^\dagger (\mathbf{k}_\parallel) h_{\mathbf{r},-\mathbf{k}}^{\beta\alpha} \hat{\gamma}_\beta (\mathbf{k}_\parallel) .
\end{equation}
This implies that $h_{\mathbf{r},-\mathbf{k}_\parallel}^{\beta\alpha} = -h_{\mathbf{r},\mathbf{k}_\parallel}^{\alpha\beta}$, or using the Hermiticity of $h_{\mathbf{r},\mathbf{k}_\parallel}^{}$, that
\begin{equation}
 h_{\mathbf{r},\mathbf{k}_\parallel} = -\mathcal{K} h_{\mathbf{r},-\mathbf{k}_\parallel} \mathcal{K},
\end{equation}
i.e., PH symmetry simply reduces to complex conjugation when projected to an edge.

\section{Other Rotational Symmetries}
\label{sec:other_symmetries}

We now outline how our construction introduced in the main text is applied to different rotational symmetries, namely for $ C_2 $- and $ C_6 $-symmetric systems.
The $ C_2 $-symmetric BZ contains only twofold fixed points, and so is contained within the $ C_4 $ construction, but the $ C_6 $ BZ contains threefold and sixfold fixed points that are not contained in the $ C_4 $ case and require some further discussion.

\begin{figure}
 \includegraphics{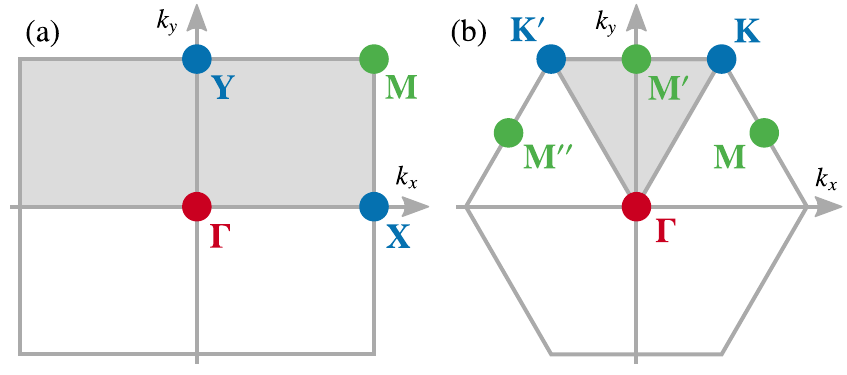}
 \caption{(a) Brillouin zone for $C_2$-symmetric models. All HSPs are twofold fixed points, i.e., they map to themselves under a $C_2$ rotation.
 (b) Brillouin zone for $C_6$-symmetric models. Only $\boldsymbol\Gamma$ is a sixfold fixed point, whereas the threefold fixed point $\mathbf{K}$ and $\mathbf{K'}$ map to each other under a sixfold rotation. The twofold fixed points $\mathbf{M}$, $\mathbf{M'}$, and $\mathbf{M''}$ form an orbit $\mathbf{M}\to\mathbf{M'} \to \mathbf{M''}$ under sixfold rotation.}
 \label{fig:brillouin_zone_C2C6}
\end{figure}

\subsection{Twofold Symmetry}

The $ C_2 $ case is simpler in some respects than the $ C_4 $ case because all the HSPs are twofold fixed points and so we keep this discussion brief.
The $ \Delta \Ch = 0 $ condition is now:
\begin{align}
\begin{split}
0 = \Delta \Ch =\, & \Delta\#X_1^+ - \Delta\#X_1^- + \Delta\#Y_1^+ - \Delta\#Y_1^- \\
& + \Delta\#M_1^+ - \Delta\#M_1^- + \Delta\#\Gamma_1^+ - \Delta\#\Gamma_1^-,
\end{split}
\end{align}
where we again split contributions according to $ \eta_{\alpha} $, the sign of the twofold rotation operators $ r_2^{\alpha} $.
This can be rewritten in terms of the $ C_2 $ rotation invariants (now $ [X] = \# X_1 - \# \Gamma_1 $, $ [Y] = \# Y_1 - \# \Gamma_1 $ and $ [M] = \# M_1 - \# \Gamma_1 $):
\begin{align}
\begin{split}
0 = & - \Delta [X] + \Delta [Y] + \Delta [M] - 4\Delta\#\Gamma_1 \\
& + 2 (\Delta\#X_1^- + \Delta\#Y_1^- + \Delta\#M_1^- + \Delta\#\Gamma_1^-),
\end{split}
\label{eq:counting_modes_invariant_C2}
\end{align}
which reproduces the $ \Ch \mod 2 $ relation of Ref.~\onlinecite{Benalcazar:2014hb}
Counting the parity of negative representations, we have
\begin{equation}
\Upsilon_{\mathbf{0}}^{(2)} = \Delta \#X_1^- + \Delta \#Y_1^- + \Delta \#M_1^- + \Delta\#\Gamma_1^- \mod 2,
\end{equation}
which we may combine with Eq.~\eqref{eq:counting_modes_invariant_C2} to write
\begin{equation}
\Upsilon_{\mathbf{0}}^{(2)} = \frac{1}{2} (\Delta [X] + \Delta [Y] + \Delta [M]) \mod 2
\end{equation}
when the physical rotation center is at the center of the unit cell ($\mathbf{c}=\mathbf{0}$).
In $C_2$-symmetric systems, there are more choices for $\mathbf{c}$ than with $C_4$ symmetry:
both $ \mathbf{c} = \mathbf{a}_1 / 2 $ and $ \mathbf{c} = \mathbf{a}_2 / 2 $ in addition to $ \mathbf{c} = (\mathbf{a}_1 + \mathbf{a}_2) / 2$.
The $C_2$ case also has two independent weak invariants $\nu_1 =[X]+[M] \mod 2$ and $\nu_2 = [Y]+[M] \mod 2$, combined into the weak invariant vector $ \Delta \mathbf{G}_{\nu} = \Delta \nu_1 \mathbf{b}_1 + \Delta \nu_2 \mathbf{b}_2 $.
Again, the indices for systems with different rotation centers are found to be related through
\begin{equation}
\Upsilon_{\mathbf{c}}^{(2)} =\Upsilon_{\mathbf{0}}^{(2)} + \frac{1}{2\pi} \Delta \mathbf{G}_{\nu} \cdot \mathbf{c} \mod 2.
\label{eq:index_C2}
\end{equation}
To arrive at Eq.~\eqref{eq:index_C2}, we used Eq.~\eqref{eq:shifted_rotation} with $R_2^T=-1$ which shows that now the sign of the representation can change at $\mathbf{X}$, $\mathbf{Y}$, and $\mathbf{M}$ depending on $ \mathbf{G}_{\nu} $ and $\mathbf{c}$.
For example, when considering $ \mathbf{c} = \mathbf{a}_1 / 2 $, one starts by counting
\begin{equation}
\Upsilon_{\mathbf{a}_1 / 2}^{(2)} = \Delta \# X_1^{+} + \Delta \# Y_1^{-} + \Delta \# M_1^{+} + \Delta \Gamma_1^{-} \mod 2,
\end{equation}
consistent with the above.

\subsection{Sixfold Symmetry}

The BZ of a $C_6$-symmetric system has three different sets of high-symmetry points:
One sixfold fixed point at $\boldsymbol\Gamma$, two threefold fixed points at $\mathbf{K}$ and $\mathbf{K'}$, and three twofold fixed points at $\mathbf{M}$, $\mathbf{M'}$ and $\mathbf{M''}$; cf.\ Fig.~\ref{fig:brillouin_zone_C2C6}(b).
Within a $ C_6 $-symmetric lattice, there is only one center of sixfold rotation at $ \mathbf{c}=\mathbf{0} $ so it need not be specified.
The bulk is characterized by the Chern number and the two rotational invariants~\cite{Benalcazar:2014hb}
\begin{align}
 [ M ] &= \# M_1 - ( \# \Gamma_1 + \# \Gamma_3 + \# \Gamma_5 ) \\
 [ K ] &= \# K_1 - ( \# \Gamma_1 + \# \Gamma_4 ) .
 \label{eq:C6_invariants}
\end{align}

Any gap closing away from $\boldsymbol\Gamma$ can be implemented analogously to the previously established description of stacked Dirac models at $\mathbf{X}$/$\mathbf{X}'$; cf.\ Sec.~\ref{sec:momenta_transforming_into_each_other}.
Gap closings at $\mathbf{K}$/$\mathbf{K'}$ require a stack of two Dirac Hamiltonians, and gap closings at $\mathbf{M}$/$\mathbf{M'}$/$\mathbf{M''}$ require a stack of three Dirac Hamiltonians.

We must proceed slightly differently with gap closings at $\boldsymbol\Gamma$, however.
While a simple Dirac Hamiltonian [Eq.~\eqref{eq:Dirac_single}] is sufficient to describe gap closings that change $\#\Gamma_1$ or $\# \Gamma_3$ (and accordingly $\# \Gamma_6$ or $\# \Gamma_4$), a $2\times 2$ Hamiltonian describing changes to $\#\Gamma_2$ (and hence $\#\Gamma_5$) with rotation representation $r_6^\alpha = \pm i \sigma_3$ requires cubic momentum terms~\cite{Fang:2019kk}:
\begin{equation}
 \mathcal{H}^\alpha_{\boldsymbol\Gamma} (\mathbf{k}) = (v_\alpha  k)^3 (\cos (3\theta) \sigma_1 + \sin (3\theta) \sigma_2 ) + m_\alpha \sigma_3.
 \label{eq:special_points}
\end{equation}
Deriving an edge theory in the same way as in Sec.~\ref{sec:effective_boundary_hamiltonian} is impeded by the presence of these non-linear terms, although we can conclude from the Chern number that a transition $m_\alpha \to - m_\alpha$ would harbor three gapless modes.

Our strategy will instead be to add trivial bands such that the transition for the whole system has $ \Delta \# \Gamma_5 = 0 $, which can be modeled with only linear Dirac Hamiltonians.
Specifically, a trivial band with sixfold eigenvalue $ \Gamma_5 = -i $ has threefold eigenvalue $ (\Gamma_5)^2=(\Gamma_5^{*})^2=-1=K_2 $ and twofold eigenvalue $ (\Gamma_5)^3 = i = M_1 $.
Thus, there exists a trivial superconductor that, when transitioning to its PH-conjugate, changes $ \Delta \# \Gamma_5 = 1 $ and $ \Delta \# M_1 = 1 $ [and $ \Delta \# K_2 = 0 $ in accordance with Eq.~\eqref{eq:delta_relations}].
Adding multiples of this trivial superconductor allows us to trade a description involving Eq.~\eqref{eq:special_points} at $\boldsymbol{\Gamma}$ for one with three Dirac models at $\mathbf{M}/\mathbf{M'}/\mathbf{M''}$.

By associating the masses and rotation representations of Dirac Hamiltonians with occupied rotation eigenvalues as in the main text, we find
\begin{align}
\begin{split}
0 = \Delta \Ch =\,& -\Delta \# \Gamma_1 + \Delta \# \Gamma_3 + 3\Delta \# \Gamma_5^{+} - 3 \Delta \# \Gamma_5^{-}\\
&- 2 \Delta \# K_1 + 3\Delta \# M_1^{-} - 3 \Delta \# M_1^{+} ,
\label{eq:Chern_C6}
\end{split}
\end{align}
consistent with the $\Ch \mod 6$ relation for the invariants in Ref~\onlinecite{Benalcazar:2014hb}.
The need to split contributions according to $ \eta_{\alpha} $ again originates from $ i\sigma_3 $ being a traceless representation (see Sections~\ref{sec:momenta_transforming_into_each_other} and \ref{subsec:explicit_Dirac}).

Counting negative representations and considering $ \Delta \# \Gamma_5 = 0 $, the index is
\begin{equation}
\Upsilon_{\mathbf{0}}^{(6)} = \Delta \# K_1 + \Delta \# \Gamma_3 + \Delta \# M_1^{-} \mod 2,
\end{equation}
where $ | \Delta \# K_1 | $ is counted because $ r_6^{\oplus} = e^{-i \sigma_3 \pi / 6} \otimes \tau_1 $ is the only rotation representation for Dirac Hamiltonians at $ \mathbf{K}/\mathbf{K'} $, which becomes $ u_6^{\oplus} = \tau_1 $ on the edge.
Gap closings at $\mathbf{M}/\mathbf{M'}/\mathbf{M''}$ permit rotation representations
\begin{equation}
r_6^{\oplus} = \pm e^{-i \sigma_3 \pi / 6} \otimes
\begin{pmatrix}
0 & 0 & 1 \\
1 & 0 & 0 \\
0 & 1 & 0 
\end{pmatrix},
\end{equation}
which have $ \det u_6^{\oplus} = \pm 1 $ and hence $ | \Delta \# M_1^{-} | $ is also counted.
Upon substitution of Eq.~\eqref{eq:Chern_C6} for $ \Delta \# K_1 $, the index is simply
\begin{equation}
\Upsilon_{\mathbf{0}}^{(6)} = \frac{1}{2} \Delta [M] \mod 2.
\end{equation}
We have thus derived the second-order bulk-boundary correspondence for the $C_6$-symmetric case using our stacked Dirac framework.
This index coincides with the index for Majoranas bound to disclinations in a $C_6$-symmetric crystal~\cite{Benalcazar:2014hb}.

\subsection{Other Rotation Representations for the Superconducting Order Parameter}
\label{subsec:pairing_symmetry}

When determining the rotation operator of the BdG Hamiltonian from the symmetry of the underlying crystal, one must also consider the symmetry of the superconducting order parameter.
A BCS pairing term $ \hat{\Delta} + \hat{\Delta}^{\dagger} $ is different to the normal-state part of the Hamiltonian because it need only be invariant \emph{up to a gauge transformation} under the act of rotation $ \hat{R}^{}_n $, so that $ \hat{R}_n^{} \hat{\Delta} \hat{R}_n^{-1} = e^{i \Theta_n} \hat{\Delta} $~\cite{Fang2017}.
In the first-quantized picture, the off-diagonal pairing term $ \Delta(\mathbf{k}) = - \Delta^{T}(- \mathbf{k}) $ transforms under rotation as~\cite{Ono2019}
\begin{equation}
\mathcal{R}(\mathbf{k}) \Delta(\mathbf{k}) \mathcal{R}^{T}(-\mathbf{k})=e^{i\Theta_n} \Delta(R_n \mathbf{k}),
\end{equation}
where the action of rotation on momentum space operators
\begin{equation}
\hat{R}_n^{} \hat{c}_{\alpha}^\dagger (\mathbf{k}) \hat{R}_n^{-1} = \hat{c}_{\beta}^{\dagger}(R_n \mathbf{k}) \mathcal{R}_{\beta \alpha}(\mathbf{k})
\end{equation}
was derived in Sec.~\ref{subsec:rotation_center}.
The symmetry of the pairing is therefore defined by a one-dimensional rotation representation $ e^{i\Theta_n} $ where $ \Theta_n = 2 \pi \ell_n / n $ and $ \ell_n \in \mathbb{Z}_n $, arising from the angular momentum of Cooper pairs.
The rotation operator for the BdG Hamiltonian should therefore be
\begin{equation}
r_n(\mathbf{k}) =
\begin{pmatrix}
\mathcal{R}(\mathbf{k}) & \\
& e^{i\Theta_n} \mathcal{R}^{*}(-\mathbf{k})
\end{pmatrix} 
\end{equation}
to reproduce the symmetry relation Eq.~\eqref{eq:bulk_symmetry_momentum}.
(The ``rotation'' operator $ r_n(\mathbf{k}) $ is now the composition of physical rotation with a gauge transformation $ U = e^{i\Theta_n/2}e^{-i\Theta_n\sigma_3/2} $ \cite{Fang2017}.)
This implies that the algebraic relation between PH and rotation operators is~\cite{Ono2019}
\begin{equation}
\Xi r_n(\mathbf{k}) \Xi^{-1} = e^{-i\Theta_n} r_n(-\mathbf{k}),
\end{equation}
which is different to the $ \ell_n = 0 $ relation that reduced to commutation at HSPs.
We now discuss the consequences of $ \ell_n \neq 0 $ for the bulk-boundary correspondence.

One can always define an alternative operator $ \tilde{r}_n(\mathbf{k}) = e^{-i\Theta_n / 2} r_{n}(\mathbf{k})$ that commutes with PH symmetry at HSPs:
\begin{equation}
\Xi \tilde{r}_n(\mathbf{k}) \Xi^{-1} = \tilde{r}_n(-\mathbf{k}).
\end{equation}
When $ \ell_n $ is even, $\tilde{r}_n(\mathbf{k})$ is a spinful operator satisfying $ \tilde{r}_{n}^{n} = -1 $ for which we can use our stacked Dirac model construction to derive a valid corner mode index in terms of the eigenvalues of $ \tilde{r}_{n}(\mathbf{k}) $.
This may be translated to an index in terms of the eigenvalues of $ r_{n}(\mathbf{k}) $ by recalling the labeling convention in Eq~\eqref{eq:label_eigenvalues}, which gives an equivalence 
\begin{equation}
\Pi_p^{(n)} \iff \tilde{\Pi}_{p-\ell_n/2}^{(n)}
\end{equation}
between the eigenvalues of a state under both operators.
Thus the case of even $ \ell_n $ is qualitatively identical to the $ \ell_n = 0 $ case.

When $ \ell_n $ is odd, however, it has been argued by Geier \textit{et al.\@}~\cite{Geier2019} that the boundary classification does not permit a second-order (nor weak) phase.
This qualitative difference arises because $\tilde{r}_n(\mathbf{k})$ now behaves like a spinless operator satisfying $\tilde{r}_n^{n} = +1 $.
The real eigenvalues of $\tilde{r}_n(\mathbf{k})$ are mapped onto themselves under PH symmetry, rather forming complex conjugate pairs (as in Figure~\ref{fig:rotation_eigenvalues}), leading to a different bulk classification.

\bibliography{hotsc}

\begin{thebibliography}{69}%
\makeatletter
\providecommand \@ifxundefined [1]{%
 \@ifx{#1\undefined}
}%
\providecommand \@ifnum [1]{%
 \ifnum #1\expandafter \@firstoftwo
 \else \expandafter \@secondoftwo
 \fi
}%
\providecommand \@ifx [1]{%
 \ifx #1\expandafter \@firstoftwo
 \else \expandafter \@secondoftwo
 \fi
}%
\providecommand \natexlab [1]{#1}%
\providecommand \enquote  [1]{``#1''}%
\providecommand \bibnamefont  [1]{#1}%
\providecommand \bibfnamefont [1]{#1}%
\providecommand \citenamefont [1]{#1}%
\providecommand \href@noop [0]{\@secondoftwo}%
\providecommand \href [0]{\begingroup \@sanitize@url \@href}%
\providecommand \@href[1]{\@@startlink{#1}\@@href}%
\providecommand \@@href[1]{\endgroup#1\@@endlink}%
\providecommand \@sanitize@url [0]{\catcode `\\12\catcode `\$12\catcode
  `\&12\catcode `\#12\catcode `\^12\catcode `\_12\catcode `\%12\relax}%
\providecommand \@@startlink[1]{}%
\providecommand \@@endlink[0]{}%
\providecommand \url  [0]{\begingroup\@sanitize@url \@url }%
\providecommand \@url [1]{\endgroup\@href {#1}{\urlprefix }}%
\providecommand \urlprefix  [0]{URL }%
\providecommand \Eprint [0]{\href }%
\providecommand \doibase [0]{https://doi.org/}%
\providecommand \selectlanguage [0]{\@gobble}%
\providecommand \bibinfo  [0]{\@secondoftwo}%
\providecommand \bibfield  [0]{\@secondoftwo}%
\providecommand \translation [1]{[#1]}%
\providecommand \BibitemOpen [0]{}%
\providecommand \bibitemStop [0]{}%
\providecommand \bibitemNoStop [0]{.\EOS\space}%
\providecommand \EOS [0]{\spacefactor3000\relax}%
\providecommand \BibitemShut  [1]{\csname bibitem#1\endcsname}%
\let\auto@bib@innerbib\@empty
\bibitem [{\citenamefont {Schnyder}\ \emph {et~al.}(2008)\citenamefont
  {Schnyder}, \citenamefont {Ryu}, \citenamefont {Furusaki},\ and\
  \citenamefont {Ludwig}}]{Schnyder:2008ez}%
  \BibitemOpen
  \bibfield  {author} {\bibinfo {author} {\bibfnamefont {A.~P.}\ \bibnamefont
  {Schnyder}}, \bibinfo {author} {\bibfnamefont {S.}~\bibnamefont {Ryu}},
  \bibinfo {author} {\bibfnamefont {A.}~\bibnamefont {Furusaki}},\ and\
  \bibinfo {author} {\bibfnamefont {A.~W.~W.}\ \bibnamefont {Ludwig}},\
  }\bibfield  {title} {\bibinfo {title} {{Classification of topological
  insulators and superconductors}},\ }\href
  {https://doi.org/10.1103/PhysRevB.78.195125} {\bibfield  {journal} {\bibinfo
  {journal} {Phys. Rev. B}\ }\textbf {\bibinfo {volume} {78}},\ \bibinfo
  {pages} {195125} (\bibinfo {year} {2008})}\BibitemShut {NoStop}%
\bibitem [{\citenamefont {Kitaev}(2009)}]{Kitaev:2009bg}%
  \BibitemOpen
  \bibfield  {author} {\bibinfo {author} {\bibfnamefont {A.}~\bibnamefont
  {Kitaev}},\ }\bibfield  {title} {\bibinfo {title} {{Periodic table for
  topological insulators and superconductors}},\ }\href
  {https://doi.org/10.1063/1.3149495} {\bibfield  {journal} {\bibinfo
  {journal} {AIP Conf. Proc.}\ }\textbf {\bibinfo {volume} {1134}},\ \bibinfo
  {pages} {22} (\bibinfo {year} {2009})}\BibitemShut {NoStop}%
\bibitem [{\citenamefont {Ryu}\ \emph {et~al.}(2012)\citenamefont {Ryu},
  \citenamefont {Moore},\ and\ \citenamefont {Ludwig}}]{Ryu:2012en}%
  \BibitemOpen
  \bibfield  {author} {\bibinfo {author} {\bibfnamefont {S.}~\bibnamefont
  {Ryu}}, \bibinfo {author} {\bibfnamefont {J.~E.}\ \bibnamefont {Moore}},\
  and\ \bibinfo {author} {\bibfnamefont {A.~W.}\ \bibnamefont {Ludwig}},\
  }\bibfield  {title} {\bibinfo {title} {{Electromagnetic and gravitational
  responses and anomalies in topological insulators and superconductors}},\
  }\href {https://doi.org/10.1103/PhysRevB.85.045104} {\bibfield  {journal}
  {\bibinfo  {journal} {Phys. Rev. B}\ }\textbf {\bibinfo {volume} {85}},\
  \bibinfo {pages} {045104} (\bibinfo {year} {2012})}\BibitemShut {NoStop}%
\bibitem [{\citenamefont {Fu}\ and\ \citenamefont {Kane}(2007)}]{Fu:2007ei}%
  \BibitemOpen
  \bibfield  {author} {\bibinfo {author} {\bibfnamefont {L.}~\bibnamefont
  {Fu}}\ and\ \bibinfo {author} {\bibfnamefont {C.~L.}\ \bibnamefont {Kane}},\
  }\bibfield  {title} {\bibinfo {title} {{Topological insulators with inversion
  symmetry}},\ }\href {https://doi.org/10.1103/PhysRevB.76.045302} {\bibfield
  {journal} {\bibinfo  {journal} {Phys. Rev. B}\ }\textbf {\bibinfo {volume}
  {76}},\ \bibinfo {pages} {045302} (\bibinfo {year} {2007})}\BibitemShut
  {NoStop}%
\bibitem [{\citenamefont {Fu}(2011)}]{Fu:2011ia}%
  \BibitemOpen
  \bibfield  {author} {\bibinfo {author} {\bibfnamefont {L.}~\bibnamefont
  {Fu}},\ }\bibfield  {title} {\bibinfo {title} {{Topological Crystalline
  Insulators}},\ }\href {https://doi.org/10.1103/PhysRevLett.106.106802}
  {\bibfield  {journal} {\bibinfo  {journal} {Phys. Rev. Lett.}\ }\textbf
  {\bibinfo {volume} {106}},\ \bibinfo {pages} {106802} (\bibinfo {year}
  {2011})}\BibitemShut {NoStop}%
\bibitem [{\citenamefont {Alexandradinata}\ \emph
  {et~al.}(2014{\natexlab{a}})\citenamefont {Alexandradinata}, \citenamefont
  {Fang}, \citenamefont {Gilbert},\ and\ \citenamefont
  {Bernevig}}]{Alexandradinata:2014jd}%
  \BibitemOpen
  \bibfield  {author} {\bibinfo {author} {\bibfnamefont {A.}~\bibnamefont
  {Alexandradinata}}, \bibinfo {author} {\bibfnamefont {C.}~\bibnamefont
  {Fang}}, \bibinfo {author} {\bibfnamefont {M.~J.}\ \bibnamefont {Gilbert}},\
  and\ \bibinfo {author} {\bibfnamefont {B.~A.}\ \bibnamefont {Bernevig}},\
  }\bibfield  {title} {\bibinfo {title} {{Spin-Orbit-Free Topological
  Insulators without Time-Reversal Symmetry}},\ }\href
  {https://doi.org/10.1103/PhysRevLett.113.116403} {\bibfield  {journal}
  {\bibinfo  {journal} {Phys. Rev. Lett.}\ }\textbf {\bibinfo {volume} {113}},\
  \bibinfo {pages} {116403} (\bibinfo {year} {2014}{\natexlab{a}})}\BibitemShut
  {NoStop}%
\bibitem [{\citenamefont {Slager}\ \emph {et~al.}(2013)\citenamefont {Slager},
  \citenamefont {Mesaros}, \citenamefont {Juri{\v{c}}i{\'{c}}},\ and\
  \citenamefont {Zaanen}}]{Slager:2013iv}%
  \BibitemOpen
  \bibfield  {author} {\bibinfo {author} {\bibfnamefont {R.-J.}\ \bibnamefont
  {Slager}}, \bibinfo {author} {\bibfnamefont {A.}~\bibnamefont {Mesaros}},
  \bibinfo {author} {\bibfnamefont {V.}~\bibnamefont {Juri{\v{c}}i{\'{c}}}},\
  and\ \bibinfo {author} {\bibfnamefont {J.}~\bibnamefont {Zaanen}},\
  }\bibfield  {title} {\bibinfo {title} {{The space group classification of
  topological band-insulators}},\ }\href {https://doi.org/10.1038/nphys2513}
  {\bibfield  {journal} {\bibinfo  {journal} {Nat. Phys.}\ }\textbf {\bibinfo
  {volume} {9}},\ \bibinfo {pages} {98} (\bibinfo {year} {2013})}\BibitemShut
  {NoStop}%
\bibitem [{\citenamefont {Morimoto}\ and\ \citenamefont
  {Furusaki}(2013)}]{Morimoto:2013cw}%
  \BibitemOpen
  \bibfield  {author} {\bibinfo {author} {\bibfnamefont {T.}~\bibnamefont
  {Morimoto}}\ and\ \bibinfo {author} {\bibfnamefont {A.}~\bibnamefont
  {Furusaki}},\ }\bibfield  {title} {\bibinfo {title} {{Topological
  classification with additional symmetries from Clifford algebras}},\ }\href
  {https://doi.org/10.1103/PhysRevB.88.125129} {\bibfield  {journal} {\bibinfo
  {journal} {Phys. Rev. B}\ }\textbf {\bibinfo {volume} {88}},\ \bibinfo
  {pages} {125129} (\bibinfo {year} {2013})}\BibitemShut {NoStop}%
\bibitem [{\citenamefont {Chiu}\ \emph {et~al.}(2015)\citenamefont {Chiu},
  \citenamefont {Teo}, \citenamefont {Schnyder},\ and\ \citenamefont
  {Ryu}}]{Chiu:2015ex}%
  \BibitemOpen
  \bibfield  {author} {\bibinfo {author} {\bibfnamefont {C.-K.}\ \bibnamefont
  {Chiu}}, \bibinfo {author} {\bibfnamefont {J.~C.~Y.}\ \bibnamefont {Teo}},
  \bibinfo {author} {\bibfnamefont {A.~P.}\ \bibnamefont {Schnyder}},\ and\
  \bibinfo {author} {\bibfnamefont {S.}~\bibnamefont {Ryu}},\ }\bibfield
  {title} {\bibinfo {title} {{Classification of topological quantum matter with
  symmetries}},\ }\href {https://doi.org/10.1103/RevModPhys.88.035005}
  {\bibfield  {journal} {\bibinfo  {journal} {Rev. Mod. Phys.}\ }\textbf
  {\bibinfo {volume} {88}},\ \bibinfo {pages} {035005} (\bibinfo {year}
  {2015})}\BibitemShut {NoStop}%
\bibitem [{\citenamefont {Kruthoff}\ \emph {et~al.}(2016)\citenamefont
  {Kruthoff}, \citenamefont {de~Boer}, \citenamefont {van Wezel}, \citenamefont
  {Kane},\ and\ \citenamefont {Slager}}]{Kruthoff:2017jj}%
  \BibitemOpen
  \bibfield  {author} {\bibinfo {author} {\bibfnamefont {J.}~\bibnamefont
  {Kruthoff}}, \bibinfo {author} {\bibfnamefont {J.}~\bibnamefont {de~Boer}},
  \bibinfo {author} {\bibfnamefont {J.}~\bibnamefont {van Wezel}}, \bibinfo
  {author} {\bibfnamefont {C.~L.}\ \bibnamefont {Kane}},\ and\ \bibinfo
  {author} {\bibfnamefont {R.-J.}\ \bibnamefont {Slager}},\ }\bibfield  {title}
  {\bibinfo {title} {{Topological Classification of Crystalline Insulators
  through Band Structure Combinatorics}},\ }\href
  {https://doi.org/10.1103/PhysRevX.7.041069} {\bibfield  {journal} {\bibinfo
  {journal} {Phys. Rev. X}\ }\textbf {\bibinfo {volume} {7}},\ \bibinfo {pages}
  {041069} (\bibinfo {year} {2016})}\BibitemShut {NoStop}%
\bibitem [{\citenamefont {Bradlyn}\ \emph {et~al.}(2017)\citenamefont
  {Bradlyn}, \citenamefont {Elcoro}, \citenamefont {Cano}, \citenamefont
  {Vergniory}, \citenamefont {Wang}, \citenamefont {Felser}, \citenamefont
  {Aroyo},\ and\ \citenamefont {Bernevig}}]{Bradlyn:2017fy}%
  \BibitemOpen
  \bibfield  {author} {\bibinfo {author} {\bibfnamefont {B.}~\bibnamefont
  {Bradlyn}}, \bibinfo {author} {\bibfnamefont {L.}~\bibnamefont {Elcoro}},
  \bibinfo {author} {\bibfnamefont {J.}~\bibnamefont {Cano}}, \bibinfo {author}
  {\bibfnamefont {M.~G.}\ \bibnamefont {Vergniory}}, \bibinfo {author}
  {\bibfnamefont {Z.}~\bibnamefont {Wang}}, \bibinfo {author} {\bibfnamefont
  {C.}~\bibnamefont {Felser}}, \bibinfo {author} {\bibfnamefont {M.~I.}\
  \bibnamefont {Aroyo}},\ and\ \bibinfo {author} {\bibfnamefont {B.~A.}\
  \bibnamefont {Bernevig}},\ }\bibfield  {title} {\bibinfo {title}
  {{Topological quantum chemistry}},\ }\href
  {https://doi.org/10.1038/nature23268} {\bibfield  {journal} {\bibinfo
  {journal} {Nature}\ }\textbf {\bibinfo {volume} {547}},\ \bibinfo {pages}
  {298} (\bibinfo {year} {2017})}\BibitemShut {NoStop}%
\bibitem [{\citenamefont {Po}\ \emph {et~al.}(2017)\citenamefont {Po},
  \citenamefont {Vishwanath},\ and\ \citenamefont {Watanabe}}]{Po:2017ci}%
  \BibitemOpen
  \bibfield  {author} {\bibinfo {author} {\bibfnamefont {H.~C.}\ \bibnamefont
  {Po}}, \bibinfo {author} {\bibfnamefont {A.}~\bibnamefont {Vishwanath}},\
  and\ \bibinfo {author} {\bibfnamefont {H.}~\bibnamefont {Watanabe}},\
  }\bibfield  {title} {\bibinfo {title} {{Symmetry-based indicators of band
  topology in the 230 space groups}},\ }\href
  {https://doi.org/10.1038/s41467-017-00133-2} {\bibfield  {journal} {\bibinfo
  {journal} {Nat. Commun.}\ }\textbf {\bibinfo {volume} {8}},\ \bibinfo {pages}
  {50} (\bibinfo {year} {2017})}\BibitemShut {NoStop}%
\bibitem [{\citenamefont {Teo}\ and\ \citenamefont
  {Hughes}(2013)}]{Teo:2013cp}%
  \BibitemOpen
  \bibfield  {author} {\bibinfo {author} {\bibfnamefont {J.~C.~Y.}\
  \bibnamefont {Teo}}\ and\ \bibinfo {author} {\bibfnamefont {T.~L.}\
  \bibnamefont {Hughes}},\ }\bibfield  {title} {\bibinfo {title} {{Existence of
  Majorana-Fermion Bound States on Disclinations and the Classification of
  Topological Crystalline Superconductors in Two Dimensions}},\ }\href
  {https://doi.org/10.1103/PhysRevLett.111.047006} {\bibfield  {journal}
  {\bibinfo  {journal} {Phys. Rev. Lett.}\ }\textbf {\bibinfo {volume} {111}},\
  \bibinfo {pages} {047006} (\bibinfo {year} {2013})}\BibitemShut {NoStop}%
\bibitem [{\citenamefont {Benalcazar}\ \emph {et~al.}(2014)\citenamefont
  {Benalcazar}, \citenamefont {Teo},\ and\ \citenamefont
  {Hughes}}]{Benalcazar:2014hb}%
  \BibitemOpen
  \bibfield  {author} {\bibinfo {author} {\bibfnamefont {W.~A.}\ \bibnamefont
  {Benalcazar}}, \bibinfo {author} {\bibfnamefont {J.~C.}\ \bibnamefont
  {Teo}},\ and\ \bibinfo {author} {\bibfnamefont {T.~L.}\ \bibnamefont
  {Hughes}},\ }\bibfield  {title} {\bibinfo {title} {{Classification of
  two-dimensional topological crystalline superconductors and Majorana bound
  states at disclinations}},\ }\href
  {https://doi.org/10.1103/PhysRevB.89.224503} {\bibfield  {journal} {\bibinfo
  {journal} {Phys. Rev. B}\ }\textbf {\bibinfo {volume} {89}},\ \bibinfo
  {pages} {224503} (\bibinfo {year} {2014})}\BibitemShut {NoStop}%
\bibitem [{\citenamefont {Liu}\ \emph {et~al.}(2014)\citenamefont {Liu},
  \citenamefont {He},\ and\ \citenamefont {Law}}]{Liu:2014kk}%
  \BibitemOpen
  \bibfield  {author} {\bibinfo {author} {\bibfnamefont {X.-J.}\ \bibnamefont
  {Liu}}, \bibinfo {author} {\bibfnamefont {J.~J.}\ \bibnamefont {He}},\ and\
  \bibinfo {author} {\bibfnamefont {K.~T.}\ \bibnamefont {Law}},\ }\bibfield
  {title} {\bibinfo {title} {{Demonstrating lattice symmetry protection in
  topological crystalline superconductors}},\ }\href
  {https://doi.org/10.1103/PhysRevB.90.235141} {\bibfield  {journal} {\bibinfo
  {journal} {Phys. Rev. B}\ }\textbf {\bibinfo {volume} {90}},\ \bibinfo
  {pages} {235141} (\bibinfo {year} {2014})}\BibitemShut {NoStop}%
\bibitem [{\citenamefont {Fang}\ \emph {et~al.}(2012)\citenamefont {Fang},
  \citenamefont {Gilbert},\ and\ \citenamefont {Bernevig}}]{Fang:2012dn}%
  \BibitemOpen
  \bibfield  {author} {\bibinfo {author} {\bibfnamefont {C.}~\bibnamefont
  {Fang}}, \bibinfo {author} {\bibfnamefont {M.~J.}\ \bibnamefont {Gilbert}},\
  and\ \bibinfo {author} {\bibfnamefont {B.~A.}\ \bibnamefont {Bernevig}},\
  }\bibfield  {title} {\bibinfo {title} {{Bulk topological invariants in
  noninteracting point group symmetric insulators}},\ }\href
  {https://doi.org/10.1103/PhysRevB.86.115112} {\bibfield  {journal} {\bibinfo
  {journal} {Phys. Rev. B}\ }\textbf {\bibinfo {volume} {86}},\ \bibinfo
  {pages} {115112} (\bibinfo {year} {2012})}\BibitemShut {NoStop}%
\bibitem [{\citenamefont {Fang}\ \emph {et~al.}(2013)\citenamefont {Fang},
  \citenamefont {Gilbert},\ and\ \citenamefont {Bernevig}}]{Fang:2013jk}%
  \BibitemOpen
  \bibfield  {author} {\bibinfo {author} {\bibfnamefont {C.}~\bibnamefont
  {Fang}}, \bibinfo {author} {\bibfnamefont {M.~J.}\ \bibnamefont {Gilbert}},\
  and\ \bibinfo {author} {\bibfnamefont {B.~A.}\ \bibnamefont {Bernevig}},\
  }\bibfield  {title} {\bibinfo {title} {{Entanglement spectrum classification
  of $C_n$-invariant noninteracting topological insulators in two
  dimensions}},\ }\href {https://doi.org/10.1103/PhysRevB.87.035119} {\bibfield
   {journal} {\bibinfo  {journal} {Phys. Rev. B}\ }\textbf {\bibinfo {volume}
  {87}},\ \bibinfo {pages} {035119} (\bibinfo {year} {2013})}\BibitemShut
  {NoStop}%
\bibitem [{\citenamefont {Fang}\ \emph {et~al.}()\citenamefont {Fang},
  \citenamefont {Bernevig},\ and\ \citenamefont {Gilbert}}]{Fang2017}%
  \BibitemOpen
  \bibfield  {author} {\bibinfo {author} {\bibfnamefont {C.}~\bibnamefont
  {Fang}}, \bibinfo {author} {\bibfnamefont {B.~A.}\ \bibnamefont {Bernevig}},\
  and\ \bibinfo {author} {\bibfnamefont {M.~J.}\ \bibnamefont {Gilbert}},\
  }\bibfield  {title} {\bibinfo {title} {{Topological crystalline
  superconductors with linearly and projectively represented $C_{n}$
  symmetry}},\ }\Eprint {https://arxiv.org/abs/1701.01944} {arXiv:1701.01944}
  \BibitemShut {NoStop}%
\bibitem [{\citenamefont {Hatsugai}(1993)}]{Hatsugai:1993fc}%
  \BibitemOpen
  \bibfield  {author} {\bibinfo {author} {\bibfnamefont {Y.}~\bibnamefont
  {Hatsugai}},\ }\bibfield  {title} {\bibinfo {title} {{Chern number and edge
  states in the integer quantum Hall effect}},\ }\href
  {https://doi.org/10.1103/PhysRevLett.71.3697} {\bibfield  {journal} {\bibinfo
   {journal} {Phys. Rev. Lett.}\ }\textbf {\bibinfo {volume} {71}},\ \bibinfo
  {pages} {3697} (\bibinfo {year} {1993})}\BibitemShut {NoStop}%
\bibitem [{\citenamefont {Hasan}\ and\ \citenamefont
  {Kane}(2010)}]{Hasan:2010ku}%
  \BibitemOpen
  \bibfield  {author} {\bibinfo {author} {\bibfnamefont {M.~Z.}\ \bibnamefont
  {Hasan}}\ and\ \bibinfo {author} {\bibfnamefont {C.~L.}\ \bibnamefont
  {Kane}},\ }\bibfield  {title} {\bibinfo {title} {{Colloquium: Topological
  insulators}},\ }\href {https://doi.org/10.1103/RevModPhys.82.3045} {\bibfield
   {journal} {\bibinfo  {journal} {Rev. Mod. Phys.}\ }\textbf {\bibinfo
  {volume} {82}},\ \bibinfo {pages} {3045} (\bibinfo {year}
  {2010})}\BibitemShut {NoStop}%
\bibitem [{\citenamefont {Qi}\ and\ \citenamefont {Zhang}(2011)}]{Qi:2011hb}%
  \BibitemOpen
  \bibfield  {author} {\bibinfo {author} {\bibfnamefont {X.~L.}\ \bibnamefont
  {Qi}}\ and\ \bibinfo {author} {\bibfnamefont {S.~C.}\ \bibnamefont {Zhang}},\
  }\bibfield  {title} {\bibinfo {title} {{Topological insulators and
  superconductors}},\ }\href {https://doi.org/10.1103/RevModPhys.83.1057}
  {\bibfield  {journal} {\bibinfo  {journal} {Rev. Mod. Phys.}\ }\textbf
  {\bibinfo {volume} {83}},\ \bibinfo {pages} {1057} (\bibinfo {year}
  {2011})}\BibitemShut {NoStop}%
\bibitem [{\citenamefont {Benalcazar}\ \emph
  {et~al.}(2017{\natexlab{a}})\citenamefont {Benalcazar}, \citenamefont
  {Bernevig},\ and\ \citenamefont {Hughes}}]{Benalcazar:2017ks}%
  \BibitemOpen
  \bibfield  {author} {\bibinfo {author} {\bibfnamefont {W.~A.}\ \bibnamefont
  {Benalcazar}}, \bibinfo {author} {\bibfnamefont {B.~A.}\ \bibnamefont
  {Bernevig}},\ and\ \bibinfo {author} {\bibfnamefont {T.~L.}\ \bibnamefont
  {Hughes}},\ }\bibfield  {title} {\bibinfo {title} {{Quantized electric
  multipole insulators}},\ }\href {https://doi.org/10.1126/science.aah6442}
  {\bibfield  {journal} {\bibinfo  {journal} {Science}\ }\textbf {\bibinfo
  {volume} {357}},\ \bibinfo {pages} {61} (\bibinfo {year}
  {2017}{\natexlab{a}})}\BibitemShut {NoStop}%
\bibitem [{\citenamefont {Benalcazar}\ \emph
  {et~al.}(2017{\natexlab{b}})\citenamefont {Benalcazar}, \citenamefont
  {Bernevig},\ and\ \citenamefont {Hughes}}]{Benalcazar:2017cn}%
  \BibitemOpen
  \bibfield  {author} {\bibinfo {author} {\bibfnamefont {W.~A.}\ \bibnamefont
  {Benalcazar}}, \bibinfo {author} {\bibfnamefont {B.~A.}\ \bibnamefont
  {Bernevig}},\ and\ \bibinfo {author} {\bibfnamefont {T.~L.}\ \bibnamefont
  {Hughes}},\ }\bibfield  {title} {\bibinfo {title} {{Electric multipole
  moments, topological multipole moment pumping, and chiral hinge states in
  crystalline insulators}},\ }\href
  {https://doi.org/10.1103/PhysRevB.96.245115} {\bibfield  {journal} {\bibinfo
  {journal} {Phys. Rev. B}\ }\textbf {\bibinfo {volume} {96}},\ \bibinfo
  {pages} {245115} (\bibinfo {year} {2017}{\natexlab{b}})}\BibitemShut
  {NoStop}%
\bibitem [{\citenamefont {Langbehn}\ \emph {et~al.}(2017)\citenamefont
  {Langbehn}, \citenamefont {Peng}, \citenamefont {Trifunovic}, \citenamefont
  {{Von Oppen}},\ and\ \citenamefont {Brouwer}}]{Langbehn:2017jn}%
  \BibitemOpen
  \bibfield  {author} {\bibinfo {author} {\bibfnamefont {J.}~\bibnamefont
  {Langbehn}}, \bibinfo {author} {\bibfnamefont {Y.}~\bibnamefont {Peng}},
  \bibinfo {author} {\bibfnamefont {L.}~\bibnamefont {Trifunovic}}, \bibinfo
  {author} {\bibfnamefont {F.}~\bibnamefont {{Von Oppen}}},\ and\ \bibinfo
  {author} {\bibfnamefont {P.~W.}\ \bibnamefont {Brouwer}},\ }\bibfield
  {title} {\bibinfo {title} {{Reflection-Symmetric Second-Order Topological
  Insulators and Superconductors}},\ }\href
  {https://doi.org/10.1103/PhysRevLett.119.246401} {\bibfield  {journal}
  {\bibinfo  {journal} {Phys. Rev. Lett.}\ }\textbf {\bibinfo {volume} {119}},\
  \bibinfo {pages} {246401} (\bibinfo {year} {2017})}\BibitemShut {NoStop}%
\bibitem [{\citenamefont {Schindler}\ \emph
  {et~al.}(2018{\natexlab{a}})\citenamefont {Schindler}, \citenamefont {Cook},
  \citenamefont {Vergniory}, \citenamefont {Wang}, \citenamefont {Parkin},
  \citenamefont {Bernevig},\ and\ \citenamefont {Neupert}}]{Schindler:2018hi}%
  \BibitemOpen
  \bibfield  {author} {\bibinfo {author} {\bibfnamefont {F.}~\bibnamefont
  {Schindler}}, \bibinfo {author} {\bibfnamefont {A.~M.}\ \bibnamefont {Cook}},
  \bibinfo {author} {\bibfnamefont {M.~G.}\ \bibnamefont {Vergniory}}, \bibinfo
  {author} {\bibfnamefont {Z.}~\bibnamefont {Wang}}, \bibinfo {author}
  {\bibfnamefont {S.~S.~P.}\ \bibnamefont {Parkin}}, \bibinfo {author}
  {\bibfnamefont {B.~A.}\ \bibnamefont {Bernevig}},\ and\ \bibinfo {author}
  {\bibfnamefont {T.}~\bibnamefont {Neupert}},\ }\bibfield  {title} {\bibinfo
  {title} {{Higher-order topological insulators}},\ }\href
  {https://doi.org/10.1126/sciadv.aat0346} {\bibfield  {journal} {\bibinfo
  {journal} {Sci. Adv.}\ }\textbf {\bibinfo {volume} {4}},\ \bibinfo {pages}
  {eaat0346} (\bibinfo {year} {2018}{\natexlab{a}})}\BibitemShut {NoStop}%
\bibitem [{\citenamefont {Kunst}\ \emph {et~al.}(2018)\citenamefont {Kunst},
  \citenamefont {van Miert},\ and\ \citenamefont {Bergholtz}}]{Kunst:2018fi}%
  \BibitemOpen
  \bibfield  {author} {\bibinfo {author} {\bibfnamefont {F.~K.}\ \bibnamefont
  {Kunst}}, \bibinfo {author} {\bibfnamefont {G.}~\bibnamefont {van Miert}},\
  and\ \bibinfo {author} {\bibfnamefont {E.~J.}\ \bibnamefont {Bergholtz}},\
  }\bibfield  {title} {\bibinfo {title} {{Lattice models with exactly solvable
  topological hinge and corner states}},\ }\href
  {https://doi.org/10.1103/PhysRevB.97.241405} {\bibfield  {journal} {\bibinfo
  {journal} {Phys. Rev. B}\ }\textbf {\bibinfo {volume} {97}},\ \bibinfo
  {pages} {241405} (\bibinfo {year} {2018})}\BibitemShut {NoStop}%
\bibitem [{\citenamefont {Geier}\ \emph {et~al.}(2018)\citenamefont {Geier},
  \citenamefont {Trifunovic}, \citenamefont {Hoskam},\ and\ \citenamefont
  {Brouwer}}]{Geier:2018ev}%
  \BibitemOpen
  \bibfield  {author} {\bibinfo {author} {\bibfnamefont {M.}~\bibnamefont
  {Geier}}, \bibinfo {author} {\bibfnamefont {L.}~\bibnamefont {Trifunovic}},
  \bibinfo {author} {\bibfnamefont {M.}~\bibnamefont {Hoskam}},\ and\ \bibinfo
  {author} {\bibfnamefont {P.~W.}\ \bibnamefont {Brouwer}},\ }\bibfield
  {title} {\bibinfo {title} {{Second-order topological insulators and
  superconductors with an order-two crystalline symmetry}},\ }\href
  {https://doi.org/10.1103/PhysRevB.97.205135} {\bibfield  {journal} {\bibinfo
  {journal} {Phys. Rev. B}\ }\textbf {\bibinfo {volume} {97}},\ \bibinfo
  {pages} {205135} (\bibinfo {year} {2018})}\BibitemShut {NoStop}%
\bibitem [{\citenamefont {Trifunovic}\ and\ \citenamefont
  {Brouwer}(2019)}]{Trifunovic:2019hi}%
  \BibitemOpen
  \bibfield  {author} {\bibinfo {author} {\bibfnamefont {L.}~\bibnamefont
  {Trifunovic}}\ and\ \bibinfo {author} {\bibfnamefont {P.~W.}\ \bibnamefont
  {Brouwer}},\ }\bibfield  {title} {\bibinfo {title} {{Higher-Order
  Bulk-Boundary Correspondence for Topological Crystalline Phases}},\ }\href
  {https://doi.org/10.1103/PhysRevX.9.011012} {\bibfield  {journal} {\bibinfo
  {journal} {Phys. Rev. X}\ }\textbf {\bibinfo {volume} {9}},\ \bibinfo {pages}
  {11012} (\bibinfo {year} {2019})}\BibitemShut {NoStop}%
\bibitem [{\citenamefont {Song}\ \emph {et~al.}(2017)\citenamefont {Song},
  \citenamefont {Fang},\ and\ \citenamefont {Fang}}]{Song:2017ev}%
  \BibitemOpen
  \bibfield  {author} {\bibinfo {author} {\bibfnamefont {Z.}~\bibnamefont
  {Song}}, \bibinfo {author} {\bibfnamefont {Z.}~\bibnamefont {Fang}},\ and\
  \bibinfo {author} {\bibfnamefont {C.}~\bibnamefont {Fang}},\ }\bibfield
  {title} {\bibinfo {title} {{$(d-2)$-Dimensional Edge States of Rotation
  Symmetry Protected Topological States}},\ }\href
  {https://doi.org/10.1103/PhysRevLett.119.246402} {\bibfield  {journal}
  {\bibinfo  {journal} {Phys. Rev. Lett.}\ }\textbf {\bibinfo {volume} {119}},\
  \bibinfo {pages} {246402} (\bibinfo {year} {2017})}\BibitemShut {NoStop}%
\bibitem [{\citenamefont {Benalcazar}\ \emph {et~al.}(2019)\citenamefont
  {Benalcazar}, \citenamefont {Li},\ and\ \citenamefont
  {Hughes}}]{Benalcazar:2019bs}%
  \BibitemOpen
  \bibfield  {author} {\bibinfo {author} {\bibfnamefont {W.~A.}\ \bibnamefont
  {Benalcazar}}, \bibinfo {author} {\bibfnamefont {T.}~\bibnamefont {Li}},\
  and\ \bibinfo {author} {\bibfnamefont {T.~L.}\ \bibnamefont {Hughes}},\
  }\bibfield  {title} {\bibinfo {title} {{Quantization of fractional corner
  charge in $C_n$-symmetric higher-order topological crystalline insulators}},\
  }\href {https://doi.org/10.1103/PhysRevB.99.245151} {\bibfield  {journal}
  {\bibinfo  {journal} {Phys. Rev. B}\ }\textbf {\bibinfo {volume} {99}},\
  \bibinfo {pages} {245151} (\bibinfo {year} {2019})}\BibitemShut {NoStop}%
\bibitem [{\citenamefont {{Van Miert}}\ and\ \citenamefont
  {Ortix}(2018)}]{VanMiert:2018cb}%
  \BibitemOpen
  \bibfield  {author} {\bibinfo {author} {\bibfnamefont {G.}~\bibnamefont {{Van
  Miert}}}\ and\ \bibinfo {author} {\bibfnamefont {C.}~\bibnamefont {Ortix}},\
  }\bibfield  {title} {\bibinfo {title} {{Higher-order topological insulators
  protected by inversion and rotoinversion symmetries}},\ }\href
  {https://doi.org/10.1103/PhysRevB.98.081110} {\bibfield  {journal} {\bibinfo
  {journal} {Phys. Rev. B}\ }\textbf {\bibinfo {volume} {98}},\ \bibinfo
  {pages} {081110} (\bibinfo {year} {2018})}\BibitemShut {NoStop}%
\bibitem [{\citenamefont {Bultinck}\ \emph {et~al.}(2019)\citenamefont
  {Bultinck}, \citenamefont {Bernevig},\ and\ \citenamefont
  {Zaletel}}]{Bultinck:2019cn}%
  \BibitemOpen
  \bibfield  {author} {\bibinfo {author} {\bibfnamefont {N.}~\bibnamefont
  {Bultinck}}, \bibinfo {author} {\bibfnamefont {B.~A.}\ \bibnamefont
  {Bernevig}},\ and\ \bibinfo {author} {\bibfnamefont {M.~P.}\ \bibnamefont
  {Zaletel}},\ }\bibfield  {title} {\bibinfo {title} {{Three-dimensional
  superconductors with hybrid higher-order topology}},\ }\href
  {https://doi.org/10.1103/PhysRevB.99.125149} {\bibfield  {journal} {\bibinfo
  {journal} {Phys. Rev. B}\ }\textbf {\bibinfo {volume} {99}},\ \bibinfo
  {pages} {125149} (\bibinfo {year} {2019})}\BibitemShut {NoStop}%
\bibitem [{\citenamefont {Schindler}\ \emph
  {et~al.}(2018{\natexlab{b}})\citenamefont {Schindler}, \citenamefont {Wang},
  \citenamefont {Vergniory}, \citenamefont {Cook}, \citenamefont {Murani},
  \citenamefont {Sengupta}, \citenamefont {Kasumov}, \citenamefont {Deblock},
  \citenamefont {Jeon}, \citenamefont {Drozdov}, \citenamefont {Bouchiat},
  \citenamefont {Gu{\'{e}}ron}, \citenamefont {Yazdani}, \citenamefont
  {Bernevig},\ and\ \citenamefont {Neupert}}]{Schindler:2018hl}%
  \BibitemOpen
  \bibfield  {author} {\bibinfo {author} {\bibfnamefont {F.}~\bibnamefont
  {Schindler}}, \bibinfo {author} {\bibfnamefont {Z.}~\bibnamefont {Wang}},
  \bibinfo {author} {\bibfnamefont {M.~G.}\ \bibnamefont {Vergniory}}, \bibinfo
  {author} {\bibfnamefont {A.~M.}\ \bibnamefont {Cook}}, \bibinfo {author}
  {\bibfnamefont {A.}~\bibnamefont {Murani}}, \bibinfo {author} {\bibfnamefont
  {S.}~\bibnamefont {Sengupta}}, \bibinfo {author} {\bibfnamefont {A.~Y.}\
  \bibnamefont {Kasumov}}, \bibinfo {author} {\bibfnamefont {R.}~\bibnamefont
  {Deblock}}, \bibinfo {author} {\bibfnamefont {S.}~\bibnamefont {Jeon}},
  \bibinfo {author} {\bibfnamefont {I.}~\bibnamefont {Drozdov}}, \bibinfo
  {author} {\bibfnamefont {H.}~\bibnamefont {Bouchiat}}, \bibinfo {author}
  {\bibfnamefont {S.}~\bibnamefont {Gu{\'{e}}ron}}, \bibinfo {author}
  {\bibfnamefont {A.}~\bibnamefont {Yazdani}}, \bibinfo {author} {\bibfnamefont
  {B.~A.}\ \bibnamefont {Bernevig}},\ and\ \bibinfo {author} {\bibfnamefont
  {T.}~\bibnamefont {Neupert}},\ }\bibfield  {title} {\bibinfo {title}
  {{Higher-order topology in bismuth}},\ }\href
  {https://doi.org/10.1038/s41567-018-0224-7} {\bibfield  {journal} {\bibinfo
  {journal} {Nat. Phys.}\ }\textbf {\bibinfo {volume} {14}},\ \bibinfo {pages}
  {918} (\bibinfo {year} {2018}{\natexlab{b}})}\BibitemShut {NoStop}%
\bibitem [{\citenamefont {Dwivedi}\ \emph {et~al.}(2018)\citenamefont
  {Dwivedi}, \citenamefont {Hickey}, \citenamefont {Eschmann},\ and\
  \citenamefont {Trebst}}]{Dwivedi:2018kp}%
  \BibitemOpen
  \bibfield  {author} {\bibinfo {author} {\bibfnamefont {V.}~\bibnamefont
  {Dwivedi}}, \bibinfo {author} {\bibfnamefont {C.}~\bibnamefont {Hickey}},
  \bibinfo {author} {\bibfnamefont {T.}~\bibnamefont {Eschmann}},\ and\
  \bibinfo {author} {\bibfnamefont {S.}~\bibnamefont {Trebst}},\ }\bibfield
  {title} {\bibinfo {title} {{Majorana corner modes in a second-order Kitaev
  spin liquid}},\ }\href {https://doi.org/10.1103/PhysRevB.98.054432}
  {\bibfield  {journal} {\bibinfo  {journal} {Phys. Rev. B}\ }\textbf {\bibinfo
  {volume} {98}},\ \bibinfo {pages} {054432} (\bibinfo {year}
  {2018})}\BibitemShut {NoStop}%
\bibitem [{\citenamefont {You}\ \emph {et~al.}(2019)\citenamefont {You},
  \citenamefont {Litinski},\ and\ \citenamefont {von Oppen}}]{You:2019kr}%
  \BibitemOpen
  \bibfield  {author} {\bibinfo {author} {\bibfnamefont {Y.}~\bibnamefont
  {You}}, \bibinfo {author} {\bibfnamefont {D.}~\bibnamefont {Litinski}},\ and\
  \bibinfo {author} {\bibfnamefont {F.}~\bibnamefont {von Oppen}},\ }\bibfield
  {title} {\bibinfo {title} {{Higher-order topological superconductors as
  generators of quantum codes}},\ }\href
  {https://doi.org/10.1103/PhysRevB.100.054513} {\bibfield  {journal} {\bibinfo
   {journal} {Phys. Rev. B}\ }\textbf {\bibinfo {volume} {100}},\ \bibinfo
  {pages} {054513} (\bibinfo {year} {2019})}\BibitemShut {NoStop}%
\bibitem [{\citenamefont {Rodriguez-Vega}\ \emph {et~al.}(2019)\citenamefont
  {Rodriguez-Vega}, \citenamefont {Kumar},\ and\ \citenamefont
  {Seradjeh}}]{Rodriguez:2019dj}%
  \BibitemOpen
  \bibfield  {author} {\bibinfo {author} {\bibfnamefont {M.}~\bibnamefont
  {Rodriguez-Vega}}, \bibinfo {author} {\bibfnamefont {A.}~\bibnamefont
  {Kumar}},\ and\ \bibinfo {author} {\bibfnamefont {B.}~\bibnamefont
  {Seradjeh}},\ }\bibfield  {title} {\bibinfo {title} {{Higher-order Floquet
  topological phases with corner and bulk bound states}},\ }\href
  {https://doi.org/10.1103/PhysRevB.100.085138} {\bibfield  {journal} {\bibinfo
   {journal} {Phys. Rev. B}\ }\textbf {\bibinfo {volume} {100}},\ \bibinfo
  {pages} {085138} (\bibinfo {year} {2019})}\BibitemShut {NoStop}%
\bibitem [{\citenamefont {Chaudhary}\ \emph {et~al.}()\citenamefont
  {Chaudhary}, \citenamefont {Haim}, \citenamefont {Peng},\ and\ \citenamefont
  {Refael}}]{Chaudhary2019}%
  \BibitemOpen
  \bibfield  {author} {\bibinfo {author} {\bibfnamefont {S.}~\bibnamefont
  {Chaudhary}}, \bibinfo {author} {\bibfnamefont {A.}~\bibnamefont {Haim}},
  \bibinfo {author} {\bibfnamefont {Y.}~\bibnamefont {Peng}},\ and\ \bibinfo
  {author} {\bibfnamefont {G.}~\bibnamefont {Refael}},\ }\bibfield  {title}
  {\bibinfo {title} {Phonon-induced floquet second-order topological phases
  protected by space-time symmetries},\ }\Eprint
  {https://arxiv.org/abs/1911.07892} {arXiv:1911.07892} \BibitemShut {NoStop}%
\bibitem [{\citenamefont {Ghorashi}\ \emph {et~al.}(2019)\citenamefont
  {Ghorashi}, \citenamefont {Hu}, \citenamefont {Hughes},\ and\ \citenamefont
  {Rossi}}]{Ghorashi:2019cj}%
  \BibitemOpen
  \bibfield  {author} {\bibinfo {author} {\bibfnamefont {S.~A.~A.}\
  \bibnamefont {Ghorashi}}, \bibinfo {author} {\bibfnamefont {X.}~\bibnamefont
  {Hu}}, \bibinfo {author} {\bibfnamefont {T.~L.}\ \bibnamefont {Hughes}},\
  and\ \bibinfo {author} {\bibfnamefont {E.}~\bibnamefont {Rossi}},\ }\bibfield
   {title} {\bibinfo {title} {{Second-order Dirac superconductors and magnetic
  field induced Majorana hinge modes}},\ }\href
  {https://doi.org/10.1103/PhysRevB.100.020509} {\bibfield  {journal} {\bibinfo
   {journal} {Phys. Rev. B}\ }\textbf {\bibinfo {volume} {100}},\ \bibinfo
  {pages} {020509} (\bibinfo {year} {2019})}\BibitemShut {NoStop}%
\bibitem [{\citenamefont {Agarwala}\ \emph {et~al.}()\citenamefont {Agarwala},
  \citenamefont {Juricic},\ and\ \citenamefont {Roy}}]{Agarwala:2019}%
  \BibitemOpen
  \bibfield  {author} {\bibinfo {author} {\bibfnamefont {A.}~\bibnamefont
  {Agarwala}}, \bibinfo {author} {\bibfnamefont {V.}~\bibnamefont {Juricic}},\
  and\ \bibinfo {author} {\bibfnamefont {B.}~\bibnamefont {Roy}},\ }\bibfield
  {title} {\bibinfo {title} {{Higher Order Topological Insulators in Amorphous
  Solids}},\ }\Eprint {https://arxiv.org/abs/1902.00507} {arXiv:1902.00507}
  \BibitemShut {NoStop}%
\bibitem [{\citenamefont {Varjas}\ \emph {et~al.}(2019)\citenamefont {Varjas},
  \citenamefont {Lau}, \citenamefont {P{\"{o}}yh{\"{o}}nen}, \citenamefont
  {Akhmerov}, \citenamefont {Pikulin},\ and\ \citenamefont
  {Fulga}}]{Varjas:2019jd}%
  \BibitemOpen
  \bibfield  {author} {\bibinfo {author} {\bibfnamefont {D.}~\bibnamefont
  {Varjas}}, \bibinfo {author} {\bibfnamefont {A.}~\bibnamefont {Lau}},
  \bibinfo {author} {\bibfnamefont {K.}~\bibnamefont {P{\"{o}}yh{\"{o}}nen}},
  \bibinfo {author} {\bibfnamefont {A.~R.}\ \bibnamefont {Akhmerov}}, \bibinfo
  {author} {\bibfnamefont {D.~I.}\ \bibnamefont {Pikulin}},\ and\ \bibinfo
  {author} {\bibfnamefont {I.~C.}\ \bibnamefont {Fulga}},\ }\bibfield  {title}
  {\bibinfo {title} {{Topological Phases without Crystalline Counterparts}},\
  }\href {https://doi.org/10.1103/PhysRevLett.123.196401} {\bibfield  {journal}
  {\bibinfo  {journal} {Phys. Rev. Lett.}\ }\textbf {\bibinfo {volume} {123}},\
  \bibinfo {pages} {196401} (\bibinfo {year} {2019})}\BibitemShut {NoStop}%
\bibitem [{\citenamefont {Chen}\ \emph {et~al.}(2020)\citenamefont {Chen},
  \citenamefont {Chen}, \citenamefont {Gao}, \citenamefont {Zhou},\ and\
  \citenamefont {Xu}}]{Chen:2020jo}%
  \BibitemOpen
  \bibfield  {author} {\bibinfo {author} {\bibfnamefont {R.}~\bibnamefont
  {Chen}}, \bibinfo {author} {\bibfnamefont {C.-Z.}\ \bibnamefont {Chen}},
  \bibinfo {author} {\bibfnamefont {J.-H.}\ \bibnamefont {Gao}}, \bibinfo
  {author} {\bibfnamefont {B.}~\bibnamefont {Zhou}},\ and\ \bibinfo {author}
  {\bibfnamefont {D.-H.}\ \bibnamefont {Xu}},\ }\bibfield  {title} {\bibinfo
  {title} {{Higher-Order Topological Insulators in Quasicrystals}},\ }\href
  {https://doi.org/10.1103/PhysRevLett.124.036803} {\bibfield  {journal}
  {\bibinfo  {journal} {Phys. Rev. Lett.}\ }\textbf {\bibinfo {volume} {124}},\
  \bibinfo {pages} {036803} (\bibinfo {year} {2020})}\BibitemShut {NoStop}%
\bibitem [{\citenamefont {Serra-Garcia}\ \emph {et~al.}(2018)\citenamefont
  {Serra-Garcia}, \citenamefont {Peri}, \citenamefont {S{\"{u}}sstrunk},
  \citenamefont {Bilal}, \citenamefont {Larsen}, \citenamefont {Villanueva},\
  and\ \citenamefont {Huber}}]{Serra:2018dj}%
  \BibitemOpen
  \bibfield  {author} {\bibinfo {author} {\bibfnamefont {M.}~\bibnamefont
  {Serra-Garcia}}, \bibinfo {author} {\bibfnamefont {V.}~\bibnamefont {Peri}},
  \bibinfo {author} {\bibfnamefont {R.}~\bibnamefont {S{\"{u}}sstrunk}},
  \bibinfo {author} {\bibfnamefont {O.~R.}\ \bibnamefont {Bilal}}, \bibinfo
  {author} {\bibfnamefont {T.}~\bibnamefont {Larsen}}, \bibinfo {author}
  {\bibfnamefont {L.~G.}\ \bibnamefont {Villanueva}},\ and\ \bibinfo {author}
  {\bibfnamefont {S.~D.}\ \bibnamefont {Huber}},\ }\bibfield  {title} {\bibinfo
  {title} {{Observation of a phononic quadrupole topological insulator}},\
  }\href {https://doi.org/10.1038/nature25156} {\bibfield  {journal} {\bibinfo
  {journal} {Nature}\ }\textbf {\bibinfo {volume} {555}},\ \bibinfo {pages}
  {342} (\bibinfo {year} {2018})}\BibitemShut {NoStop}%
\bibitem [{\citenamefont {Kempkes}\ \emph {et~al.}(2019)\citenamefont
  {Kempkes}, \citenamefont {Slot}, \citenamefont {van~den Broeke},
  \citenamefont {Capiod}, \citenamefont {Benalcazar}, \citenamefont
  {Vanmaekelbergh}, \citenamefont {Bercioux}, \citenamefont {Swart},\ and\
  \citenamefont {{Morais Smith}}}]{Kempkes:2019cl}%
  \BibitemOpen
  \bibfield  {author} {\bibinfo {author} {\bibfnamefont {S.~N.}\ \bibnamefont
  {Kempkes}}, \bibinfo {author} {\bibfnamefont {M.~R.}\ \bibnamefont {Slot}},
  \bibinfo {author} {\bibfnamefont {J.~J.}\ \bibnamefont {van~den Broeke}},
  \bibinfo {author} {\bibfnamefont {P.}~\bibnamefont {Capiod}}, \bibinfo
  {author} {\bibfnamefont {W.~A.}\ \bibnamefont {Benalcazar}}, \bibinfo
  {author} {\bibfnamefont {D.}~\bibnamefont {Vanmaekelbergh}}, \bibinfo
  {author} {\bibfnamefont {D.}~\bibnamefont {Bercioux}}, \bibinfo {author}
  {\bibfnamefont {I.}~\bibnamefont {Swart}},\ and\ \bibinfo {author}
  {\bibfnamefont {C.}~\bibnamefont {{Morais Smith}}},\ }\bibfield  {title}
  {\bibinfo {title} {{Robust zero-energy modes in an electronic higher-order
  topological insulator}},\ }\href {https://doi.org/10.1038/s41563-019-0483-4}
  {\bibfield  {journal} {\bibinfo  {journal} {Nat. Mater.}\ }\textbf {\bibinfo
  {volume} {18}},\ \bibinfo {pages} {1292} (\bibinfo {year}
  {2019})}\BibitemShut {NoStop}%
\bibitem [{\citenamefont {Imhof}\ \emph {et~al.}(2018)\citenamefont {Imhof},
  \citenamefont {Berger}, \citenamefont {Bayer}, \citenamefont {Brehm},
  \citenamefont {Molenkamp}, \citenamefont {Kiessling}, \citenamefont
  {Schindler}, \citenamefont {Lee}, \citenamefont {Greiter}, \citenamefont
  {Neupert},\ and\ \citenamefont {Thomale}}]{Imhof:2018gs}%
  \BibitemOpen
  \bibfield  {author} {\bibinfo {author} {\bibfnamefont {S.}~\bibnamefont
  {Imhof}}, \bibinfo {author} {\bibfnamefont {C.}~\bibnamefont {Berger}},
  \bibinfo {author} {\bibfnamefont {F.}~\bibnamefont {Bayer}}, \bibinfo
  {author} {\bibfnamefont {J.}~\bibnamefont {Brehm}}, \bibinfo {author}
  {\bibfnamefont {L.~W.}\ \bibnamefont {Molenkamp}}, \bibinfo {author}
  {\bibfnamefont {T.}~\bibnamefont {Kiessling}}, \bibinfo {author}
  {\bibfnamefont {F.}~\bibnamefont {Schindler}}, \bibinfo {author}
  {\bibfnamefont {C.~H.}\ \bibnamefont {Lee}}, \bibinfo {author} {\bibfnamefont
  {M.}~\bibnamefont {Greiter}}, \bibinfo {author} {\bibfnamefont
  {T.}~\bibnamefont {Neupert}},\ and\ \bibinfo {author} {\bibfnamefont
  {R.}~\bibnamefont {Thomale}},\ }\bibfield  {title} {\bibinfo {title}
  {{Topolectrical-circuit realization of topological corner modes}},\ }\href
  {https://doi.org/10.1038/s41567-018-0246-1} {\bibfield  {journal} {\bibinfo
  {journal} {Nat. Phys.}\ }\textbf {\bibinfo {volume} {14}},\ \bibinfo {pages}
  {925} (\bibinfo {year} {2018})}\BibitemShut {NoStop}%
\bibitem [{\citenamefont {Peterson}\ \emph {et~al.}(2018)\citenamefont
  {Peterson}, \citenamefont {Benalcazar}, \citenamefont {Hughes},\ and\
  \citenamefont {Bahl}}]{Peterson:2018kz}%
  \BibitemOpen
  \bibfield  {author} {\bibinfo {author} {\bibfnamefont {C.~W.}\ \bibnamefont
  {Peterson}}, \bibinfo {author} {\bibfnamefont {W.~A.}\ \bibnamefont
  {Benalcazar}}, \bibinfo {author} {\bibfnamefont {T.~L.}\ \bibnamefont
  {Hughes}},\ and\ \bibinfo {author} {\bibfnamefont {G.}~\bibnamefont {Bahl}},\
  }\bibfield  {title} {\bibinfo {title} {{A quantized microwave quadrupole
  insulator with topologically protected corner states}},\ }\href
  {https://doi.org/10.1038/nature25777} {\bibfield  {journal} {\bibinfo
  {journal} {Nature}\ }\textbf {\bibinfo {volume} {555}},\ \bibinfo {pages}
  {346} (\bibinfo {year} {2018})}\BibitemShut {NoStop}%
\bibitem [{\citenamefont {Khalaf}\ \emph {et~al.}(2018)\citenamefont {Khalaf},
  \citenamefont {Po}, \citenamefont {Vishwanath},\ and\ \citenamefont
  {Watanabe}}]{Khalaf:2018hq}%
  \BibitemOpen
  \bibfield  {author} {\bibinfo {author} {\bibfnamefont {E.}~\bibnamefont
  {Khalaf}}, \bibinfo {author} {\bibfnamefont {H.~C.}\ \bibnamefont {Po}},
  \bibinfo {author} {\bibfnamefont {A.}~\bibnamefont {Vishwanath}},\ and\
  \bibinfo {author} {\bibfnamefont {H.}~\bibnamefont {Watanabe}},\ }\bibfield
  {title} {\bibinfo {title} {{Symmetry Indicators and Anomalous Surface States
  of Topological Crystalline Insulators}},\ }\href
  {https://doi.org/10.1103/PhysRevX.8.031070} {\bibfield  {journal} {\bibinfo
  {journal} {Phys. Rev. X}\ }\textbf {\bibinfo {volume} {8}},\ \bibinfo {pages}
  {031070} (\bibinfo {year} {2018})}\BibitemShut {NoStop}%
\bibitem [{\citenamefont {Schindler}\ \emph {et~al.}(2019)\citenamefont
  {Schindler}, \citenamefont {Brzezi{\'{n}}ska}, \citenamefont {Benalcazar},
  \citenamefont {Iraola}, \citenamefont {Bouhon}, \citenamefont {Tsirkin},
  \citenamefont {Vergniory},\ and\ \citenamefont {Neupert}}]{Schindler:2019jk}%
  \BibitemOpen
  \bibfield  {author} {\bibinfo {author} {\bibfnamefont {F.}~\bibnamefont
  {Schindler}}, \bibinfo {author} {\bibfnamefont {M.}~\bibnamefont
  {Brzezi{\'{n}}ska}}, \bibinfo {author} {\bibfnamefont {W.~A.}\ \bibnamefont
  {Benalcazar}}, \bibinfo {author} {\bibfnamefont {M.}~\bibnamefont {Iraola}},
  \bibinfo {author} {\bibfnamefont {A.}~\bibnamefont {Bouhon}}, \bibinfo
  {author} {\bibfnamefont {S.~S.}\ \bibnamefont {Tsirkin}}, \bibinfo {author}
  {\bibfnamefont {M.~G.}\ \bibnamefont {Vergniory}},\ and\ \bibinfo {author}
  {\bibfnamefont {T.}~\bibnamefont {Neupert}},\ }\bibfield  {title} {\bibinfo
  {title} {{Fractional corner charges in spin-orbit coupled crystals}},\ }\href
  {https://doi.org/10.1103/PhysRevResearch.1.033074} {\bibfield  {journal}
  {\bibinfo  {journal} {Phys. Rev. Research}\ }\textbf {\bibinfo {volume}
  {1}},\ \bibinfo {pages} {033074} (\bibinfo {year} {2019})}\BibitemShut
  {NoStop}%
\bibitem [{\citenamefont {Shiozaki}()}]{Shiozaki2019}%
  \BibitemOpen
  \bibfield  {author} {\bibinfo {author} {\bibfnamefont {K.}~\bibnamefont
  {Shiozaki}},\ }\bibfield  {title} {\bibinfo {title} {{Variants of the
  symmetry-based indicator}},\ }\href@noop {} {\ }\Eprint
  {https://arxiv.org/abs/1907.13632} {arXiv:1907.13632} \BibitemShut {NoStop}%
\bibitem [{\citenamefont {Geier}\ \emph {et~al.}()\citenamefont {Geier},
  \citenamefont {Brouwer},\ and\ \citenamefont {Trifunovic}}]{Geier2019}%
  \BibitemOpen
  \bibfield  {author} {\bibinfo {author} {\bibfnamefont {M.}~\bibnamefont
  {Geier}}, \bibinfo {author} {\bibfnamefont {P.~W.}\ \bibnamefont {Brouwer}},\
  and\ \bibinfo {author} {\bibfnamefont {L.}~\bibnamefont {Trifunovic}},\
  }\bibfield  {title} {\bibinfo {title} {{Symmetry-based indicators for
  topological Bogoliubov-de Gennes Hamiltonians}},\ }\Eprint
  {https://arxiv.org/abs/1910.11271} {arXiv:1910.11271} \BibitemShut {NoStop}%
\bibitem [{\citenamefont {Mondragon-Shem}\ and\ \citenamefont
  {Hughes}()}]{Mondragon-Shem2019}%
  \BibitemOpen
  \bibfield  {author} {\bibinfo {author} {\bibfnamefont {I.}~\bibnamefont
  {Mondragon-Shem}}\ and\ \bibinfo {author} {\bibfnamefont {T.~L.}\
  \bibnamefont {Hughes}},\ }\bibfield  {title} {\bibinfo {title} {{Robust
  topological invariants of topological crystalline phases in the presence of
  impurities}},\ }\Eprint {https://arxiv.org/abs/1906.11847} {arXiv:1906.11847}
  \BibitemShut {NoStop}%
\bibitem [{Note1()}]{Note1}%
  \BibitemOpen
  \bibinfo {note} {This Fourier transform convention implies a certain choice
  of basis functions~\cite {Alexandradinata:2014ju,Alexandradinata:2016kb}. For
  details on the basis choice cf.\ Appendix~\ref
  {sec:unit_cell_restriction}.}\BibitemShut {Stop}%
\bibitem [{\citenamefont {Ono}\ \emph {et~al.}()\citenamefont {Ono},
  \citenamefont {Po},\ and\ \citenamefont {Watanabe}}]{Ono2019}%
  \BibitemOpen
  \bibfield  {author} {\bibinfo {author} {\bibfnamefont {S.}~\bibnamefont
  {Ono}}, \bibinfo {author} {\bibfnamefont {H.~C.}\ \bibnamefont {Po}},\ and\
  \bibinfo {author} {\bibfnamefont {H.}~\bibnamefont {Watanabe}},\ }\bibfield
  {title} {\bibinfo {title} {{Refined symmetry indicators for topological
  superconductors in all space groups}},\ }\Eprint
  {https://arxiv.org/abs/1909.09634} {arXiv:1909.09634} \BibitemShut {NoStop}%
\bibitem [{\citenamefont {Jackiw}\ and\ \citenamefont
  {Rebbi}(1976)}]{Jackiw:1976ky}%
  \BibitemOpen
  \bibfield  {author} {\bibinfo {author} {\bibfnamefont {R.}~\bibnamefont
  {Jackiw}}\ and\ \bibinfo {author} {\bibfnamefont {C.}~\bibnamefont {Rebbi}},\
  }\bibfield  {title} {\bibinfo {title} {{Solitons with fermion number
  \textonehalf{}}},\ }\href {https://doi.org/10.1103/PhysRevD.13.3398}
  {\bibfield  {journal} {\bibinfo  {journal} {Phys. Rev. D}\ }\textbf {\bibinfo
  {volume} {13}},\ \bibinfo {pages} {3398} (\bibinfo {year}
  {1976})}\BibitemShut {NoStop}%
\bibitem [{\citenamefont {Su}\ \emph {et~al.}(1979)\citenamefont {Su},
  \citenamefont {Schrieffer},\ and\ \citenamefont {Heeger}}]{Su:1979cb}%
  \BibitemOpen
  \bibfield  {author} {\bibinfo {author} {\bibfnamefont {W.~P.}\ \bibnamefont
  {Su}}, \bibinfo {author} {\bibfnamefont {J.~R.}\ \bibnamefont {Schrieffer}},\
  and\ \bibinfo {author} {\bibfnamefont {A.~J.}\ \bibnamefont {Heeger}},\
  }\bibfield  {title} {\bibinfo {title} {{Solitons in Polyacetylene}},\ }\href
  {https://doi.org/10.1103/PhysRevLett.42.1698} {\bibfield  {journal} {\bibinfo
   {journal} {Phys. Rev. Lett.}\ }\textbf {\bibinfo {volume} {42}},\ \bibinfo
  {pages} {1698} (\bibinfo {year} {1979})}\BibitemShut {NoStop}%
\bibitem [{Note2()}]{Note2}%
  \BibitemOpen
  \bibinfo {note} {The $ \protect \mathbf {k} \cdot \protect \boldsymbol
  {\sigma }$ term may always be brought to this form because the relative sign
  of $ k_1 $ and $ k_2 $ is altered by a basis rotation $ \protect \mathcal
  {H}^{\alpha }(\protect \mathbf {k}) \to \sigma _1 \protect \mathcal
  {H}^{\alpha }(\protect \mathbf {k}) \sigma _1 $ for which we would change the
  sign assigned to $ m_{\alpha } $.}\BibitemShut {Stop}%
\bibitem [{Note3()}]{Note3}%
  \BibitemOpen
  \bibinfo {note} {Allowing symmetric terms of the form $ \sigma _3 \otimes M
  $, for example, where $ M = M^T $ and $ OMO^T = \DOTSB \bigoplus@ \slimits@
  _{\alpha } m_{\alpha } $ does not change the resulting edge theory but its
  derivation (Appendix~\ref {sec:derivation_edge_theory}) requires a different
  ansatz.}\BibitemShut {Stop}%
\bibitem [{\citenamefont {Fang}\ and\ \citenamefont {Fu}(2019)}]{Fang:2019kk}%
  \BibitemOpen
  \bibfield  {author} {\bibinfo {author} {\bibfnamefont {C.}~\bibnamefont
  {Fang}}\ and\ \bibinfo {author} {\bibfnamefont {L.}~\bibnamefont {Fu}},\
  }\bibfield  {title} {\bibinfo {title} {{New classes of topological
  crystalline insulators having surface rotation anomaly}},\ }\href
  {https://doi.org/10.1126/sciadv.aat2374} {\bibfield  {journal} {\bibinfo
  {journal} {Sci. Adv.}\ }\textbf {\bibinfo {volume} {5}},\ \bibinfo {pages}
  {eaat2374} (\bibinfo {year} {2019})}\BibitemShut {NoStop}%
\bibitem [{Note4()}]{Note4}%
  \BibitemOpen
  \bibinfo {note} {That the sign of the mass at $ \protect \mathbf
  {X}^{(\protect \mathbf {\prime })}$ is not set by the occupied rotation
  eigenvalues is also one source of the ``surface-state ambiguity'' tabulated
  in Ref.~\protect \rev@citealp {Khalaf:2018hq}, which is where the surface
  signature of a nontrivial bulk is not uniquely determined from the symmetry
  indicators. Once we specify that $ \Delta \protect \mathrm {Ch}= 0 $ (using
  information beyond symmetry indicators alone), there will be no ambiguity in
  the surface-state of this system.}\BibitemShut {Stop}%
\bibitem [{Note5()}]{Note5}%
  \BibitemOpen
  \bibinfo {note} {This $ 2 \times 2 $ block-diagonal structure is possible by
  choosing the original stacking order in Eq.~\protect \textup {\hbox
  {\mathsurround \z@ \protect \normalfont (\ignorespaces \ref
  {eq:Dirac_stack}\unskip \@@italiccorr )}} to be such that Dirac Hamiltonians
  with $ m_{\alpha } < 0 $ in the bulk appear first.}\BibitemShut {Stop}%
\bibitem [{\citenamefont {Fu}\ and\ \citenamefont {Kane}(2008)}]{Fu:2008gu}%
  \BibitemOpen
  \bibfield  {author} {\bibinfo {author} {\bibfnamefont {L.}~\bibnamefont
  {Fu}}\ and\ \bibinfo {author} {\bibfnamefont {C.~L.}\ \bibnamefont {Kane}},\
  }\bibfield  {title} {\bibinfo {title} {{Superconducting Proximity Effect and
  Majorana Fermions at the Surface of a Topological Insulator}},\ }\href
  {https://doi.org/10.1103/PhysRevLett.100.096407} {\bibfield  {journal}
  {\bibinfo  {journal} {Phys. Rev. Lett.}\ }\textbf {\bibinfo {volume} {100}},\
  \bibinfo {pages} {096407} (\bibinfo {year} {2008})}\BibitemShut {NoStop}%
\bibitem [{\citenamefont {Kitaev}(2001)}]{Kitaev:2001gb}%
  \BibitemOpen
  \bibfield  {author} {\bibinfo {author} {\bibfnamefont {A.~Y.}\ \bibnamefont
  {Kitaev}},\ }\bibfield  {title} {\bibinfo {title} {{Unpaired Majorana
  fermions in quantum wires}},\ }\href
  {https://doi.org/10.1070/1063-7869/44/10S/S29} {\bibfield  {journal}
  {\bibinfo  {journal} {Physics-Uspekhi}\ }\textbf {\bibinfo {volume} {44}},\
  \bibinfo {pages} {131} (\bibinfo {year} {2001})}\BibitemShut {NoStop}%
\bibitem [{\citenamefont {Budich}\ and\ \citenamefont
  {Ardonne}(2013)}]{Budich:2013it}%
  \BibitemOpen
  \bibfield  {author} {\bibinfo {author} {\bibfnamefont {J.~C.}\ \bibnamefont
  {Budich}}\ and\ \bibinfo {author} {\bibfnamefont {E.}~\bibnamefont
  {Ardonne}},\ }\bibfield  {title} {\bibinfo {title} {{Equivalent topological
  invariants for one-dimensional Majorana wires in symmetry class D}},\ }\href
  {https://doi.org/10.1103/PhysRevB.88.075419} {\bibfield  {journal} {\bibinfo
  {journal} {Phys. Rev. B}\ }\textbf {\bibinfo {volume} {88}},\ \bibinfo
  {pages} {075419} (\bibinfo {year} {2013})}\BibitemShut {NoStop}%
\bibitem [{Note6()}]{Note6}%
  \BibitemOpen
  \bibinfo {note} {$ \protect \qopname \relax m{det}u_{n,\protect \mathbf
  {c}}^{} =+ 1 $ corresponding to the case without boundary Majorana bound
  states is also consistent with the fact that if $ \protect \qopname \relax
  m{det}u_{n,\protect \mathbf {c}}^{} =+ 1 $, then for a circular boundary the
  transformation $ h_{\protect \mathbf {r},\protect \mathbf {k}} \to h_{R_n
  \protect \mathbf {r},R_n \protect \mathbf {k}} $ could be achieved via many
  infinitesimal orthogonal transformations (connected to the identity). There
  would therefore exist a continuous deformation between the two Hamiltonians
  that does not close any gaps and respects all the symmetries, thereby making
  them topologically equivalent.}\BibitemShut {Stop}%
\bibitem [{\citenamefont {C\ifmmode \u{a}\else \u{a}\fi{}lug\ifmmode~\u{a}\else
  \u{a}\fi{}ru}\ \emph {et~al.}(2019)\citenamefont {C\ifmmode \u{a}\else
  \u{a}\fi{}lug\ifmmode~\u{a}\else \u{a}\fi{}ru}, \citenamefont {Juri\ifmmode
  \check{c}\else \v{c}\fi{}i\ifmmode~\acute{c}\else \'{c}\fi{}},\ and\
  \citenamefont {Roy}}]{Juricic:2019}%
  \BibitemOpen
  \bibfield  {author} {\bibinfo {author} {\bibfnamefont {D.}~\bibnamefont
  {C\ifmmode \u{a}\else \u{a}\fi{}lug\ifmmode~\u{a}\else \u{a}\fi{}ru}},
  \bibinfo {author} {\bibfnamefont {V.}~\bibnamefont {Juri\ifmmode
  \check{c}\else \v{c}\fi{}i\ifmmode~\acute{c}\else \'{c}\fi{}}},\ and\
  \bibinfo {author} {\bibfnamefont {B.}~\bibnamefont {Roy}},\ }\bibfield
  {title} {\bibinfo {title} {{Higher-order topological phases: A general
  principle of construction}},\ }\href
  {https://doi.org/10.1103/PhysRevB.99.041301} {\bibfield  {journal} {\bibinfo
  {journal} {Phys. Rev. B}\ }\textbf {\bibinfo {volume} {99}},\ \bibinfo
  {pages} {041301} (\bibinfo {year} {2019})}\BibitemShut {NoStop}%
\bibitem [{Note7()}]{Note7}%
  \BibitemOpen
  \bibinfo {note} {Furthermore, a set $ \Omega $ will not always be closed
  under rotation $ \Omega _i \to u_{n,\protect \mathbf {c}}^{} \Omega _i
  u_{n,\protect \mathbf {c}}^T \notin \Omega $ when all the Dirac Hamiltonians
  in the stack are allowed to be different.}\BibitemShut {Stop}%
\bibitem [{\citenamefont {Goringe}\ \emph {et~al.}(1997)\citenamefont
  {Goringe}, \citenamefont {Bowler},\ and\ \citenamefont
  {Hern{\'{a}}ndez}}]{Goringe:1997bx}%
  \BibitemOpen
  \bibfield  {author} {\bibinfo {author} {\bibfnamefont {C.~M.}\ \bibnamefont
  {Goringe}}, \bibinfo {author} {\bibfnamefont {D.~R.}\ \bibnamefont
  {Bowler}},\ and\ \bibinfo {author} {\bibfnamefont {E.}~\bibnamefont
  {Hern{\'{a}}ndez}},\ }\bibfield  {title} {\bibinfo {title} {{Tight-binding
  modelling of materials}},\ }\href
  {https://doi.org/10.1088/0034-4885/60/12/001} {\bibfield  {journal} {\bibinfo
   {journal} {Reports Prog. Phys.}\ }\textbf {\bibinfo {volume} {60}},\
  \bibinfo {pages} {1447} (\bibinfo {year} {1997})}\BibitemShut {NoStop}%
\bibitem [{\citenamefont {Alexandradinata}\ \emph {et~al.}(2016)\citenamefont
  {Alexandradinata}, \citenamefont {Wang},\ and\ \citenamefont
  {Bernevig}}]{Alexandradinata:2016kb}%
  \BibitemOpen
  \bibfield  {author} {\bibinfo {author} {\bibfnamefont {A.}~\bibnamefont
  {Alexandradinata}}, \bibinfo {author} {\bibfnamefont {Z.}~\bibnamefont
  {Wang}},\ and\ \bibinfo {author} {\bibfnamefont {B.~A.}\ \bibnamefont
  {Bernevig}},\ }\bibfield  {title} {\bibinfo {title} {{Topological Insulators
  from Group Cohomology}},\ }\href {https://doi.org/10.1103/PhysRevX.6.021008}
  {\bibfield  {journal} {\bibinfo  {journal} {Phys. Rev. X}\ }\textbf {\bibinfo
  {volume} {6}},\ \bibinfo {pages} {021008} (\bibinfo {year}
  {2016})}\BibitemShut {NoStop}%
\bibitem [{\citenamefont {Alexandradinata}\ \emph
  {et~al.}(2014{\natexlab{b}})\citenamefont {Alexandradinata}, \citenamefont
  {Dai},\ and\ \citenamefont {Bernevig}}]{Alexandradinata:2014ju}%
  \BibitemOpen
  \bibfield  {author} {\bibinfo {author} {\bibfnamefont {A.}~\bibnamefont
  {Alexandradinata}}, \bibinfo {author} {\bibfnamefont {X.}~\bibnamefont
  {Dai}},\ and\ \bibinfo {author} {\bibfnamefont {B.~A.}\ \bibnamefont
  {Bernevig}},\ }\bibfield  {title} {\bibinfo {title} {{Wilson-loop
  characterization of inversion-symmetric topological insulators}},\ }\href
  {https://doi.org/10.1103/PhysRevB.89.155114} {\bibfield  {journal} {\bibinfo
  {journal} {Phys. Rev. B}\ }\textbf {\bibinfo {volume} {89}},\ \bibinfo
  {pages} {155114} (\bibinfo {year} {2014}{\natexlab{b}})}\BibitemShut
  {NoStop}%
\bibitem [{\citenamefont {Khalaf}\ \emph {et~al.}()\citenamefont {Khalaf},
  \citenamefont {Benalcazar}, \citenamefont {Hughes},\ and\ \citenamefont
  {Queiroz}}]{Khalaf2019}%
  \BibitemOpen
  \bibfield  {author} {\bibinfo {author} {\bibfnamefont {E.}~\bibnamefont
  {Khalaf}}, \bibinfo {author} {\bibfnamefont {W.~A.}\ \bibnamefont
  {Benalcazar}}, \bibinfo {author} {\bibfnamefont {T.~L.}\ \bibnamefont
  {Hughes}},\ and\ \bibinfo {author} {\bibfnamefont {R.}~\bibnamefont
  {Queiroz}},\ }\bibfield  {title} {\bibinfo {title} {Boundary-obstructed
  topological phases},\ }\Eprint {https://arxiv.org/abs/1908.00011}
  {arXiv:1908.00011} \BibitemShut {NoStop}%
\end{thebibliography}%

\end{document}